\RequirePackage{fix-cm}
\documentclass[twocolumn]{svjour3}          
\smartqed  
\usepackage[usenames]{color}
\usepackage[table]{xcolor}
\usepackage{soul}
\usepackage{textcomp}
\usepackage{multicol} 
\usepackage{multirow} 
\usepackage{xspace}
\usepackage{amsmath} 
\usepackage{graphicx}
\usepackage{breqn}
\usepackage{algorithm}
\usepackage{algorithmic}
\usepackage{bbm}
\newcommand{\data}[1]{{\textsl{#1}}}
\newcommand{\algo}[1]{{\textsf{#1}}}

\newcommand{\HL}[1]{\bfseries{#1}}

\def\G{\mbox{$\mathcal{G}$}} 
\def\V{\mbox{$\mathcal{V}$}} 
\def\E{\mbox{$\mathcal{E}$}} 
\def\A{\mbox{$\mathbf{A}$}}  
\def\Dout{\mbox{$\mathbf{D}_{out}$}}  
\def\e{\mbox{$\mathbf{e}$}}  

\def\d{\mbox{$\alpha$}}  
\def\r{\mbox{$\mathbf{r}$}}  
\def\M{\mbox{$\mathbf{S}$}}  

\def\L{\mbox{$\mathcal{L}$}} 

\journalname{SNAM}
\begin{document}

\title{Lurking in Social Networks: Topology-based Analysis and Ranking Methods
\thanks{An abridged version of this paper appeared in~\cite{ASONAM}.}}


\author{Andrea Tagarelli \and Roberto Interdonato}


\institute{Dept. Computer Engineering, Modeling, Electronics,   and Systems Sciences. 
 University of Calabria, Italy\\ 
 \email{\{tagarelli,rinterdonato\}@dimes.unical.it}
}

\date{Received: date / Accepted: date}

\maketitle

\begin{abstract} 
The massive presence of  silent members in online communities, the so-called \textit{lurkers}, 
 has long attracted the attention of researchers in social science, 
cognitive psychology, and computer-human interaction.   
However, the study of lurking phenomena  represents an unexplored opportunity of research in 
data mining, information retrieval and related fields.  
In this paper, we take a first step towards 
the formal specification and  analysis of lurking   in social networks.   
We address the new problem of \textit{lurker ranking} 
and propose the first centrality methods specifically conceived for ranking lurkers in social networks. 
Our approach utilizes only the network topology without probing into text contents or user relationships related to media. 
Using   Twitter, Flickr, FriendFeed and GooglePlus as cases in point, 
our methods' performance was evaluated against 
data-driven rankings as well as  existing centrality methods, including 
the classic PageRank and alpha-centrality. 
Empirical evidence has shown the significance    
of our lurker ranking approach, 
and its uniqueness in 
effectively identifying and ranking  lurkers in an online social network.  

\keywords{lurker ranking \and lurking coefficient \and LurkerRank \and delurking}
\end{abstract}

\section{Introduction}
\label{sec:intro}
 The majority of members of online communities play a passive or silent role as individuals that do not readily contribute to the shared online space. Such individuals are   called \emph{lurkers}, since they belong to a community but remain quite unnoticed while watching, reading or, in general, benefiting from others' information or services without significantly giving back to the community.  

Lurking characterization in online communities  has been a controversial issue from a  social science and computer-human interaction perspective~\cite{Edelmann13}. Since the early works on social motivations and implications of lurking~\cite{NonneckeP00,PreeceNA04},    
one common perception of lurking is that based on the infrequency of active participation   to the community life, 
but other definitions have been given under the hypotheses of  free-riding~\cite{KollockS96}, 
legitimate peripheral participation~\cite{LaveW91,HalfakerKT13}, individual information strategy of microlearning~\cite{Kahnwald2006}, and knowledge sharing barriers (e.g., interpersonal or technological barriers)~\cite{Ardichvili08}. 
Lurkers might also be perceived as a menace for the cyberspace as they maliciously feed on others' intellects. For instance,  in P2P file sharing systems~\cite{DhungelWSR08}, lurking may correspond to a leeching behavior 
whenever a user wastes valuable bandwidth by downloading much more than what s/he uploads.  
In the realm of online social networks (OSNs), negative views of the lurkers have been however supplanted with a neutral or even marginally positive view. A neutral perception of lurkers is related to the fact that their silent presence is seen as harmless and reflects a subjective reticence (rather than malicious motivations) to contribute to the community wisdom; half of times, a lurker simply feels that gathering information by browsing is enough without the need of being further involved in the community~\cite{PreeceNA04}. However, lurking can be expected or even encouraged because it allows users (especially newcomers) to learn or improve their understanding of the etiquette of an online community before they can decide to provide a valuable contribution over time.

Lurking is responsible for a \textit{participation inequality phenomenon} that is shared by all large-scale online communities. This phenomenon is explained by the so-called ``1:9:90'' rule, which states that while 90\% of users do not actively contribute,  9\% of users may contribute (i.e., comment, like or edit)  from time to time,  
and only 1\% of users create the vast majority of social content~\cite{NonneckeP00,PreeceNA04}. 
Consequently,  such inequities lead to a biased understanding of the community,  
whereby a major  risk is that we will never hear from the silent majority of lurkers. 
Therefore, a challenge is to attract, or \emph{de-lurk}, the crowd of lurkers, whereby 
 online advertising strategies should be tailored to the lurkers' behavioral profile. 
Moreover, since lurkers have knowledge about the online community (as a result of the substantial time they dedicate towards learning from the community), delurking can mainly be  seen as a mix of strategies aimed at encouraging lurkers to return their acquired social capital, through a more active participation to the community life.

Understanding user behaviors  has long been studied in online social networks. 
A key element that is shared by all studies is the use of a social graph model as the basic tool to represent relationships among users~\cite{WassermanF94}. Relationships, or \textit{ties}~\cite{Granovetter73}, can vary over a spectrum that include friendships and followships~\cite{KumarNT06,AhnHKMJ07,MisloveMGDB07,KwakLPM10,ChaHBG10,WilsonSPZ12}, visible interactions~\cite{ChunKEAMJ08,LeskovecH08,ViswanathMCG09,WilsonSPZ12,de2013analyzing}, and  latent interactions (based on, e.g., browsing profiles or  clickstream data)~\cite{SchneiderFKW09,BenevenutoRCA09,JiangWWHSDZ13}.

Surprisingly, despite the fact that  lurking  has been recognized and surveyed 
in social sciences,  
we are not aware of any previous study  
on lurking in social networks from a graph data management or mining perspective. 
Particularly, no computational method has been so far conceived to determine, and eventually, rank lurkers in an OSN graph. 
Note that, beyond the frequent yet trivial case of users that exhibit a peripheral unstructured  membership,   hidden forms of lurking are massively present in OSNs, which make  it challenging to mine  lurkers.  
While lurking is hard  to track from a personal dispositional viewpoint, 
it appears that ranking lurkers is still possible by handling the 
situational variables that are related to the network of  relationships between members.  
Moreover, a well-founded principle of eigenvector centrality, which is adopted in this work, will enable 
the determination of each node's lurking score in function of the lurking scores of the nodes that it is connected to.

One may notice that ranking influential people is clearly valuable as we  naturally  tend  to follow leaders and learn from them, and conversely wonder \emph{``why ranking lurkers?''}.  
We argue that scoring  community members  as lurkers, rather than limiting to solely recognize (potential or actual) lurkers, should be seen as essential to determine the contingencies in the network under which different lurking behaviors occur, and ultimately  to aid devising both generic and ad-hoc  de-lurking plans and strategies.  
In effect, ordering members by decreasing lurking score would enable to manage priority in de-lurking applications, to identify the sub-communities particularly affected by lurkers, 
and to define personalized triggers of active participation. For example, lurkers of a given sub-community developed around an entity of interest (e.g., a person, or theme)    
would welcome  messages that highlight the key topics (a service that is already delivered to its users by Twitter, for example),   social events that describe how to approach a discussion in a forum or to start off your own project in a collaboration network, or introduce the role of forum moderators  or  team leaders. Moreover, in order  to alleviate  information overload, which is recognized as a major negative factor for participation, various mechanisms of filtering (e.g., recommending threads of discussion, providing visual maps of the   categories of activities) 
or promotion of lightweight contribution tasks (e.g.,~\cite{FarzanDB10}) could be applied with the ultimate goal of revealing the lurker's value (i.e., ideas, opinions, expertise) to the community.

\textbf{Contributions.\ } 
This paper extends our previous work~\cite{ASONAM}, in which we took a 
first step towards mining  lurkers in OSNs.    
We scrutinize the concept of lurking in OSNs to determine the essential criteria 
that can be taken  as the basis for mining  lurkers.  
We lay out a    \textit{topology-driven lurking}  definition upon a network  representation 
modeling  the directed relationships from information-producer to information-consumer.  
Our lurking definition is based on three principles that respectively express in/out-degree related properties 
of a given node, its in-neighborhood, and its out-neighbor\-hood. 
 We also define a lurking coefficient  to characterize 
 the topology of a network in terms of lurking degree. 

The proposed lurking definition  lends itself naturally to score  the  users in an OSN according to their lurking behavior, thus enabling the development of ranking mechanisms. We hence focus on the problem of \textit{lurker ranking}, and define  three formulations of it that rely on the different  aspects of our topology-driven lurking concept. By  resorting to classic link-analysis ranking algorithms,  
 PageRank and alpha-centrality, we provide a complete specification of 
  lurker ranking  methods.    
 We also propose a randomization-like model that simulates a 
 mechanism of  ``self-delurking'' of a network, and a lurking-oriented percolation analysis 
 to unveil possible relations between lurkers and users that act as bridges over subnetworks.

We conducted  experiments on   Twitter, Flickr,   FriendFeed, and GooglePlus networks.  
Quantitative and qualitative results have shown    
the effectiveness of our lurker ranking approach,   
highlighting superior  performance a\-gainst 
  PageRank, alpha-centrality and  the Fair-Bets model, which conversely    
might fail to correctly identify and rank   presumed lurkers.  
We have finally provided a preliminary exploration of relations between lurking and  trustworthiness in an OSN.  
 
 The remainder of this paper is organized as follows.  
 Section~\ref{sec:problem-statement} introduces our definitions of topology-driven lurking and lurking coefficient of a network.  
 Our lurker ranking methods are described in Section~\ref{sec:lurkerranking}. 
 Section~\ref{sec:evaluation} and 
 Section~\ref{sec:results} present experimental methodology and results. 
    Section~\ref{sec:related} discusses  related work.  
 Pointers for future research are provided in  Section~\ref{sec:challenges},  
and  Section~\ref{sec:conclusion} concludes the paper.

\section{In-degree, Out-degree, and Lurking} 
\label{sec:problem-statement}

\begin{figure}[t!]
\centering
\includegraphics[width=0.3\textwidth]{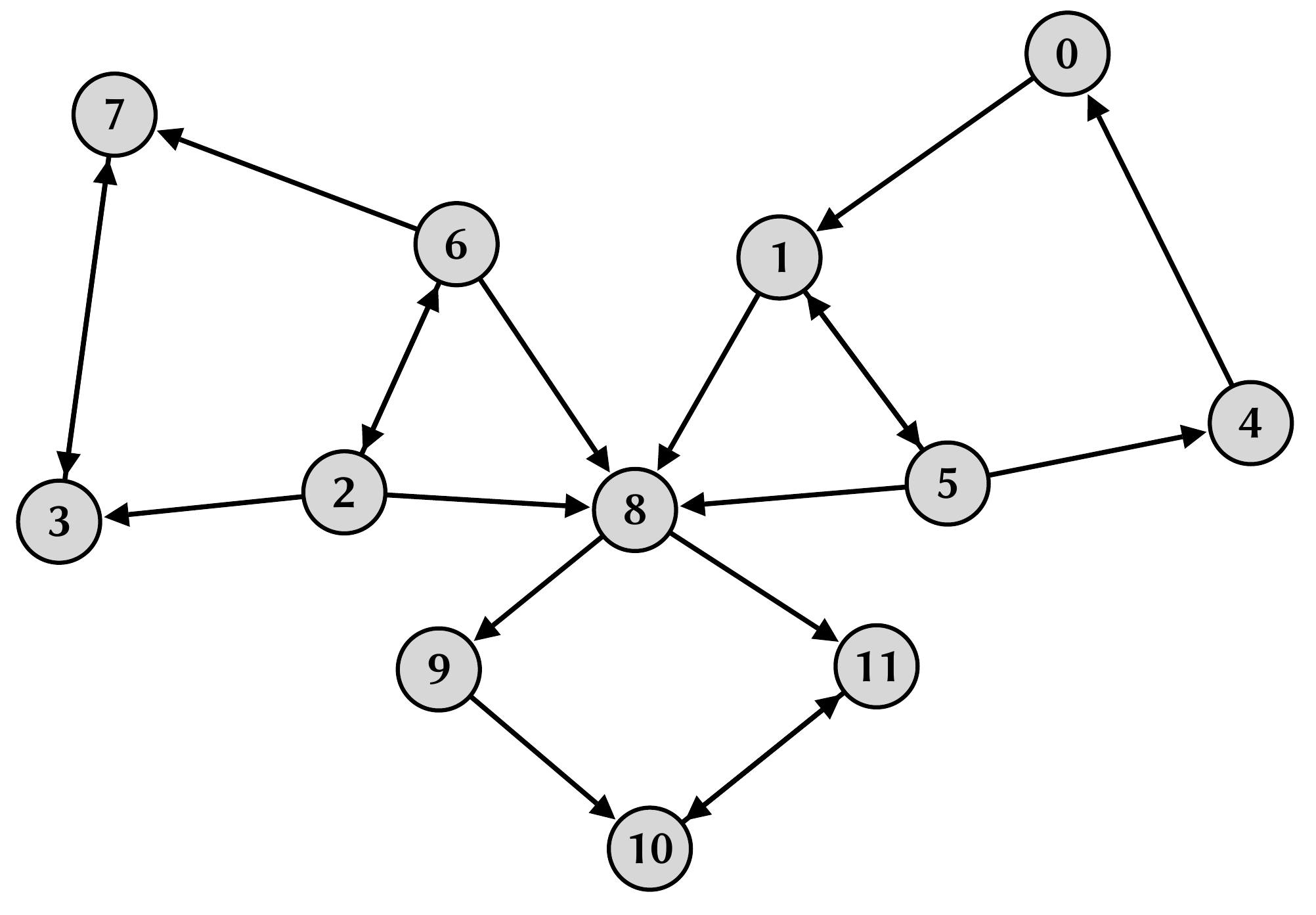} 
\caption{An example OSN graph for our lurking-oriented ranking analysis.} 
\label{fig:example}
\end{figure}

User interactions in an OSN  are typically  modeled as   influence  relationships,   
whose varying strengths  are used to determine and rank the influential users.     
In effect, ranking methods, such as PageRank, follow the conventional model 
of \emph{influence graph}, which  implies that the more 
incoming links a node has the more important or authoritative  it is;   
for example, translated to Twitter terms, the more followers a user has, the more interesting his/her published tweets might be.  
Actually,  as is well-known in spam detection, a node's in-degree can easily be affected by malicious manipulation, and hence the number of incoming links is not to be trusted as unique estimator of the node's importance score. Rather, as discussed in~\cite{Gayo-Avello13}  in the Twitter scenario,  the follower-to-followee ratio should in principle be considered: if the number of followers exceeds those of followees then the user is likely to be an opinion-maker, otherwise  her/his  tweets are not that interesting. 

We however observe that classic authority-based ranking methods (i.e., PageRank and related methods)   cannot  be directly applied to lurking analysis because they assume 
that links across users carry the meaning of  node influence propagation, which is related to the \textit{a\-mount of information (number of walks) a node produces}.  
By contrast, lurking behaviors build on the \textit{amount of information a node consumes};    
again, in Twitter terms, if user $v$ follows user $u$, then $v$ is benefiting from $u$'s information (i.e., $v$ is receiving $u$'s tweets).

 \begin{figure*}[t!]
\centering
\begin{tabular}{ccc}
\includegraphics[width=0.28\textwidth]{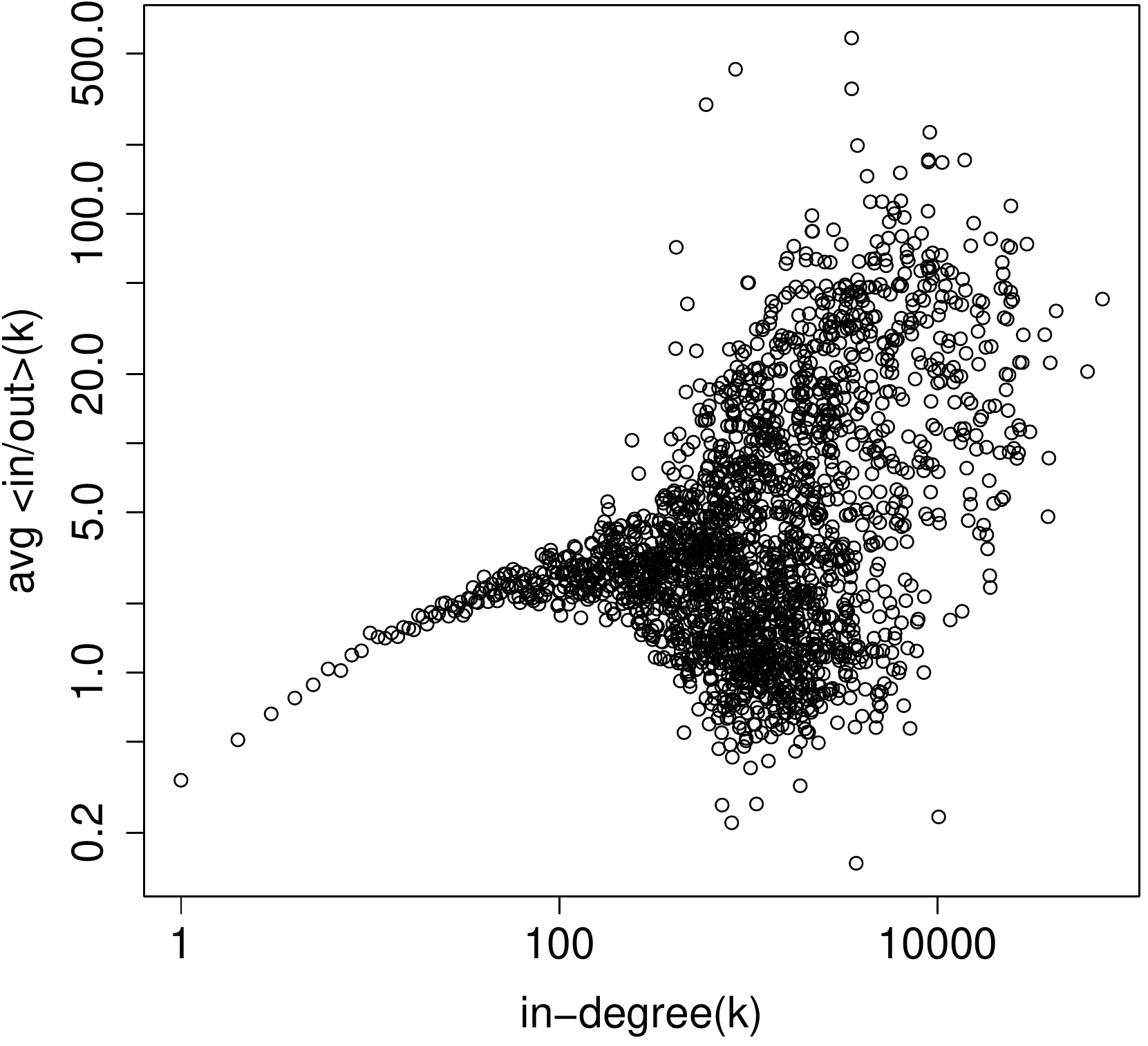} & \quad
\includegraphics[width=0.28\textwidth]{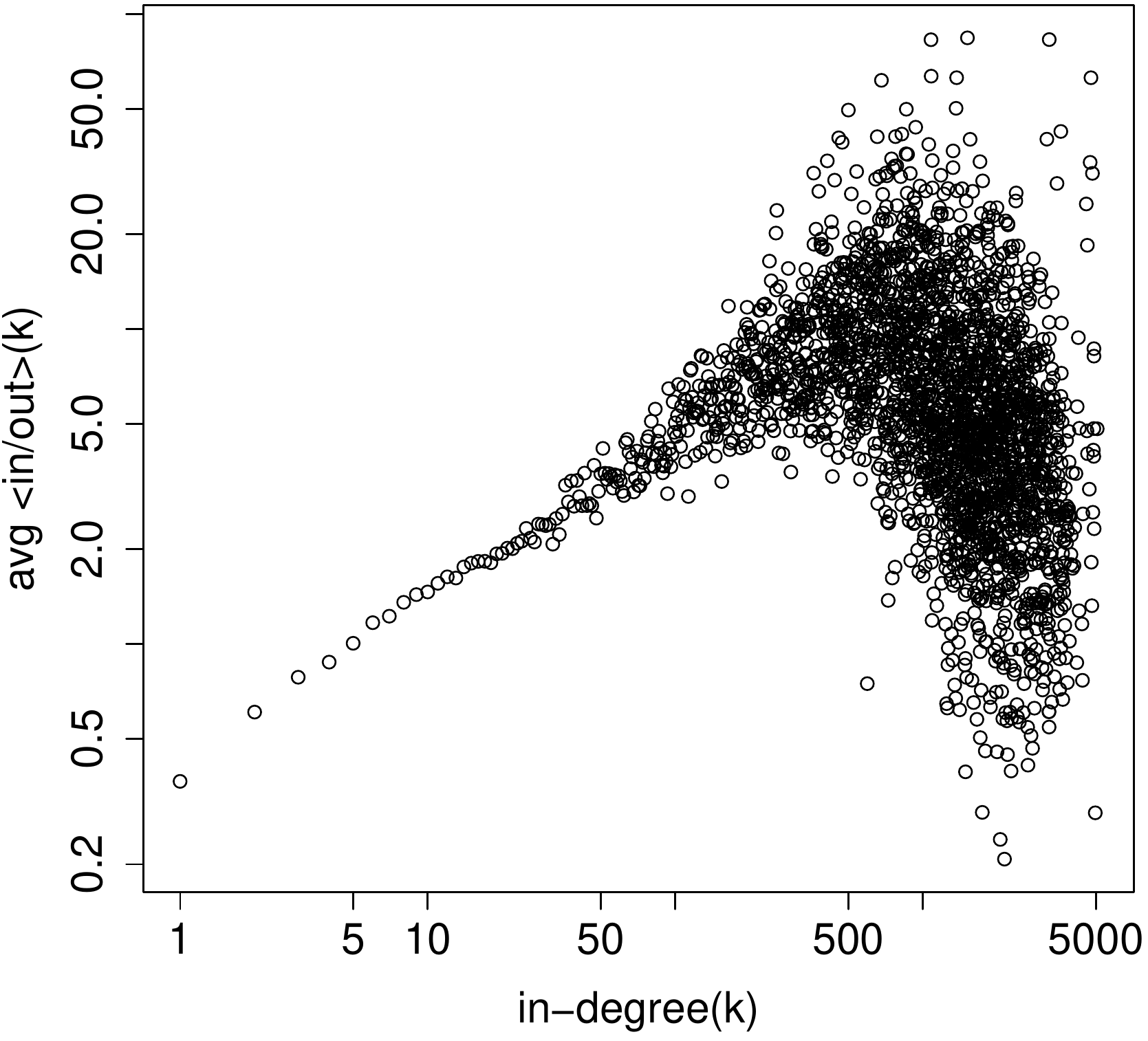} &   \quad
 \includegraphics[width=0.28\textwidth]{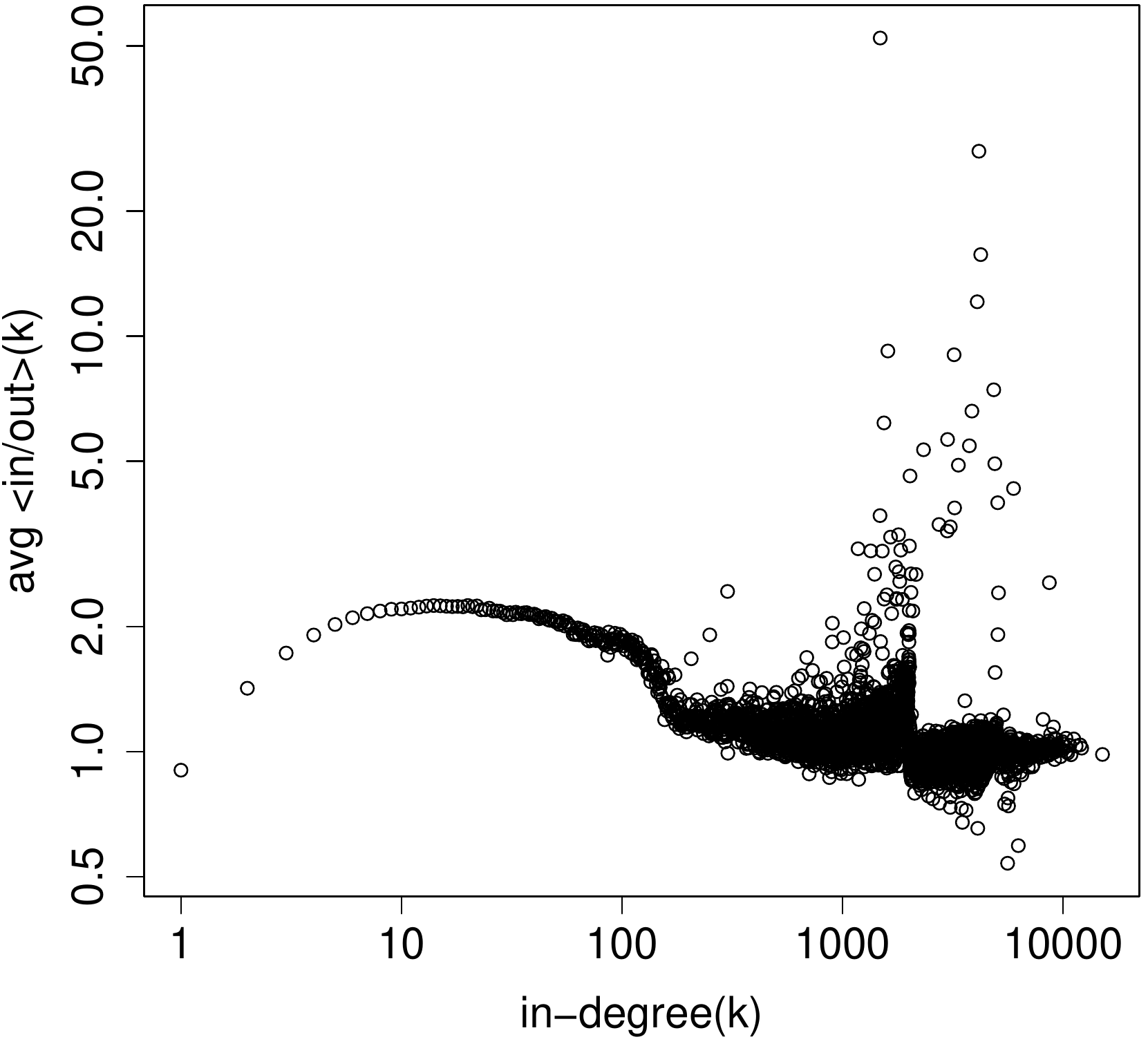} \\
(a) \data{FriendFeed} & \quad  (b) \data{GooglePlus}  & \quad  (c) \data{Twitter-UDI}
\end{tabular}
\caption{Average in/out-degree as function of the in-degree $k$, on double-logarithmic scale. Sink and source nodes are discarded.}
\label{fig:avgInOut}
\end{figure*}

A question might arise whether   there is any evident correlation between the in/out-degree ratio and the in-degree distribution in an OSN graph. To roughly answer the question, we empirically investigated this aspect on the networks we used for our experimental evaluation (cf. Section~\ref{sec:data}); 
Figure~\ref{fig:avgInOut} displays the average in/out-degree for each in-degree $k$, on some selected datasets. While the charts show substantially different trends, they all provide evidence on the poor correlation between in/out-degree ratio and the in-degree distribution. For the \data{FriendFeed} and \data{GooglePlus} cases, it can be observed a slightly upward trend for low in-degree values, while for  \data{Twitter-UDI}, the initial uptrend rapidly   decreases  for low-mid in-degrees. All cases however present high dispersion of in/out-degrees for mid-high in-degrees.

\subsection{Topology-driven Lurking}

Upon the    in/out-degree ratio intuition, we now provide a  basic definition of lurking  
which aims to lay out the essential hypotheses of a lurking status   based solely on the topology information 
available in an OSN.

\begin{definition}[Topology-driven lurking]\label{def:base}
Let   $\G = \langle \V, \E \rangle$ denote the directed graph representing  an OSN, with 
set of nodes (members) $\V$ and set of edges $\E$, whereby the semantics of any edge $(u, v)$ is that 
 $v$ is consuming information produced by $u$.  
A node $v$ with infinite in/out-degree ratio (i.e., a sink node) is trivially regarded as a lurker. 
A node $v$ with  in/out-degree ratio not below 1 shows a lurking status, 
whose strength is determined  based on:
\begin{description}
\item[\bf Principle I:] \textbf{Overconsumption}. The excess of informa\-tion-consumption over information-production.   The strength of $v$'s lurking status is   proportional  to its in/out-degree ratio. 
\item[\bf Principle II:] \textbf{Authoritativeness of the information received.}  The valuable amount of information received from  its in-neighbors. The strength of $v$'s  lurking status is     proportional to the influential (non-lurking) status of the $v$'s in-neighbors.
\item[\bf Principle III:] \textbf{Non-authoritativeness of the information   produced.}  The non-valuable amount      of  information sent to   its out-neighbors. The strength of $v$'s lurking status is     proportional to the  lurking status of the $v$'s out-neighbors.
\end{description}   
\end{definition} 
To support this intuition, let us consider the example of network in Figure~\ref{fig:example}.  
Nodes $3, 7, 8, 10, 11$ have the   highest in/out-degree ratio (i.e., 2), and as such they are candidate lurkers in the network. However, node $8$ should be scored higher than others, since it benefits from information coming from two   connected components, which are likely to contain influential nodes in the network (i.e., $5$, $6$). By contrast, nodes $10, 11$ should be scored as lurkers lower than node   $8$, since they are mainly fed by $8$ itself; similarly, nodes $3, 7$ should be scored higher than $10, 11$ but lower than $8$, since they receive information that propagates from a smaller subgraph. 
Note that the example allows us to shed light on a crucial aspect   related to the role that node   $8$ has in the network. In effect, one may say that $8$ is a ``bridge'' as it allows readers $9$, $10$, and $11$ to peek into two otherwise separated communities. However, in our network model oriented to   information consumption, the  notion of bridge is also revised:  the communication received from $9$, $10$, and $11$ is likely to be less significant (in terms of amount and/or quality) than the bandwidth of information flow originated from the two largest components and received from $8$.  
In Section~\ref{sec:percolation}, we shall investigate the relationship between lurkers and bridges, which will confirm that it's correct to regard node $8$ as top-lurker.

\subsection{Lurking Coefficient of a network}
\label{sec:lurkiness}
The participation inequality ``1:9:90'' rule  loosely tells us that the majority of users shows a potential lurking behavior, in any generic online community.  But, can we have a more precise indication of the presence of lurkers given a particular network? 
  To answer this question we introduce here a measure, named  \textit{Lurking Coefficient}, as a basic    lurking-related property of the topology of a network. 
 
 Given the directed graph $\G = \langle \V, \E \rangle$ representing an OSN, 
for any node $i \in \V$  let $B_i = \{j | (j,i) \in \E \} $ and $R_i = \{j | (i,j) \in \E \}$ denote the set of in-neighbors 
(i.e., backward nodes) 
and out-neighbors (i.e., reference nodes)  of $i$, respectively.  
The sizes of sets $B_i$ and $R_i$ are the in-degree and the out-degree of $i$, denoted as 
$in(i)$ and $out(i)$, respectively.    
The \textit{local  Lurking Coefficient} of a node is first introduced to measure how likely any given  node $i$ is a lurker within its neighborhood. We define this quantity as:
 \begin{multline}\label{eq:lc}
 lc_i = \frac{1}{|\V_i|} \left(\sum_{j \in B_i}  \mathbbm{1}\left\{\frac{in(j)}{out(j)} < \frac{in(i)}{out(i)}\right\} + \right. \\ 
         \left.  \sum_{j \in R_i}  \mathbbm{1}\left\{\frac{in(j)}{out(j)} \geq \frac{in(i)}{out(i)}\right\}\right) 
 \end{multline} 
where $\V_i$ is the set of neighbors of $i$, and    $\mathbbm{1}\{A\}$ is the \textit{indicator} function, which is equal to 1 when the event $A$ is true,   0 otherwise. Note that the two additive terms in Eq. (\ref{eq:lc})  are in accordance with Principle II and Principle III, respectively, of Def. 1. 
The    \textit{Lurking Coefficient} of a graph $\G$ is then given by the weighted average of the local Lurking Coefficients over the nodes in $\G$: 
 \begin{equation}
 LC_{\G} = \frac{1}{|\V|} \sum_{i \in \V} p_i \cdot lc_i
 \end{equation}
where  $p_i$ is the weight of $lc_i$. This weight, unitary  by default, can be set in accordance with Principle I, hence it is defined as the  in/out-degree ratio of $i$ normalized over all  
 nodes in its neighborhood. We will refer to the variant of $LC$ with non-unitary weights as  weighted Lurking Coefficient ($wLC$).

 \section{Lurker Ranking}
 \label{sec:lurkerranking} 
 In this section we formulate  our solutions to the problem of lurker ranking.  
To this aim, we will capitalize on the three principles stated in our topology-driven lurking definition. 
Note that, as a general premise valid for all lurker ranking methods that we shall present, 
we introduce a Laplace smoothing factor in the calculation of both in-degree and out-degree of node, i.e., 
 $in(i)$ (resp. $out(i)$) is meant hereinafter 
as the actual in-degree (resp. out-degree) of   node $i$ plus one. This allows us to deal with sink nodes and avoid infinite in/out-degree ratios.

According to Principle I in Definition~\ref{def:base},  
a basic way of scoring a node as a lurker is by means of its in/out-degree ratio.  
However, this way has clearly  the disadvantage of assigning many nodes the same 
or very close ranks and, as we previously discussed, it ignores that the status of both the in-neighbors (Principle II)
and  out-neighbors (Principle III) contributes to the status of any given node. 
In the following we elaborate on each of those aspects separately. 

  \vspace{2mm}
 \noindent 
\textbf{In-neighbors-driven lurking.\ } 
According to Principle II in Definition~\ref{def:base}, 
 an in-neighbors-driven lurking measure can be defined as: 
  \begin{equation*}\label{eq:back}
 r_i = \sum_{j \in B_i}  \frac{out(j)}{in(j)} r_j  
\end{equation*}
Hence, the score of node $i$ increases with the number of its in-neighbors 
and with their likelihood of being non-lurkers, which is   expressed by 
a relatively high  out/in-degree.  The above formula   
can be enhanced by including   a factor that is inversely proportional to the $i$'s out-degree. 
Formally, we define the \textit{in-neighbors-driven lurking} score of node $i$ as: 
\begin{equation}\label{eq:back-enhanced}
r_i = \frac{1}{out(i)} \sum_{j \in B_i}  \frac{out(j)}{in(j)} r_j
\end{equation}
Note that Eq.~(\ref{eq:back-enhanced})  
accounts for both the contribution of a node's in-neighbors and its own 
in/out-degree property.

  \vspace{2mm}
 \noindent 
\textbf{Out-neighbors-driven lurking.\ } 
The exclusive contribution of out-neighbors 
for the calculation of a node's  lurking score, according to 
Principle III of Definition~\ref{def:base}, can be formalized as: 
\begin{equation*}
r_i =  \sum_{j \in R_i}  \frac{in(j)}{out(j)} r_j 
\end{equation*}
 However, this method would let the score of a node increase with 
 the tendency of its out-neighbors of being lurkers, while  
 ignoring  the status of the node  itself; as a consequence, 
 not only   reciprocal lurkers will be scored high but also 
 every node from which lurkers receive information. 
A correction factor should hence be introduced  as proportional 
to the in-degree of the target node. 
Formally, we define the \textit{out-neighbors-driven lurking} score of node $i$ as:
\begin{equation}\label{eq:ref-enhanced}
r_i =  \frac{in(i)}{\sum_{j \in R_i} in(j)} \sum_{j \in R_i}   \frac{in(j)}{out(j)}  r_j
\end{equation}
Note that in Eq.~(\ref{eq:ref-enhanced}), the in-degree of node $i$ is divided by the sum of in-degrees of its out-neighbors 
in order to score $i$ higher if it receives more than what its out-neighbors receive.

 \vspace{2mm}
 \noindent 
\textbf{In-Out-neighbors-driven lurking.\ } 
The two previous definitions of lurking can in principle be combined 
to obtain an integrated representation of all three principles  
in  Definition~\ref{def:base}.  
To this aim, we define the \textit{in-out-neighbors-driven lurking} score of node $i$ as:  
\begin{multline}\label{eq:backref}
r_i = \left(\frac{1}{out(i)} \sum_{j \in B_i}  \frac{out(j)}{in(j)} r_j \right) \\  \left(  1 + \left( \frac{in(i)}{\sum_{j \in R_i} in(j)} \sum_{j \in R_i}   \frac{in(j)}{out(j)} r_j \right)\!\right)  
\end{multline}

\noindent  
Note that in Eq.~(\ref{eq:backref}) we have emphasized the aspect related to the strength of non-lurking  
 behavior of in-neighbors, which is expected to have a better 
 fit of the hypothetical likelihood function for a given node.

\subsection{LurkerRank methods} 
\label{sec:complete-formulations}
We now define our lurker ranking methods, dubbed \textit{LurkerRank} (for short \algo{LR}), 
 upon the previously defined lurking models. 
 In order to provide a complete specification of our models,
  we resorted to the  classic eigenvector-centrality schemes offered by \textit{PageRank}~\cite{PR}  
and \textit{alpha-centrality}~\cite{AC}. Note that while being widely applied to a variety 
of application domains with the purpose of scoring the influence or prestige 
in information networks, PageRank and alpha-centrality rely on different assumptions 
which make it worth the exploration of lurker ranking through both approaches.

Let us first recall the PageRank mathematics. 
The    PageRank  vector is the unique solution of the   iterative equation $\r = \d \M \r + (1-\d) \mathbf{v}$.  
$\M$ denotes the column-stochastic transition probability matrix, which is defined as $(\Dout^{-1} \A)^{\mathrm{T}} + \e\mathbf{a}^{\mathrm{T}}/|\V|$, where 
  $\A$ is the adjacency matrix of the network graph $\G = \langle \V, \E \rangle$, with $A_{ij}=1$ if $(v_i,v_j) \in \E$, and $A_{ij}=0$ otherwise;    
$\Dout = diag(\A\e)$ is the out-degree diagonal matrix;  
 $\e$ denotes a $|\V|$-dimensional column vector of ones; and   $\mathbf{a}$ is  defined  
such that $a_i=1$ if node $i$ has zero out-degree, and 0 otherwise.  
Vector $\mathbf{v}$ is typically defined as $(1/|\V|) \e$, but 
can  be modeled to bias 
the PageRank to boost a specific subset of nodes in the graph.  
Term $\d$ is a real-valued coefficient ($\d \in [0, 1]$, commonly set to 0.85), which acts as a
  damping factor so that   
the random surfer is expected to discontinue the chain with probability $1-\d$, and hence 
to randomly select a page each with relevance $1/|\V|$ (teleportation).   

We formulate three of our methods according to a PageRank-like scheme, i.e., at a high level, 
according to a combination of  a random walk term  with a random teleportation term. 
Our first LurkerRank method  is named \textit{in-neighbors-driven LurkerRank} (hereinafter denoted as \algo{LRin}) since it is built upon Eq.~(\ref{eq:back-enhanced}): 
 \begin{equation}\label{eq:LRin}
r_i = \d \left( \frac{1}{out(i)} \sum_{j \in B_i}  w(j,i) \frac{out(j)}{in(j)} r_j \right)   \  + \ \frac{1-\d}{|\V|}   
\end{equation}
Note that with Eq.~(\ref{eq:LRin}), we introduce  edge weights 
to deal with  weighted   graphs as well, for the sake of generality; 
although, as  in our experimental setting, they are set as unitary by default. 
Analogously, the \textit{out-neighbors-driven LurkerRank} (hereinafter denoted as 
\algo{LRout}) is defined as:
 \begin{multline} 
r_i = \d \left( \frac{in(i)}{\sum_{j \in R_i} in(j)} \sum_{j \in R_i}  w(i,j) \frac{in(j)}{out(j)} r_j \right) +\\  \frac{1-\d}{|\V|}   
\end{multline}
Finally, the \textit{in-out-neighbors-driven LurkerRank} (hereinafter denoted as 
\algo{LRin-out}) is defined as:
\begin{multline}
 r_i = \d  \left[ \left(  \frac{1}{out(i)} \sum_{j \in B_i} w(j,i) \frac{out(j)}{in(j)} r_j   \right) \Biggl( 1 +  \Biggr.  \right. \\ \hspace{-10mm}\left. \left. \left( \frac{in(i)}{\sum_{j \in R_i} in(j)} \sum_{j \in R_i}  w(i,j) \frac{in(j)}{out(j)} r_j \right) \right) \right]      +   \frac{1-\d}{|\V|}   
\end{multline}

Alpha-centrality~\cite{AC}  expresses the centrality of a node as the number of paths linking it to other nodes, exponentially attenuated by their length. 
Moreover, it takes into account the possibility that each  node's status  
may also depend on information that comes from outside the network   
or that may regard solely the member. Alpha-centrality is defined   
as  $\r = \d \mathbf{A}^{\mathrm{T}} \r + \mathbf{v}$,  
where $\mathbf{v}$ is the vector of exogenous source of information 
($\mathbf{v}=\e$ as default),  and   $\d$ here reflects the relative importance  of endogenous versus exogenous factors in the determination of centrality. High values of $\d$ (e.g., 0.85) make the close neighborhood contribute less to the centrality of a given node.    
The rank obtained using alpha-centrality can be considered as the steady state distribution of an information spread process on a network, with probability $\d$   to transmit a message or influence along a link.  

We will denote our alpha-centrality based LurkerRank methods with prefix \algo{ac-} to distinguish them from the PageRank-based counterparts. 
The alpha-centrality-based in-neighbors-driven LurkerRank (\algo{ac-LRin}) is defined as: 
 \begin{equation}
r_i = \d \left( \frac{1}{out(i)} \sum_{j \in B_i}  w(j,i) \frac{out(j)}{in(j)} r_j  \right)   \  + 1   
\end{equation}
Analogously, other two methods, denoted as   \algo{ac-LRout} and \algo{ac-LRin-out}, are defined 
according to the out-neighbors-driven and in-out-neighbors-driven lurking models, respectively.

 \begin{figure}[t!]
\centering
\includegraphics[width=0.3\textwidth]{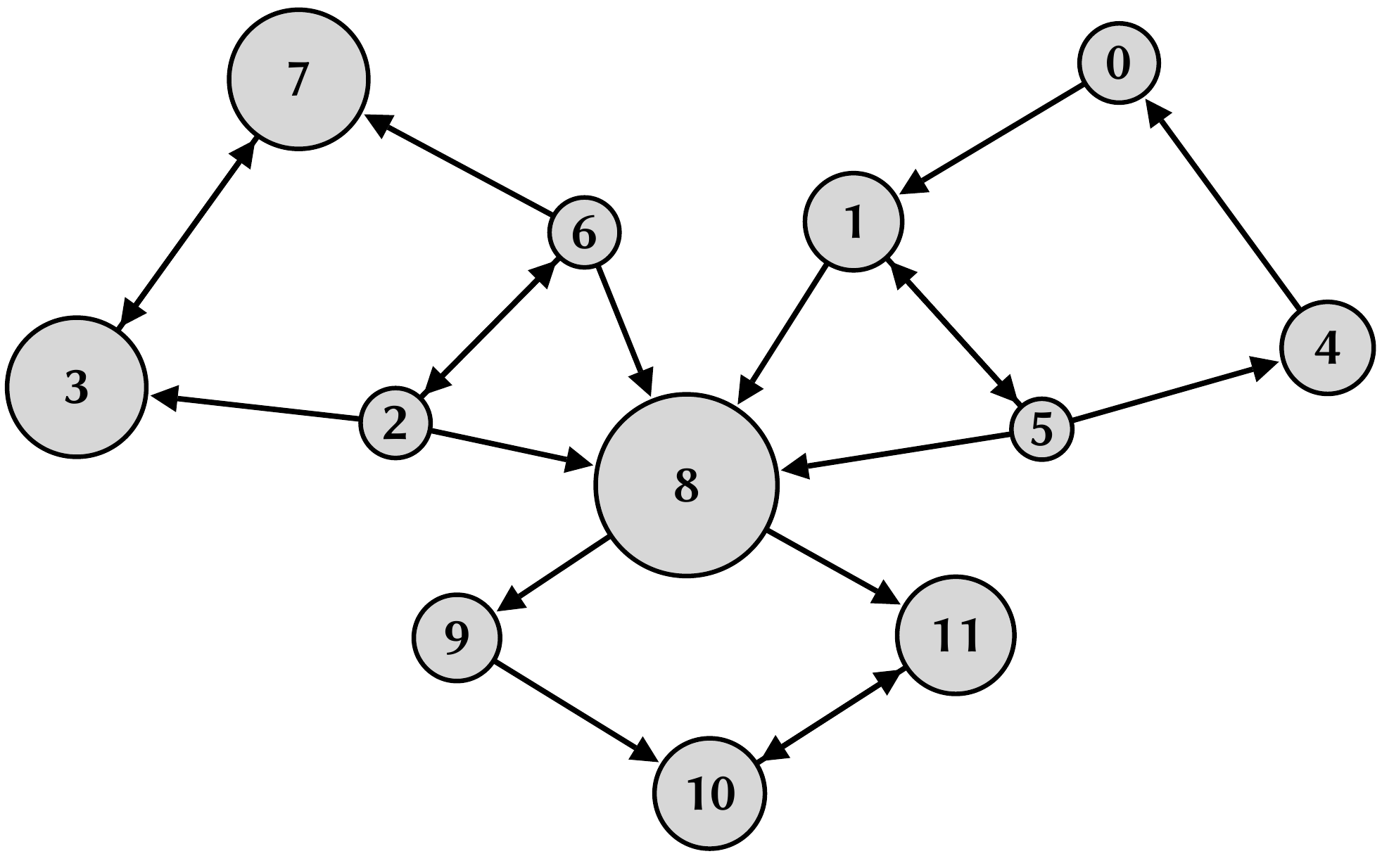} \\
\includegraphics[width=0.3\textwidth]{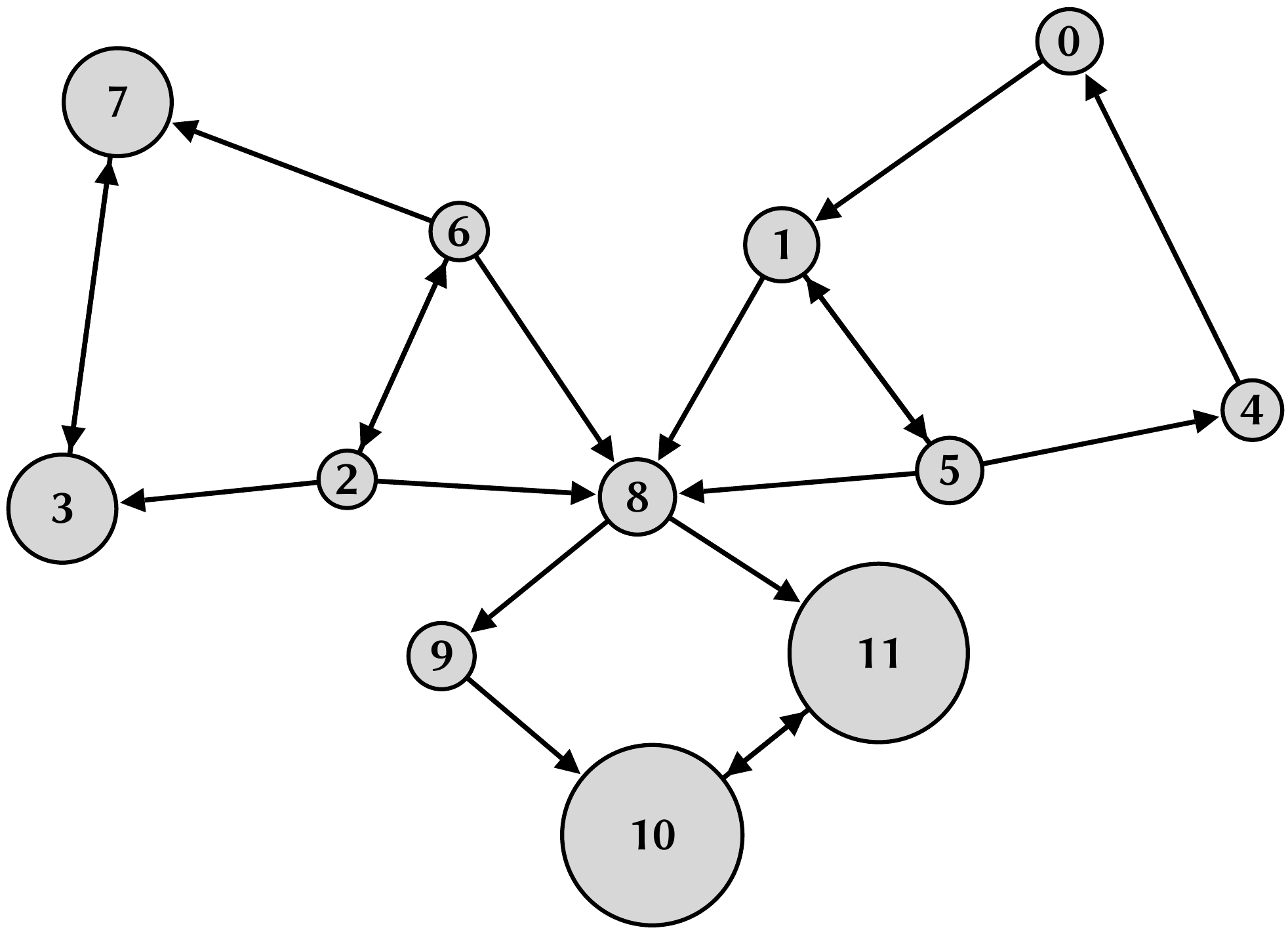} \\
\caption{Lurker ranking  in the example OSN graph of Fig.~\ref{fig:example}:  \algo{LRin} (on top) versus PageRank (on bottom). Nodes are sized proportionally to their ranking scores.}
\label{fig:example-results}
\end{figure}

Figure~\ref{fig:example-results} compares the rankings 
obtained by our \algo{LRin}  and 
basic PageRank on the example network of Figure~\ref{fig:example} ($\d$ 
set to the default  0.85).  
Using \algo{LRin}, node  8 was ranked highest (0.146), 
 followed by 3 and 7 (0.112), and then 11 (0.094), 10 (0.088): 
  this  sheds light on the ability of  \algo{LRin} 
 to match our definition of lurking (cf. discussion about  Fig.~\ref{fig:example}   
 in Section~\ref{sec:problem-statement}).  
 By contrast, PageRank ranked first nodes 10 and 11 (both around 0.256), 
 and then 3 and 7 with a significant  gap in score from the first two (0.116), 
 followed by 8 (0.052), 1 (0.048); moreover, node 5 was ranked eighth, 
 despite it is a major feeder of the lurker 8, while it 
 was correctly  ranked lowest by   
 \algo{LRin}. Similarly, alpha-centrality (results not shown)  did not fare well 
as it  ranked first nodes 11 (0.317) and 10 (0.308), before 
 ranking node 8 (0.095), and nodes 3 and 7 in ninth and tenth 
 position both with a score of 0.004.

\subsection{Limit $\d \rightarrow 0$ of the \algo{LR} functions}
We investigate the behavior of \algo{LR} functions to understand  whether \algo{LR} rank can be  reduced to either the in/out-degree or the out/in-degree rank as $\d$ approaches 0. 
We take the  \algo{LRin} functional form as case in point, while analogous conclusions can be drawn for the other \algo{LR} functions.

\begin{figure*}[t!]
\centering
\begin{tabular}{ccc}
 \includegraphics[width=0.28\textwidth]{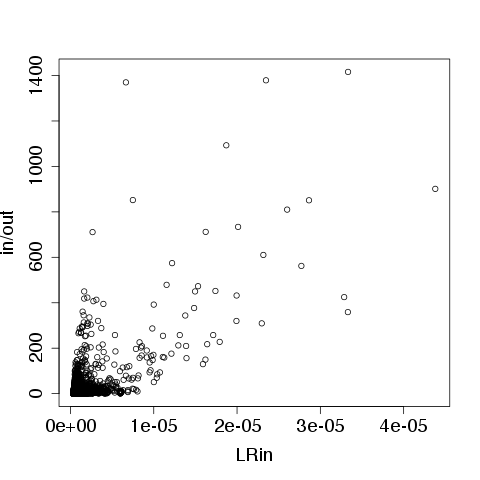} & \quad
\includegraphics[width=0.28\textwidth]{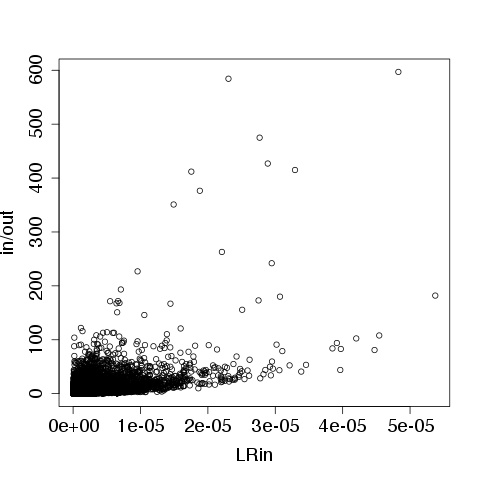} & \quad  
\includegraphics[width=0.28\textwidth]{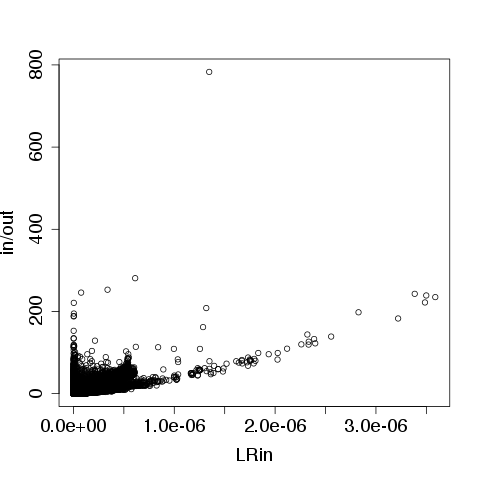} \\
(a) \data{Flickr} & \quad  (b) \data{FriendFeed}  &   \quad  (c) \data{Twitter-Kwak}
\end{tabular}
\caption{\algo{LRin} rank versus in/out-degree rank. Damping factor $\d$ is set to $0.01$. Sink and source vertices are discarded.}
\label{fig:LRIn_limit}
\end{figure*}

In the extreme case $\d = 0$, the \algo{LRin} score of each vertex is equal to $1/|\V|$. If $\d \approx 0$, then $(1-\d) \rightarrow 1$, therefore we write $1-\d = 1- \epsilon$, with $\epsilon \ll 1$, and 
$r_i \approx 1/|\V|$. Substituting these into Eq.~(\ref{eq:LRin}), with unitary edge weights for the sake of simplicity, we have:
\begin{multline}\label{eq:LRin-limit}
 r_i = \epsilon \left( \frac{1}{out(i)} \sum_{j \in B_i}   \frac{out(j)}{in(j)} r_j \right)   \  + \ \frac{1-\epsilon}{|\V|}
 \\ 
 \approx \frac{1}{|\V|} \left[ 1+  \epsilon \left(\frac{1}{out(i)} \sum_{j \in B_i}  \frac{out(j)}{in(j)} - 1 \right) \right]   
\end{multline}  
A crucial part in Eq. (\ref{eq:LRin-limit}) is the estimation of the sum. 
This term would be estimated as proportional to the in/out-degree of vertex $i$ and to the average out/in-degree $\langle\frac{out}{in}\rangle$:
\begin{equation}
 r_i  \approx \frac{1}{|\V|} \left[ 1+  \epsilon \left(\frac{in(i)}{out(i)}   \left\langle\frac{out}{in}\right\rangle - 1 \right) \right]   
\end{equation}  
The above approximation is however admissible only if a relatively small dispersion can be assumed to hold for the out/in-degree distribution.   
Unfortunately,  in all our evaluation network datasets (cf. Sect.~\ref{sec:data}), this does not seem the case since the out/in-degree distribution is always found to be less narrow than the corresponding in/out-degree distribution, as reported in Table~\ref{tab:mean-sd}.   
This would indicate that in principle \algo{LRin} rank distribution is likely not to follow exactly the same trend as that of in/out-degree as $\d \approx 0$. In effect, although  a moderate to strong positive correlation may still occur --- 0.568 on \data{FriendFeed} (Fig.~\ref{fig:LRIn_limit}(b)), 
0.674  on \data{Twitter-UDI},   
0.679 on \data{Flickr} (Fig.~\ref{fig:LRIn_limit}(a)), 
0.686  on \data{Twitter-Kwak} (Fig.~\ref{fig:LRIn_limit}(c)), 
and  0.745 on \data{GooglePlus} --- Fig.~\ref{fig:LRIn_limit} shows that top-ranked vertices by \algo{LRin} often do not correspond to top-ranked in/out.

\begin{table*}[t!]
\caption{Mean and standard deviation values of in/out-degree and out/in-degree.}
\centering 
\scalebox{0.8}{
\begin{tabular}{|l||c|c|c|c||c|c|c|c|}
\hline
  & \multicolumn{2}{|c||}{in/out $^*$}     & \multicolumn{2}{|c||}{in/out $^{**}$}   
  & \multicolumn{2}{|c|}{out/in $^*$}  & \multicolumn{2}{|c|}{out/in $^{**}$}   \\ 
  \cline{2-9}
  & \textit{mean} & \textit{sd} &  \textit{mean} & \textit{sd}      
    & \textit{mean} & \textit{sd} &  \textit{mean} & \textit{sd} 
  \\ \hline \hline
\data{Flickr}  & 1.096  &  3.377  & 2.731  & 10.557 & 1.263  & 4.583  & 5.554  &  20.085 \\
\data{FriendFeed}  &  1.664 &  5.693  &  10.359 & 15.353  & 8.682  & 71.771  & 63.269  & 219.387  \\
\data{GooglePlus}  &  3.947 & 11.200   & 24.350  & 27.665 &  3.739 &  46.235 & 27.984  & 144.051  \\
\data{Twitter-Kwak}  & 2.647  &  3.863  &  11.662 & 9.442 & 1.263  & 46.078  & 6.910  &  145.665 \\
 \data{Twitter-UDI}  & 1.541  &  1.530  & 5.517  & 3.582 & 1.202  & 15.758  & 4.946  & 50.321  \\
 \hline
\end{tabular}
}

\vspace{1 mm}
\scriptsize
$^{*}$  Sink nodes and source nodes are discarded. \hspace{3mm} $^{**}$  Like $^{*}$, but only 90th percentile is considered.
\label{tab:mean-sd}
\end{table*}

 \section{Experimental Evaluation}
\label{sec:evaluation}

\begin{table*}[t!]
\caption{Main structural characteristics of the evaluation network datasets.}
\centering
\scalebox{0.7}{
\begin{tabular}{|l|c|c|c|c|c|c|c|c|}
\hline
\emph{data} & \textit{\# nodes} & \textit{\# links} &  \textit{avg}  &  \textit{avg} & \textit{clustering} & \textit{assortativity} &  \textit{\# sources}  & \textit{LC} \\
  &   &  &  \textit{in-degree}  &  \!\!\textit{path length}\!\! & \textit{coefficient} & & \textit{\# sinks}  & \textit{wLC} \\
 \hline \hline
  \multirow{2}{*}{\data{Flickr}}    &  \multirow{2}{*}{2,302,925}   &  \multirow{2}{*}{33,140,018}  &  \multirow{2}{*}{14.39} & \multirow{2}{*}{4.36*} &  \multirow{2}{*}{0.107} & \multirow{2}{*}{0.015}  &   360,416   & 0.573\\
 				& & & &  & & & 57,424 &   0.248 \\
 \hline
  \multirow{2}{*}{\data{FriendFeed}}    &  \multirow{2}{*}{493,019}   &  \multirow{2}{*}{19,153,367}  &  \multirow{2}{*}{38.85} & \multirow{2}{*}{3.82} &  \multirow{2}{*}{0.029} & \multirow{2}{*}{-0.128}  &  41,953   & 0.955\\
 				& & & &  & & & 292,003   &  0.354 \\
 \hline 
 \multirow{2}{*}{\data{GooglePlus}}    &  \multirow{2}{*}{107,612}   &  \multirow{2}{*}{13,673,251}  &  \multirow{2}{*}{127.06} & \multirow{2}{*}{3.32} &  \multirow{2}{*}{0.154} &  \multirow{2}{*}{-0.074}  &  35,341   & 0.869 \\
 				& & & &  & & &  22      &   0.096\\
\hline 
\multirow{2}{*}{\data{Twitter-Kwak}} & \multirow{2}{*}{16,009,364}  &  \!\!\multirow{2}{*}{132,290,000}\!\! &  \multirow{2}{*}{8.26}  &  \multirow{2}{*}{5.91*}  &  \multirow{2}{*}{1.26E-4} & \multirow{2}{*}{-0.095}  & 1,067,936   & 0.914\\
				& & & & 									&	& 	& 10,298,788   &  0.435 \\
\hline 
\multirow{2}{*}{\data{Twitter-UDI}}  & \multirow{2}{*}{24,984,590}  &  \!\!\multirow{2}{*}{284,884,500}\!\! &  \multirow{2}{*}{11.40}  &  \multirow{2}{*}{5.45*}  &  \multirow{2}{*}{4.96E-3} & \multirow{2}{*}{-0.297}  &  3,380,805 & 0.790\\
				& & & & 									&	& 	& 8,065,287   & 0.470  \\
\hline  
\end{tabular}
}
 
\vspace{1 mm}
\scriptsize
 $^*$  Value estimated as $(\log(|\V|))/\log(2|\E|/|\V|)$.
\label{tab:data}
\end{table*}

\subsection{Data}\label{sec:data}
We used five OSN datasets   for our evaluation, namely 
 \data{Twitter} (with two different dumps),  
 \data{Flickr}, \data{FriendFeed},  and \data{GooglePlus}:

\begin{itemize}
\item
From the Twitter  dump studied in~\cite{KwakLPM10},   which we will refer to as \data{Twitter-Kwak}, we extracted 
  the follower-followee topology starting from a connected component of one hundred thousands of users and their complete 
  neighborhoods.  
A partial copy of the tweet data used in~\cite{KwakLPM10} was exploited to define a Twitter-based data-driven ranking and also to perform a qualitative evaluation on \data{Twitter-Kwak}, as we shall describe in  Section~\ref{sec:exp:criteria}.   
\item 
The \data{Twitter-UDI} dataset~\cite{LiWDWC12}  
was originally collected in May 2011, hence it's more recent and also larger than \data{Twitter-Kwak}.  
Tweet data  
however could not be exploited for our analysis since they are available only for a very small subset of users in \data{Twitter-UDI} (less than 0.6\%)  
and they are also upper-bounded  (limit of 500 tweets per user)~\cite{LiWDWC12}.   
\item
We used the entire  \data{Flickr} data studied in~\cite{MisloveKGDB08}, originally collected in 2006-2007.  
Information on the number of views and  number of   favorite markings every  photo had, was exploited for our definition of  Flickr-based data-driven ranking. 
\item
We used the latest version of the \data{FriendFeed}  dataset studied in~\cite{CelliLMPR10}.   
Due to the recognized presence of spambots in this OSN dataset, 
we filtered out  users with an excessive number of posts (above  
20 posts per day) as suggested in~\cite{CelliLMPR10}. 
\item
\data{GooglePlus} dataset was originally studied in~\cite{McAuleyL12}, and consists of \emph{circles} from GooglePlus. The dataset was collected from users who had manually shared their circles using the \emph{share circle} feature, and the topology was built by combining the edges from each node's ego network. 
\end{itemize}

 Beyond the complexity of their technical and sociological aspects, 
 the five networks  have been selected since  they naturally provide 
  asymmetric relationships --- recall that in our setting, 
 a link from user $i$ to user $j$ means that  $j$ is a  follower or subscriber of $i$ --- and also because they offer a variety of  topological properties, as shown in Table~\ref{tab:data}.  
The table also reports   each network's Lurking Coefficient ($LC$), in the upper row, and weighted $LC$ ($wLC$), in the bottomer row  (cf. Section~\ref{sec:lurkiness}). Notably, a high   $LC$ (ranging from about 0.8 to 0.95) was found for all networks except for \data{Flickr}: this may prompt us to suppose that lurkers would not characterize  Flickr  as much as other OSNs; in effect, differently from the other selected networks,   users would subscribe and join the Flickr community when they are willing to upload and share their photos, thus showing a normal attitude to participate.   
 Moreover, the  lower value of weighted $LC$ that characterizes  \data{GooglePlus} could be explained due to a clustering coefficient, along with  variation of in/out degree (Table~\ref{tab:mean-sd}),   exhibited by this network, which are both relatively   higher than in the other ones.  
Yet, note that the values of assortativity reported in  Table~\ref{tab:data}    are always negative or close to zero, which would indicate no tendency of vertices with similar degree to connect to each other; interestingly, \data{Twitter-UDI} which has the most negative degree of assortativity, has also the largest value of weighted $LC$.

\subsection{Assessment methodology}
\label{sec:exp:criteria} 
\paragraph*{Competing methods and notations.}\  
We compared our proposed methods against PageRank (henceforth \algo{PR}), 
alpha-centrality (henceforth \algo{AC}), and  Fair-Bets model~\cite{BudalakotiB12} 
(henceforth \algo{FB}). 
The latter method was   included in the comparative evaluation as it also  exploits the notion of in/out-degree ratio to rank users, which is seen  as a fair-bets model of social capital accumulation and expenditure; originally conceived to rank players in round-robin tournaments, the Fair-Bets model assumes that users are paying each other to accept invitations on an online community, then the fair bets score of a user is the amount she/he can afford to pay on average. Fair-Bets 
computes the score of any node $i$ as 
$$r_i = \frac{1}{out(i)}  \sum_{j \in  B_i} r_j$$  
Finally, we included in the evaluation 
the  in/out-degree distribution of the nodes in a network dataset, 
as a baseline method (henceforth \algo{IO}).

\paragraph*{Data-driven evaluation.}\  
Given the novelty of the problem at hand, we had to cope with  
  an issue relating to the lack of ground-truth data  for lurker ranking. 
In the attempt of simulating  a ground-truth evaluation, 
we generated a  \textit{data-driven ranking} (henceforth \algo{DD}) for a network  dataset  
and used it to assess  the proposed and competing methods.

On \data{Twitter-Kwak}, we calculated the score of a node 
as directly proportional to its  in/out-degree (Laplace add-one smoothed, cf. Section~\ref{sec:lurkerranking}) 
 and inversely exponentially with a Twitter-specific measure of influence:
$$r^*_i = \frac{in(i)}{out(i)} \exp(-EI(i))$$
$EI(\cdot)$ denotes the \textit{empirical measure of influence}~\cite{BakshyHMW11}
which is used to estimate the influence of a user based on the amount of information s/he posted (i.e., tweets) 
and that her/his followers have  
retweeted. For a  user  $i$,  
$$EI(i) = \frac{1}{out(i)}  \sum_{j \in R_i} \textit{nRetweets}(j)$$   
where  $\textit{nRetweets}(j)$ is the number of re\-tweets  by follower $j$. 
Note that, as found in~\cite{KwakLPM10}, a ranking based on retweets differs from that based on the number of followers, 
and this   prompted us to combine the two aspects in our data-driven ranking.

We defined an analytically similar function for the \data{FriendFeed} 
data-driven ranking, in which the \textit{empirical measure of influence} has been redefined as:
$$EI(i) = \left(\frac{1}{out(i)}  \sum_{j \in R_i} \textit{nCom}(j,i)\right) \log_{10}{(\textit{nPosts}(i)+10)}$$
 where   $\textit{nCom}(j,i)$ is  the number of comments from user $j$ to posts by user $i$, and $\textit{nPosts}(i)$ is the total number of posts by user $i$. 
Note that this combination of     indicators of   user's activity with  user's influence  
was needed since only a limited portion (below 10\%) of users in \data{FriendFeed} had 
information on the number of received  comments. 

For \data{Flickr}  we produced two data-driven rankings, dubbed \algo{DD-F} and \algo{DD-V}. While still related to the in/out degree as for the previously defined \algo{DD}, we used  
the  number of \emph{favorites} (\algo{DD-F}), or alternatively the number of \emph{views} (\algo{DD-V}), received by a user's photos to set the exponent (with negative sign) in the data-driven ranking function.

Unfortunately, for both \data{Twitter-UDI} and \data{GooglePlus} we were unable at the time of this writing to gather adequate  information to produce a data-driven ranking, also due to the restrictive usage limits of both networks APIs. Note  that the information used to generate \algo{DD} for \data{Twitter-Kwak} was substantially  incomplete and obsolete to be used for \data{Twitter-UDI}.

\paragraph*{Assessment criteria.}\ 
 In order to comparatively evaluate  our proposed methods' performance 
with respect to   the competing methods, 
we resorted to   well-known assessment criteria, namely 
\emph{Kendall tau rank correlation coefficient}~\cite{abdi07}
\emph{Fagin's intersection metric}~\cite{FaginKS03} 
and \textit{Bpref}~\cite{BuckleyV04}.

Kendall correlation evaluates the similarity between two rankings, expressed as sets of ordered pairs,  based on the number of inversions of   pairs which are needed to transform one ranking into the other. Formally: 
$$
\tau(\L',\L'') = 1 - \frac{2\Delta(\mathcal{P}(\L'),\mathcal{P}(\L''))}{M(M-1)}
$$
where $\L'$ and $\L''$ are the two rankings to be compared, $M=|\L'|=|\L''|$ and $\Delta(\mathcal{P}(\L'),\mathcal{P}(\L''))$ is the symmetric difference distance between the two rankings, calculated as number of unshared  pairs between the two lists.   
The score returned by $\tau$ is in the interval $[-1,1]$, where a value of $1$ means that the two rankings are identical and a value of $-1$ means that one ranking is the reverse of the other.    

Fagin measure allows for determining how well two ranking lists are in agreement with each other.  
This is regarded as the problem of comparing ``partial rankings'', since elements in one list may not be present in the other list. Moreover, according to~\cite{Webber10}, a ranking evaluation 
measure should consider top-weightedness, i.e., the top of the list gets higher weight than the tail.   
Applied to any two top-$k$ lists $\L',\L''$, the Fagin score is defined as:  
\begin{equation*} \label{eq:fagin}
F(\L',\L'',k) = \frac{1}{k}\sum_{q=1}^k \frac{|\L'_{:q} \cap \L''_{:q}|}{q}
\end{equation*}
where $\L_{:q}$  denotes the sets of nodes  from the 1st  to the $q$th position in the   ranking.
Therefore,  $F$ is the average over the sum of the weighted overlaps based on the first $k$ nodes  in both rankings.

Bpref~\cite{BuckleyV04}  evaluates the performance from a different view, i.e., the number of non-relevant candidates. 
It computes a preference relation of whether judged relevant candidates $R$ of a list $\L'$ 
are retrieved, i.e., occur  in a list $\L''$, ahead of judged irrelevant candidates $N$, and 
 is formulated as 
 $$
 Bpref(R, N)\!\!=\!\!\frac{1}{|R|}\!\sum_{r}\!\left(1\!-\!\frac{\# \mbox{of\ } n \mbox{\ ranked higher than\ } r}{|R|}\right)
 $$ 
 where $r$ is a relevant retrieved candidate, and $n$ is a member of the first $|R|$ irrelevant retrieved candidates. In our setting, we first determined  $N$     as the set of nodes with data-driven ranking score  below or equal to 1, and used it for comparisons with \algo{DD}, when available; 
whereas, for  comparisons among competing methods, $N$ was defined as either the bottom of the corresponding method's  ranking having the same size as $N$ in the  data-driven ranking, or (when \algo{DD} is not available) as the bottom-25\% of the method's ranking.  $R$ was selected as the set of nodes having  top-$l\%$  score from the complement of $N$.

Both $F$ and $Bpref$ are within  $[0,1]$, whereby values closer to $1$ correspond to better scores.  
For the experiments discussed in the following, we setup  the size $k$ of the top-ranked lists for Fagin evaluation 
to $k=10^2, 10^3, 10^4$, and the $l\%$ of relevant candidates for Bpref evaluation to $l=10, 25, 50$ (i.e., relevant candidates in the 90th percentile, the third quartile and the median). 
Moreover, unless otherwise specified, $F$ scores will correspond to ranking lists without     sink nodes,  
 in order to avoid biasing (presumably overstating) our evaluation with trivial lurkers.

\begin{table*}[t!]
\caption{Reciprocity and lurking. \textit{rle} is the number of reciprocal lurking edges (i.e., reciprocal edges in the lurking-induced network graph) divided by the total number of edges in the original graph.}
\centering 
\scalebox{0.7}{
\begin{tabular}{|l||c||c|c|c||c|c|c||c|c|c|}
\hline
 	&		&	  		\multicolumn{3}{|c||}{top-25\% of the \algo{LRin-out} solution}			&			\multicolumn{3}{|c||}{top-10\% of the \algo{LRin-out}  solution}		&				\multicolumn{3}{|c|}{top-5\% of the \algo{LRin-out}  solution}					\\ \cline{3-11}
	 	 &  \!\!\textit{\# recip. edges}\!\! 		&	\textit{\# edges}	&	\textit{\# reciprocal} 	&	\textit{\% rle} 	&	\textit{\# edges}	&	\textit{\# reciprocal} 	&	\textit{\% rle}   &	\textit{\# edges}	&	\textit{\# reciprocal} 	&	\textit{\% rle}  	\\
	&	\textit{(full graph)}	&	\textit{(induced graph)}	&	\textit{lurking edges}	&	 	&	\textit{(induced graph)}	&	\textit{lurking edges}	&	 	 &	\textit{(induced graph)}	&	\textit{lurking edges}	&	 \\ \hline \hline
\data{Flickr}	&	20,603,483	&	23,352,367	&	16,440,872	&	49.61	&	12,349,595	&	8,704,922	&	26.27	&	5,030,759	&	3,192,712	&	9.63	\\
\data{FriendFeed}	&	3,014,306	&	340,935	&	33,654	&	0.18	&	1,096	&	46	&	$<$0.01	&	2	&	0	&	0.00	\\
\data{GooglePlus}	&	2,870,336	&	1,413,468	&	667,422	&	4.88	&	49,481	&	23,562	&	0.17	&	5,310	&	2,624	&	0.02	\\
\data{Twitter-Kwak}	&	52,137,192	&	7,293	&	2,806	&	$<$0.01	&	216	&	52	&	$<$0.01	&	64	&	10	&	$<$0.01	\\
\data{Twitter-UDI}	&	191,858,256	&	18,839,845	&	10,078,339	&	3.54	&	3,094,341	&	1,198,615	&	0.42	&	872,332	&	271,751	&	0.10	\\\hline
\end{tabular}
}
\label{tab:reciprocation}
\end{table*}

\section{Results}\label{sec:results} 
 
We present here our experimental results, which are organized as follows.  
We begin first with an analysis of reciprocity and attachment behaviors of lurkers.   
Section~\ref{sec:ranking-performance} is devoted to present quantitative results on the ranking performance obtained by the proposed and competing methods.  
In Section~\ref{sec:delurkification}, we introduce a  ran\-domization-like model to study how to support  ``self-delurking'' of a network, whereas in Section~\ref{sec:percolation} we present a lurking-oriented percolation analysis.  
Finally, in Section~\ref{sec:qualitative}, we provide a qualitative insight into the methods' ranking behavior.

\textbf{Notations:}  Here we briefly recall main notations that will be used  throughout this section.  
\algo{LR} and \algo{ac-LR} prefixed abbreviations refer to our proposed LurkerRank methods (cf. Section~\ref{sec:complete-formulations}). 
The following notations are abbreviations for the competing methods (cf. Section~\ref{sec:exp:criteria}): 
\algo{IO} stands for in/out-degree ratio ranking; 
\algo{PR}, \algo{PR}, and \algo{FB} stand for PageRank, alpha-centrality, and  Fair-Bets model, respectively. Moreover,  \algo{DD} symbols refer to  data-driven rankings.

\subsection{Lurker reciprocity and attachment}
\label{sec:reciprocity}

We aimed at understanding two different aspects of the lurking behaviors: (1) how lurkers relate to each other, in terms of \textit{link reciprocity},  
and (2) how lurker distribution grows with respect to active users, which can be explained in terms of \textit{attachment} mechanisms. 
 
\paragraph{Reciprocity.\ } 
We examined the impact of the presence of lurkers on measures of reciprocity in the various network graphs, 
under three different settings that correspond to   the  top-25\%, top-10\% and top-5\%, respectively, of a LR ranking solution.  
Specifically, we considered four measures of reciprocity, namely (i) the number of reciprocal lurking edges  (i.e., reciprocal edges in the lurking-induced network graph), 
(ii)  the percentage of reciprocal lurking edges  to the total number of edges in the original graph (denoted as \textit{rle}), (iii) the fraction of reciprocal edges in the original network graph that connect lurkers to each other, and  (iv) the fraction of edges that connect lurkers to each other within a  lurking-induced subgraph.

Table~\ref{tab:reciprocation} reports results obtained by the \algo{LRin-out} method. 
A first remark is that  \textit{rle}   was very small or negligible regardless of the portion of LR ranking solution considered. An exception was represented by \data{Flickr}, whose \textit{rle} varied from about 50\% to 10\%; this could be explained as an effect of the crawling mechanism used to build the \data{Flickr} network dataset,  since unlike the other datasets, it was obtained  starting from a single seed user and then performing a breadth-first search  on the social network graph.   
Considering the fraction of reciprocal edges in the original network graph that connect lurkers to each other  (results not shown), again with the exception of \data{Flickr} we observed a very small value even for the case of top-25\% lurkers (around 23\% for \data{GooglePlus}, 5\% for \data{Twitter-UDI}, and below 1\% for \data{FriendFeed} and \data{Twitter-Kwak}), while approaching zero when the top-ranked solution is narrowed to 10\% or smaller.

 \begin{figure}[t!]
\centering
\includegraphics[width=0.35\textwidth]{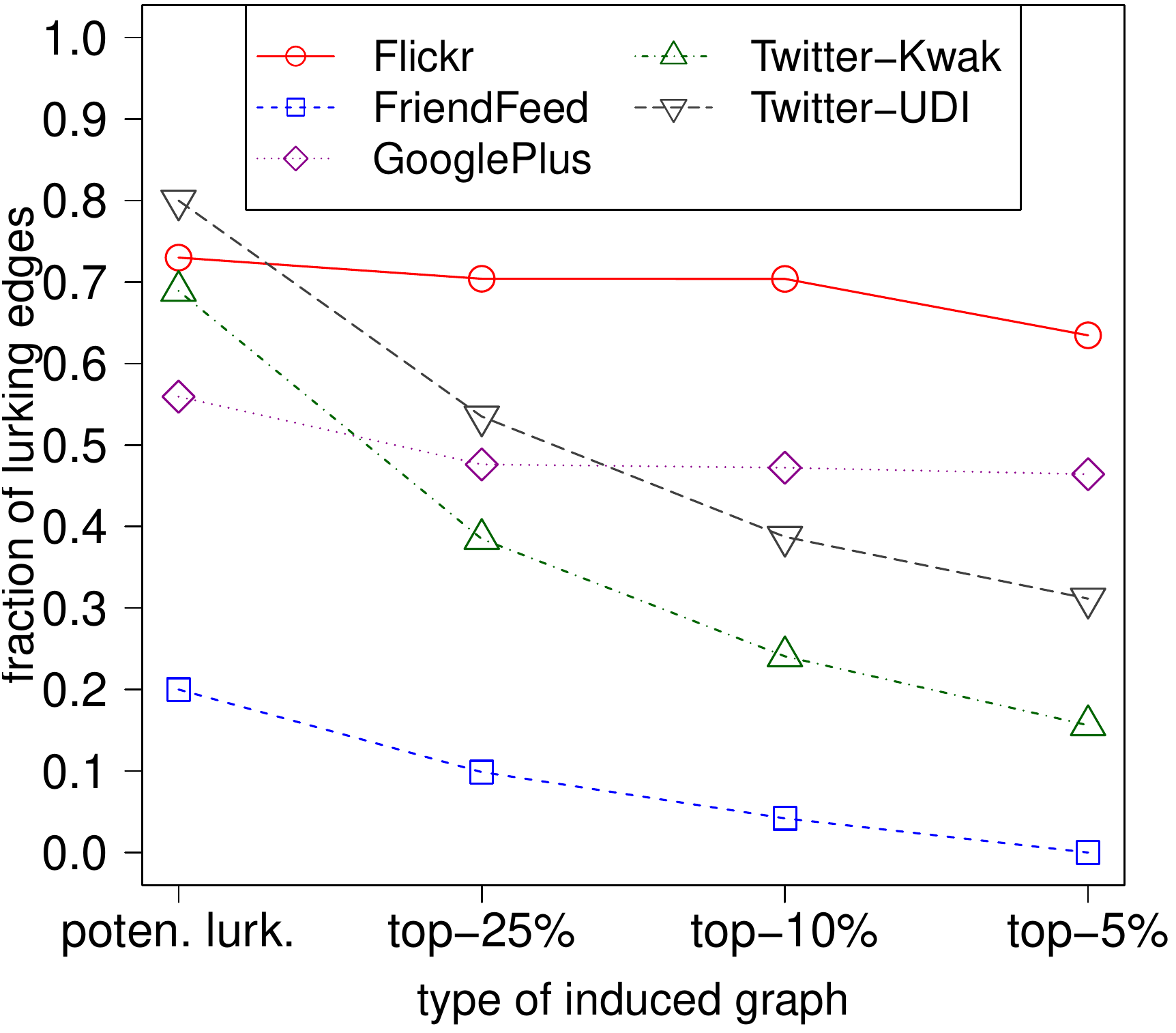} 
\caption{Fraction of reciprocal edges in the lurking-induced subnetworks.}
\label{fig:reciprocal-lurkers}
\end{figure}
 
 Note that, while  \algo{LRin} behaved very similarly to \algo{LRin-out},  results obtained by  \algo{LRout} showed that   \textit{rle} values were significantly higher than those observed in Table~\ref{tab:reciprocation}, with averages over the datasets equal to  35\% (top-25\%), 27\% (top-10\%), and 20\% (top-5\%). 
 Even higher were the values of the fraction of reciprocal edges in the original network graph   connecting lurkers, with peaks above 90\% in the top-25\% case, and averages of 85\%  (top-25\%),  63\%  (top-10\%), and   45\% (top-5\%). 
 These findings were actually  not surprising since \algo{LRout} is designed to emphasize the  lurking attitude of any node from which a target node receives information.

  \begin{figure*}[t!]
\centering
\begin{tabular}{cccc}
\hspace{-3mm}
\includegraphics[width=0.24\textwidth]{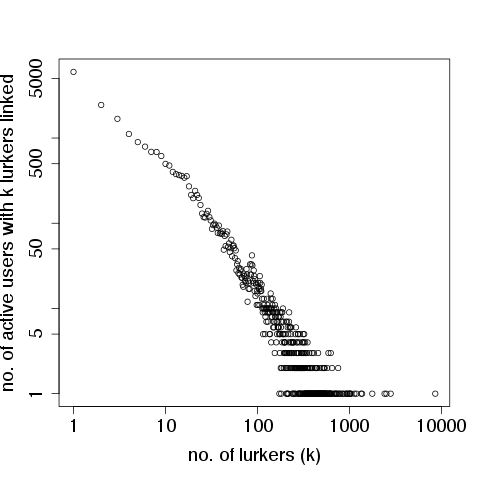} & \hspace{-3mm}
\includegraphics[width=0.24\textwidth]{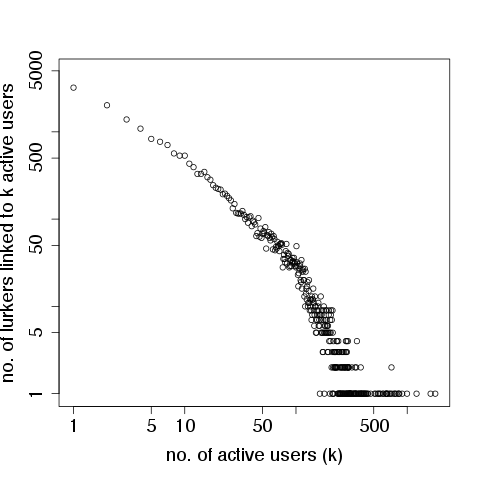} & 
\hspace{-3mm}
\includegraphics[width=0.24\textwidth]{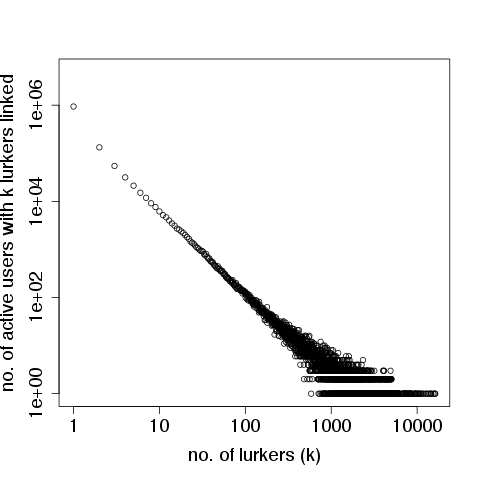} & \hspace{-3mm}
\includegraphics[width=0.24\textwidth]{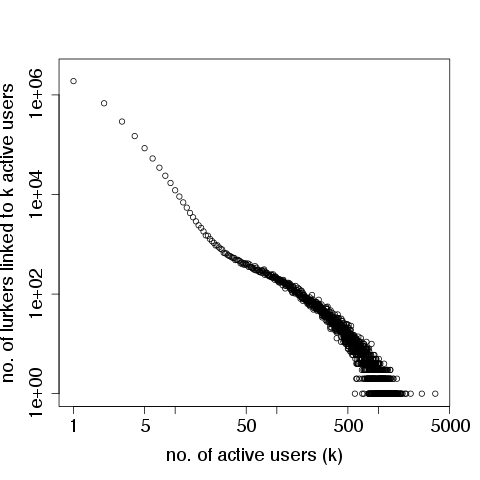}\\
(a) & (b) & (c) & (d)
\end{tabular}
\caption{Distribution of active users as a function of the lurkers-followers (a)-(c) and distribution of lurkers as a function of the active users-followees (b)-(d). (\data{GooglePlus}, two plots from the left, and \data{Twitter-UDI}, two plots from the right).}
\label{fig:powerlaw-static}
\end{figure*}
 
Figure~\ref{fig:reciprocal-lurkers}, as complementary to  Table~\ref{tab:reciprocation}, shows the fraction of edges that connect lurkers to each other within a  lurking-induced subgraph. In the figure, we also included for comparison the case of  ``potential lurkers'', regarding them as those nodes having in/out-degree ratio above 1. 
An evident  remark is that the reciprocity between lurkers   generally followed a decreasing trend varying from the ``potential lurkers'' to the top-5\% setting; this trend was quite slow or roughly stagnant on three out of five datasets (i.e.,  \data{Flickr}, \data{GooglePlus}, and \data{FriendFeed}) but much sharper  in the two largest networks (i.e., the two Twitter datasets). 
Interestingly, when considering \algo{LRout} instead of \algo{LRin-out} or \algo{LRin},  the fraction of reciprocal edges in the lurking-induced subgraph was in general not longer observed as a decreasing function by decreasing sizes of lurker sets; the trend was rather increasing for the Twitter datasets (upper values of 0.78 for \data{Twitter-UDI} and 0.60 for \data{Twitter-Kwak}) and for \data{FriendFeed} (upper value of 0.32).

\paragraph{Attachment.\ } 
We focus now on the relation between lurkers and the ``active'' users they are linked to. 
Specifically, we  analyzed the distribution of lurkers  as function of the degree of attached active users, and dually  for  the distribution of active users. 
For this analysis, we selected the same fraction (25\%) from the top and from the bottom of the \algo{LRin-out} ranking solution in order to choose the set of lurkers and the set of active users, respectively, under examination.

Our goal was to understand whether the probability of observing active users with a certain degree of attached lurkers, and vice versa, can be predicted by a power law. Therefore, for each dataset, we learned the best fit of a power law distribution to the observed data, where the statistical significance of this fitting  was assessed based on a  Kolmogorov-Smirnov test.  
The resulting plots obtained on our datasets showed  a power law behavior for both the distribution of lurkers (following   $k$ active users) and the distribution of active users (followed by $k$ lurkers); Figure~\ref{fig:powerlaw-static}  shows the plots for \data{GooglePlus} and \data{Twitter-UDI}.    
The exponent of the fitted power law distributions varied from  1.67 (\data{GooglePlus} and \data{Twitter-Kwak}) to 2 (\data{FriendFeed}, \data{Twitter-UDI}), for the distribution of active users, and from 1.36 (\data{Flickr}, \data{Twitter-Kwak}) to 1.86 (\data{GooglePlus}), for the distribution of lurkers.
Significant fitting was actually found in general for both distributions in each dataset, which would indicate that they may follow a preferential attachment mechanism: active users, who already are followed by a large number of lurkers, are likely to attract even more lurkers; analogously, lurkers, who already follow a large number of active users, are more likely to do so. 
Moreover, as smaller values of the Kolmogorov-Smirnov statistic  denote better fit, we observed a slight tendency of better explaining  the growing of the number of lurkers (rather than of active users) by preferential attachment on \data{GooglePlus} and \data{Twitter-UDI}, while an opposite situation was found on \data{Flickr}.

\subsection{Ranking evaluation}
\label{sec:ranking-performance}

\begin{table*}[t!]
\caption{Comparative performance of LurkerRank methods and competitors with respect to data-driven rankings: Kendall tau rank correlation values (with 95\% confidence intervals in parentheses).} 
\centering 
\scalebox{0.67}{
\begin{tabular}{|l||c|c|c|c||c|c|c|c|c|c|}
\hline
  \algo{dataset} & \algo{IO} &  \algo{PR} & \algo{AC}   & \algo{FB} &	\algo{LRin} &	\algo{LRout} 	&	\algo{LRin-out} 	&	\algo{ac-LRin} 	&	\algo{ac-LRout}	&	\algo{ac-LRin-out}
\\ \hline \hline
\data{FriendFeed} & .169 ($\pm$ .003) & .128 ($\pm$ .004) & .230 ($\pm$ .005) & .373 ($\pm$ .004) & .661 ($\pm$ .003) & -.169 ($\pm$ .005) & .497 ($\pm$ .003) & \textbf{.664} ($\pm$ .003) & -.189 ($\pm$ .005) & .470 ($\pm$ .003) \\
\data{Flickr} vs \algo{DD-V}\!\!& .046 ($\pm$ .008) & .043 ($\pm$ .005) & .043 ($\pm$ .008) & .047 ($\pm$ .002) & .247 ($\pm$ .007) & \textit{-.007} ($\pm$ .013) & .239 ($\pm$ .014) & .234 ($\pm$ .014) & \textit{.011} ($\pm$ .014) & \textbf{.251} ($\pm$ .013) \\
\data{Flickr} vs \algo{DD-F}\!\!& .052 ($\pm$ .007) & .049 ($\pm$ .005) & .049 ($\pm$ .008) & .053 ($\pm$ .002) & .231 ($\pm$ .006) & \textit{.003} ($\pm$ .012) & .260 ($\pm$ .013) & .255 ($\pm$ .013) & \textit{.011} ($\pm$ .014) & \textbf{.273} ($\pm$ .012) \\
\data{Twitter-Kwak} & .171 ($\pm$ .006) & \textit{.004} ($\pm$ .011) & .215 ($\pm$ .010) & .235 ($\pm$ .012) & \textbf{.671} ($\pm$ .007) & -.082 ($\pm$ .004) & .559 ($\pm$ .008) & .659 ($\pm$ .008) & -.073 ($\pm$ .004) & .560 ($\pm$ .008) \\
\hline
\end{tabular}
}

\vspace{-.5em}
\scriptsize
\begin{flushleft}
Bold values refer to the highest correlation per dataset. 
All values except those in italic  are statistically significant (under the null hypothesis of independence of two rankings).
\end{flushleft}
\label{tab:kendallDDrank}
\end{table*}

\paragraph{Correlation analysis with data-driven rankings.\ }
Table~\ref{tab:kendallDDrank} shows the Kendall tau rank correlation obtained by   our LurkerRank methods and by the competing methods with respect to the data-driven ranking (\algo{DD}) for all eligible datasets.  

The in-neighbors-driven and in-out-neighbors-driven LurkerRank methods generally obtained the highest correlation with  \algo{DD} (e.g.,  $0.67$ by \algo{LRin} on \data{Twitter-Kwak}, 
$0.66$ by  \algo{ac-LRin} on \data{FriendFeed}).   
Results confirmed that \algo{LRin} and \algo{LRin-out} (and their \algo{ac-} counterparts) significantly improved upon all competing methods, with maximum gains of $0.59$ against \algo{IO},   $0.66$ against \algo{PR},   $0.45$ against \algo{AC} and  $0.43$ against \algo{FB}.  

Note that \algo{LRout} and \algo{ac-LRout} obtained the lowest scores on all datasets: 
interestingly, this behavior  confirms our intuition that determining the strength of lurking of a given node should not depend  solely on the strength of the lurking behavior shown by the out-neighbors of that node (i.e., Principle III of  Definition~\ref{def:base}). 

Concerning correlation of each of the competing methods with \algo{DD}, 
we observed on \data{FriendFeed} and \data{Twitter-Kwak} some correlation for \algo{FB} (up to $0.37$) and \algo{AC} (up to $0.23$), while \algo{IO} and \algo{PR} showed poor correlation. However, on  \data{Flickr},   all competing methods tended to  be uncorrelated with the two \algo{DD}, with an average correlation of 0.05  over all competitors. 
More interestingly, it is worth noting that \algo{IO} generally showed poor correlation with \algo{DD}, which not only would justify the use of in/out-degree ranking as a baseline competing method, but also  gives evidence that  in/out-degree cannot be considered as a  basic  approximation of LurkerRank.

\paragraph{Comparative evaluation with LurkerRank methods.\ }
 Tables~\ref{tab:results-twitter}--\ref{tab:results-gplus} compare our LurkerRank methods 
 against PageRank, alpha-centrality, Fair-Bets (all at convergence) as well as against \algo{DD} (where possible) and \algo{IO}. 
 Note that results are organized on 3-row groups, where each row in a group corresponds to a specific variation of  the Fagin's or Bpref's parameters. 
 
On \data{Twitter-Kwak} (Table~\ref{tab:results-twitter}),    
\algo{LRin} and \algo{LRin-out}  along with   their \algo{ac-} counterparts showed 
 a relatively much higher $F$ intersection with \algo{DD} (0.516 on average)  and \algo{IO} (0.473) than with 
 \algo{FB} (0.08), and a nearly empty $F$   with respect to \algo{PR} and \algo{AC}. 
By contrast, \algo{LRout} and \algo{ac-LRout} exhibited a larger $F$ with \algo{PR}, although 
below 0.316 on average, while   scoring even lower with respect to the other methods. 
Bpref evaluation led to mostly similar remarks on the relative comparison between proposed and other methods: 
\algo{LRin}, \algo{LRin-out}  and their \algo{ac-} counterparts     highly matched 
  \algo{DD} and \algo{IO}  (around 0.97 on average), but also a moderately high 
  $Bpref$ with respect to \algo{AC}  (0.87)   and 
  mid-low $Bpref$  with respect to \algo{FB}  (0.47). 
 Again, as already observed  for both the Kendall evaluation and the Fagin evaluation, \algo{LRout} and \algo{ac-LRout}   showed no significant matches 
in practice with \algo{DD} (while scoring pretty high with respect to \algo{PR}).

\begin{table}[t!]
\caption{Comparative performances on \data{Twitter-Kwak}.}
\centering 
\scalebox{0.7}{
\begin{tabular}{|l||c||c|c|c|c||c||c|c|c|c|}
\hline
  & \multicolumn{5}{|c||}{$F$}  
  & \multicolumn{5}{|c|}{$Bpref$}  \\
    & \multicolumn{5}{|c||}{$k=10^2~//~10^3~//~10^4$}  & \multicolumn{5}{|c|}{$l=10~//~25~//~50$}  \\
  \cline{2-11}
  & \algo{DD} & \algo{IO} &  \algo{PR} & \algo{AC}   & \algo{FB}  
  & \algo{DD} & \algo{IO} &  \algo{PR} & \algo{AC}   & \algo{FB}
  \\ \hline \hline
\!\!\algo{LRin}  &  .527  &  .404  &  0.0   &  0.0  &  .112        &  \bfseries{\underline{.997}}  &  .992  &  .121   &  .790  &  .441 \\ 
                                &  .289  &  .209  &  0.0   &  0.0  &  .127        &  .995  &  .989  &  .473   &  .914  &  .704 \\  
                                &  .581  &  \HL{.617}  &  .001   &  .001  &  .068        &  .985  &  .962  &  .521   &  .866  &  .606 \\   \hline
\!\!\algo{LRout} &  .030  &  .032  &  .181   &  .010  &  .034        &  .045  &  0.0  &  .754   &  .311  &  .313 \\ 
                                 &  .008  &  .008  &  .351   &  .024  &  .015        &  .055  &  .001  &  \HL{.757}   &  .650  &  .600 \\ 
                                 &  .003  &  .002  &  \HL{.437}   &  .048  &  .005        &  .109  &  .074  &  .641   &  .678  &  .648 \\  \hline
\!\!\algo{LRin-out}\!\! &  .475  &  .364  &  0.0   &  0.0  &  .064        &  .968  &  \HL{.981}  &  .039   &  .826  &  .204 \\ 
                                       &  .314  &  .277  &  0.0   &  0.0  &  .063        &  .979  &  .977  &  .387   &  .929  &  .524 \\  
                                       &  .666  &  \HL{.688}  &  .001   &  .001  &  .032        &  .961  &  .925  &  .453   &  .878  &  .489 \\ \hline
\!\!\algo{ac-LRin}  &  .583  &  .459  &  0.0   &  0.0  &  .174        &  \HL{.993}  &  .990  &  .072   &  .808  &  .339 \\ 
                                      &  .573  &  .570  &  0.0   &  0.0  &  .122        &  .992  &  .988  &  .443   &  .921  &  .653 \\ 
                                      &  .767  &  \bfseries{\underline{.810}}  &  .001   &  .001  &  .048        &  .982  &  .967  &  .501   &  .872  &  .575 \\  \hline
\!\!\algo{ac-LRout}\!\!\!\! &  .038  &  .032  &  .244   &  .006  &  .036        &  .049  &  0.0  &  \HL{.796}   &  .339  &  .307 \\  
                                        &  .009  &  .008  &  .319   &  .017  &  .011        &  .059  &  0.0  &  .775   &  .659  &  .598 \\  
                                        &  .003  &  .002  &  \HL{.362}   &  .042  &  .004        &  .120  &  .081  &  .654   &  .687  &  .643 \\ \hline
\!\!\algo{ac-LRin-}\!\!   &  .473  &  .363  &  0.0   &  0.0  &  .062        &  .957  & \HL{.981}  &  .039   &  .828  &  .203 \\  
\!\!\algo{out}\!\!                                                      &  .278  &  .234  &  0.0   &  0.0  &  .062        &  .975  &  .976  &  .386   &  .930  &  .464 \\  
                                                      &  .663  &  \HL{.685}  &  .001   &  .001  &  .031        &  .957  &  .933  &  .453   &  .880  &  .454 \\  \hline
\end{tabular}
}

\vspace{-.5em}
\scriptsize
\begin{flushleft}
Bold values refer to the highest scores per LurkerRank method and assessment criterion. 
Underlined bold values refer to the highest scores per assessment criterion.
\end{flushleft}
\label{tab:results-twitter}
\end{table}

\begin{table}[t!]
\caption{Comparative performances on \data{Twitter-UDI}.}
\centering 
\scalebox{0.7}{
\begin{tabular}{|l||c|c|c|c||c|c|c|c|}
\hline
  & \multicolumn{4}{|c||}{$F$}  
  & \multicolumn{4}{|c|}{$Bpref$}  \\
    & \multicolumn{4}{|c||}{$k=10^2~//~10^3~//~10^4$}  & \multicolumn{4}{|c|}{$l=10~//~25~//~50$}  \\
  \cline{2-9}
  & \algo{IO} &  \algo{PR} & \algo{AC}   & \algo{FB}  
   & \algo{IO} &  \algo{PR} & \algo{AC}   & \algo{FB}
  \\ \hline \hline
\!\!\algo{LRin}    & .337 & 0.0 & 0.0 & .184  & .809 & .254 & .230 & .477 \\ 
                                  & .245 & 0.0 & 0.0 & .136  & .917 & .645 & .633 & .546 \\ 
                                  & \HL{.455} & 0.0 & 0.0 & .292  & \HL{.927} & .709 & .715 & .568 \\   \hline
\!\!\algo{LRout}   & 0.0 & 0.0 & 0.0 & 0.0  & 0.0 & .867 & .767 & .164 \\ 
                                   & 0.0 & .004 & 0.0 & .002  & 0.0 & \HL{.876} & .766 & .339 \\ 
                                   & .001 & \HL{.021} & .006 & .001  & .275 & .810 & .732 & .531 \\   \hline
\!\!\algo{LRin-out}   & \HL{.305} & 0.0 & 0.0 & .160  & .762 & .130 & .123 & .299 \\ 
                                         & .178 & 0.0 & 0.0 & .078  & .897 & .546 & .550 & .405 \\ 
                                         & .172 & 0.0 & 0.0 & .076  & \HL{.902} & .646 & .656 & .443 \\   \hline
\!\!\algo{ac-LRin}    & .343 & 0.0 & 0.0 & .186  & .825 & .216 & .202 & .454 \\ 
                                        & .267 & 0.0 & 0.0 & .152  & .924 & .617 & .617 & .524 \\ 
                                        & \bfseries{\underline{.446}} & 0.0 & 0.0 & .324  & \HL{.932} & .690 & .704 & .550 \\   \hline
\!\!\algo{ac-LRout}   & 0.0 & 0.0 & 0.0 & 0.0  & 0.0 & .861 & .765 & .159 \\ 
                                          & 0.0 & .004 & 0.0 & .002  & 0.0 & \HL{.873} & .765 & .338 \\  
                                          & .001 & \HL{.021} & .006 & .001  & .272 & .807 & .730 & .530 \\   \hline
\!\!\algo{ac-LRin-out}\!\!     & \HL{.306} & 0.0 & 0.0 & .161  & .877 & .113 & .153 & .140 \\ 
                                                        & .176 & 0.0 & 0.0 & .076  & .947 & .482 & .607 & .293 \\ 
                                                        & .153 & 0.0 & 0.0 & .060  & \bfseries{\underline{.949}} & .598 & .692 & .399 \\   \hline
\end{tabular}
}

\vspace{-.5em}
\scriptsize
\begin{flushleft}
Bold values refer to the highest scores per LurkerRank method and assessment criterion. 
Underlined bold values refer to the highest scores per assessment criterion.
\end{flushleft}
\label{tab:results-twitterudi}
\end{table}

\begin{table}[t!]
\caption{Comparative performances on \data{Flickr}.}    
\centering 
\scalebox{0.58}{
\begin{tabular}{|l||c|c||c|c|c|c||c|c||c|c|c|c|}
\hline
  & \multicolumn{6}{|c||}{$F$}  
  & \multicolumn{6}{|c|}{$Bpref$}  \\
    & \multicolumn{6}{|c||}{$k=10^2~//~10^3~//~10^4$}  & \multicolumn{6}{|c|}{$l=10~//~25~//~50$}  \\
  \cline{2-13}
  & \!\!\algo{DD-F}\!\!  & \!\!\algo{DD-V}\!\! & \algo{IO} &  \algo{PR} & \algo{AC}   & \algo{FB}  
  & \!\!\algo{DD-F}\!\!  & \!\!\algo{DD-V}\!\! & \algo{IO} &  \algo{PR} & \algo{AC}   & \algo{FB}
  \\ \hline \hline
\!\!\algo{LRin} & .576 & .574 & \HL{.639} & 0.0 & 0.0 & .552                & .361 & .327 & .921 & .465 & .769 & .502 \\
                               & .451 & .433 & .511 & .003 & .007 & .463           & .532 & .496 & .953 & .522 & .783 & .488 \\
                               & .297 & .286 & .383 & .018 & .008 & .313           & .650 & .630 & .931 & .499 & \bfseries{\underline{.987}} & .570 \\   \hline
\!\!\algo{LRout}  & .102 & .101 & .123 & .045 & 0.0 & .037          & .071 & .060 & .206 & .620 & .862 & .138  \\ 
                                & .124 & .121 & .107 & .064 & 0.0 & .008            & .252 & .218 & .509 & .503 & .868 & .229 \\ 
                                & .015 & .014 & .126 &  \HL{.237} & .007 & .033         & .460 & .446 & .645 & .411 & \HL{.878} & .392 \\   \hline
\!\!\algo{LRin-out}  & .561 & .559 &  \HL{.626} & 0.0 & 0.0 & .536 & .353 & .321 &  .878 & .441 & .761 & .520  \\
                                      & .462 & .444 & .520 & .004 & .007 & .462 & .305 & .292 &  \HL{.883} & .474 & .766 & .509 \\
                                      & .311 & .301 & .398 & .021 & .008 & .310 & .430 & .417 & .667 & .478 & .748 & .594 \\   \hline
\!\!\algo{ac-LRin} & .609 & .607 & \bfseries{\underline{.676}} & 0.0 & 0.0 & .587               & .349 & .316 & .878 & .458 & .784 & .498 \\
                                     & .535 & .513 & .604 & .004 & .007 & .538          & .523 & .487 &  \HL{.940} & .484 & .792 & .482 \\
                                     & .348 & .336 & .447 & .018 & .009 & .352           & .644 & .625 & .921 & .481 & .795 & .573 \\   \hline
\!\!\algo{ac-LRout}\!\!   & .102 & .009 & .123 & .051 & 0.0 & .037          & .071 & .060 & .209 & .622 & .660 & .138 \\
                                       & .105 & .101 & .107 & .072 & 0.0 & .008           & .256 & .220 & .514 & .510 & .670 & .232 \\
                                       & .115 & .114 & .127 &  \HL{.229} & .007 & .034         & .477 & .464 & .645 & .413 &  \HL{.675} & .392 \\   \hline
\!\!\algo{ac-LRin-}\!\!  & .443 & .440 &  \HL{.510} & 0.0 & 0.0 & .432 & .375 & .345 & .958 & .604 & .640 & .520 \\
\!\!\algo{out}\!\!  & .305 & .293 & .337 & .002 & .004 & .291 & .569 & .533 &  \HL{.970} & .675 & .677 & .466 \\
                                                     & .232 & .224 & .293 & .013 & .006 & .215 & .676 & .655 & .954 & .569 & .706 & .494 \\   \hline
\end{tabular}
}

\vspace{-.5em}
\scriptsize
\begin{flushleft}
Bold values refer to the highest scores per LurkerRank method and assessment criterion. 
Underlined bold values refer to the highest scores per assessment criterion.
\end{flushleft}
\label{tab:results-flickr}
\end{table}

\begin{table}[t!]
\caption{Comparative performances on \data{FriendFeed}.}
\centering 
\scalebox{0.7}{
\begin{tabular}{|l||c||c|c|c|c||c||c|c|c|c|}
\hline
  & \multicolumn{5}{|c||}{$F$}     
  & \multicolumn{5}{|c|}{$Bpref$}  \\ 
    & \multicolumn{5}{|c||}{$k=10^2~//~10^3~//~10^4$}  & \multicolumn{5}{|c|}{$l=10~//~25~//~50$}  \\
  \cline{2-11}
  & \algo{DD} & \algo{IO} &  \algo{PR} & \algo{AC}   & \algo{FB}  
  & \algo{DD} & \algo{IO} &  \algo{PR} & \algo{AC}   & \algo{FB}
  \\ \hline \hline 
\!\!\algo{LRin}  &  .542  &  \HL{.690}  &  .024   &  .010  &  .453        &  \bfseries{\underline{1.0}} & .980 & .331 & .606 & .985 \\ 
                                &  .488  &  .586  &  .108   &  .118  &  .384        &  .998 & .976 & .570 & .802 & .977\\  
                                &  .576  &  .628  &  .126   &  .153  &  .493        & .986 & .953 & .678 & .843 & .898 \\   \hline 
\!\!\algo{LRout} &  .015  &  .009 &  .479   &  .620  &  .011        &  .008 & 0.0 & .691 & .672 & .031 \\ 
                                 &  .138  &  .163  &  .550   &  \HL{.725}  &  .167        & .030 & .038 & \HL{.764} & .746 & .066 \\ 
                                 &  .154  &  .156  &  .498  &  .704  &  .184        &  .062 & .110 & .739 & .737 & .258\\  \hline 
\!\!\algo{LRin-out}\!\! &  .207  &  .297  &  .032   &  .042  &  .170        & \HL{.972} & .910 & .252 & .604 & .879 \\ 
                                       &  .278  &  .320  &  .061   &  .064  &  .166        &  .955 & .910 & .553 & .794 & .870 \\  
                                       &  .424  &  \HL{.455}  &  .076   &  .099  &  .338        & .914 & .874 & .642 & .815 & .813\\ \hline 
\!\!\algo{ac-LRin}  &  .575  &  \bfseries{\underline{.735}}  &  .025   &  .014  &  .467        &  \bfseries{\underline{1.0}} & .980 & .300 & .605 & .980 \\ 
                                      &  .520  &  .627  &  .118   &  .131  &  .403        &  .999 & .977 & .548 & .803 & .969  \\ 
                                      &  .603  &  .660  &  .130   &  .161  &  .503        &  .988 & .954 & .661 & .845 & .882 \\  \hline 
\!\!\algo{ac-LRout}\!\!\!\! &  .015  &  .009  &  .479   &  .620  &  .011        & .008 & 0.0 & .691 & .672 & .031 \\  
                                        &  .138  &  .163  &  .550   &  \HL{.725}  &  .167        &  .030 & 0.0 & \HL{.749} & .726 & .066\\  
                                        &  .154  &  .156  &  .498   &  .704  &  .184        & .040 & .080 & .723 & .718 & .257 \\ \hline 
\!\!\algo{ac-LRin-}\!\!   &  .169  &  .243  &  0.0   &  0.0  &  .126       & \HL{.958} & .891 & .237 & .594 & .852\\  
\!\!\algo{out}\!\!                                                      &  .240  &  .273  &  .001   &  .001  &  .122        & .942 & .892 & .546 & .785 & .836 \\  
                                                      &  .400  &  \HL{.426}  &  .041   &  .064  &  .310        &  .898 & .853 & .634 & .803 & .782 \\  \hline
\end{tabular}
}

\vspace{-.5em}
\scriptsize
\begin{flushleft}
Bold values refer to the highest scores per LurkerRank method and assessment criterion. 
Underlined bold values refer to the highest scores per assessment criterion.
\end{flushleft}
\label{tab:results-friendfeed}
\end{table}

\begin{table}[t!]
\caption{Comparative performances on \data{GooglePlus}.}    
\centering 
\scalebox{0.7}{
\begin{tabular}{|l||c|c|c|c||c|c|c|c|c|}
\hline
  & \multicolumn{4}{|c||}{$F$}  
  & \multicolumn{4}{|c|}{$Bpref$}  \\
    & \multicolumn{4}{|c||}{$k=10^2~//~10^3~//~10^4$}  & \multicolumn{4}{|c|}{$l=10~//~25~//~50$}  \\
  \cline{2-9}
   & \algo{IO} &  \algo{PR} & \algo{AC}   & \algo{FB}  
    & \algo{IO} &  \algo{PR} & \algo{AC}   & \algo{FB}
  \\ \hline \hline
\!\!\algo{LRin}    & .742 & 0.0 & 0.0 & .363  &  \bfseries{\underline{1.0}} & .434 & .582 & .976 \\ 
                                  & .850 & .001 & 0.0 & .480  & .993 & .584 & .695 & .962 \\ 
                                  & \HL{.881} & .063 & .144 & .592  & .987 & .684 & .722 & .937 \\   \hline
\!\!\algo{LRout}   & .011 & .079 & 0.0 & .015  & \HL{.972} & .796 & .796 & .686 \\ 
                                   & .015 & .107 & .012 & .015  & .971 & .793 & .790 & .815 \\ 
                                   & .223 & \HL{.322} & .144 & .213  & .964 & .782 & .774 & .807 \\   \hline
\!\!\algo{LRin-out}   & .629 & 0.0 & 0.0 & .318  &  \bfseries{\underline{1.0}} & .462 & .587 & .907 \\ 
                                         & .721 & 0.0 & 0.0 & .419  & .991 & .572 & .688 & .910 \\ 
                                         & \HL{.799} & .045 & .130 & .547  & .989 & .677 & .731 & .886 \\   \hline
\!\!\algo{ac-LRin}    & .747 & 0.0 & 0.0 & .361  &  \bfseries{\underline{1.0}} & .456 & .578 & .976 \\ 
                                        & .851 & .001 & 0.0 & .477  & .992 & .546 & .702 & .963 \\ 
                                        &  \bfseries{\underline{.882}} & .063 & .143 & .591  & .988 & .699 & .724 & .937 \\   \hline
\!\!\algo{ac-LRout}   & .011 & .077 & 0.0 & .015  & \HL{.972} & .796 & .796 & .687 \\ 
                                          & .015 & .107 & .012 & .015  & .971 & .793 & .790 & .815 \\ 
                                          & .223 & \HL{.322} & .145 & .212  & .965 & .782 & .774 & .807 \\   \hline
\!\!\algo{ac-LRin-out}\!\!     & .647 & 0.0 & 0.0 & .328  &  \bfseries{\underline{1.0}} & .489 & .586 & .896 \\ 
                                                        & .729 & 0.0 & 0.0 & .422  & .994 & .612 & .675 & .899 \\ 
                                                        & \HL{.795} & .042 & .125 & .543  & .983 & .702 & .727 & .875 \\   \hline
\end{tabular}
}

\vspace{-.5em}
\scriptsize
\begin{flushleft}
Bold values refer to the highest scores per LurkerRank method and assessment criterion. 
Underlined bold values refer to the highest scores per assessment criterion.
\end{flushleft}
\label{tab:results-gplus}
\end{table}

Results on \data{Twitter-UDI} (Table \ref{tab:results-twitterudi}) corroborated  the advantage of \algo{LRin} and \algo{ac-LRin} with respect to the other \algo{LR} methods.  
 \algo{LRin-out} and \algo{ac-LRin-out} achieved  lower $F$ than \algo{LRin} and \algo{ac-LRin}, respectively, with respect to \algo{IO} and \algo{FB}, especially for higher $k$.  
Compared to the \data{Twitter-Kwak} case,  $Bpref$ values were relatively higher (respectively, lower)  with respect to  \algo{PR}  (respectively, \algo{AC}),  except   for \algo{LRout} and \algo{ac-LRout}  which  had higher $Bpref$ with respect to \algo{AC} than in  \data{Twitter-Kwak}.

On \data{Flickr} (Table~\ref{tab:results-flickr}), once again the best performance against the data-driven ranking (\algo{DD-F} and \algo{DD-V}) was obtained by \algo{LRin} and \algo{LRin-out} along with their \algo{ac-} counterparts, and also roughly similar $F$ values were obtained with respect to  \algo{IO} and \algo{FB}. 
Note that both data-driven ranking (the \textit{favorites}-based one, \algo{DD-F}, and the \textit{views}-based one, \algo{DD-V}) corresponded 
to nearly identical results, with a slightly better agreement of the \algo{LR} algorithms with respect to  \algo{DD-F}.  
In terms of $Bpref$, \algo{LRin}, \algo{LRin-out}  and their \algo{ac-} counterparts     highly matched \algo{IO}.  
$Bpref$ values were also moderately high with respect to \algo{AC} and mid-low with respect to \algo{PR} and \algo{FB}.

Looking at  \data{FriendFeed} results (Table~\ref{tab:results-friendfeed}),  
\algo{LRin} and \algo{LRin-out}  along with   their \algo{ac-} counterparts were again 
the best-performing methods against  \algo{DD} (0.42 $F$ and  0.97  $Bpref$), and 
also showed mid $F$ (0.34) and high $Bpref$ (0.89) with respect to \algo{FB}.    
Yet,   \algo{LRout} and \algo{ac-LRout}  were moderately in agreement  with \algo{PR} and \algo{AC} in terms of $F$, 
whereas all \algo{LR} generally achieved mid $Bpref$ with both \algo{PR} and \algo{AC}. 

\data{GooglePlus} evaluation results (Table~\ref{tab:results-gplus}) led us to  draw conclusions similar to the  other network datasets in terms of  $F$ values: \algo{in-} and \algo {in-out}-based algorithms outperformed the  \algo{out}-based ones when comparing with \algo{IO} and \algo{FB}, while nearly empty intersection was found with respect to \algo{PR} and \algo{AC}. 
\algo{LRin}, \algo{LRin-out}  and their \algo{ac-} counterparts   achieved very high $Bpref$ with respect to  \algo{IO}, and also showed good agreement with \algo{FB}.

\paragraph{Statistical significance testing.\ }
We  also  determined the statistical significance of the better performance of
LurkerRank methods with respect to the competing ones, through two stages of statistical testing analysis; 
in both cases, we fixed the Fagin parameter as $k=10^4$ (which ensured a larger overlap between the 
ranking lists to be compared)   and the Bpref parameter as $l=25$ (for which $|R|$ was always   smaller than $|N|$).  
Results  refer  here to \data{Twitter-Kwak} and \data{FriendFeed}, nevertheless  
similar conclusions were actually reached for the other evaluation networks.

Tables~\ref{tab:ttest-periteration-twitter}--\ref{tab:ttest-periteration-friendfeed} show the p-values 
resulting from an unpaired  two-tail t-test, in which  the performance scores obtained for each iteration by a ranking method with respect to \algo{DD} were regarded as the statistical samples, under the null    hypothesis of no difference in  performance  with respect to  \algo{DD} 
between  a LurkerRank method and a competing method. Note that in all cases, 
the number of iterations (samples) was adequate to perform a t-test (generally above 50).   
Looking at the two tables and both $F$ and $Bpref$ evaluation, 
the p-values turned out to be  extremely low in most cases,  
thus giving a strong evidence that the null hypothesis was always rejected,   at 1\% significance level.  
This finding was  useful to confirm that  a certain difference (actually, the improvement) in performance between 
the \algo{LR} methods and the competing ones, also  on \data{FriendFeed} for which relatively high $Bpref$ scores were observed in the previous analysis.

 \begin{table}[t!]
\caption{\data{Twitter-Kwak} t-test on the  per-iteration   performances.}
\centering 
\scalebox{0.7}{
\begin{tabular}{|l||c|c|c||c|c|c|}
\hline
  & \multicolumn{3}{|c||}{\textit{Fagin evaluation}}  & \multicolumn{3}{|c|}{\textit{Bpref evaluation}}  \\ 
  \cline{2-7}
  &   \algo{PR} & \algo{AC}   & \algo{FB}  
  &   \algo{PR} & \algo{AC}   & \algo{FB}
  \\ \hline \hline
\!\!\algo{LRin} &	4.4E-65	&	4.4E-65	&	8.4E-11	&	5.2E-110	&	1.1E-25	&	2.1E-65	\\ \hline
\!\!\algo{LRout} &	2.8E-41	&	2.7E-41	&	1.8E-04	&	3.2E-50	&	5.5E-79	&	9.2E-71	\\ \hline
\!\!\algo{LRin-out} &	4.3E-277	&	4.4E-277	&	2.9E-12	&	1.5E-89	&	6.7E-21	&	7.6E-65	\\ \hline
\!\!\algo{ac-LRin}&	5.6E-228	&	5.6E-228	&	4.8E-14	&	1.2E-91	&	2.1E-25	&	2.7E-65	\\ \hline
\!\!\algo{ac-LRout}&	6.5E-34	&	6.2E-34	&	1.8E-04	&	4.1E-54	&	1.8E-71	&	2.3E-73	\\ \hline
\!\!\algo{ac-LRin-out}\!\!&	3.8E-213	&	3.3E-265	&	3.4E-12	&	5.8E-85	&	2.1E-21	&	1.0E-64	\\ \hline
\end{tabular}
}
\label{tab:ttest-periteration-twitter}
\end{table}

  \begin{table}[t!] 
\caption{\data{FriendFeed} t-test on the  per-iteration   performances.}
\centering 
\scalebox{0.7}{
\begin{tabular}{|l||c|c|c||c|c|c|}
\hline
  & \multicolumn{3}{|c||}{\textit{Fagin evaluation}}  & \multicolumn{3}{|c|}{\textit{Bpref evaluation}}  \\ 
  \cline{2-7}
  &   \algo{PR} & \algo{AC}   & \algo{FB}  
  &   \algo{PR} & \algo{AC}   & \algo{FB}
  \\ \hline \hline
\!\!\algo{LRin} &	1.3E-116	&	1.3E-103      &	2.6E-10  &	4.5E-195	&	5.9E-197 &	6.1E-10	\\  \hline
\!\!\algo{LRout}  &	8.5E-12	&	1.6E-101	   &	1.5E-38	       &	6.8E-252	&	1.3E-264	&	2.5E-271	\\  \hline
\!\!\algo{LRin-out}  &	6.0E-193	&	2.4E-166	&	2.1E-24	       &	1.3E-298	&		2.1E-212	&	2.2E-116	\\  \hline
\!\!\algo{ac-LRin} &	1.0E-195	&	1.0E-172	    &	4.4E-13	       &	5.0E-298	&	 3.9E-189	&	7.8E-10	\\  \hline
\!\!\algo{ac-LRout} &	2.6E-12	&	5.1E-88	&	1.3E-38	      &	4.1E-99	&	5.9E-299	&	1.4E-282	\\  \hline
\!\!\algo{ac-LRin-out}\!\! &8.1E-63	&	1.3E-96  & 2.1E-25	&	8.3E-82	&	5.1E-226	&	1.5E-75	\\  \hline
\end{tabular}
}
\label{tab:ttest-periteration-friendfeed}
\end{table}

In the second stage of statistical testing, we analogously performed a    paired two-tail t-test  in which the samples corresponded 
to the $F$ scores   respectively obtained by two ranking methods with respect to \algo{DD} over 
the same  randomly generated subgraph. 
For each of the network datasets, we  extracted 100 subgraphs, each time 
starting from a randomly picked seed node and roughly covering a  fixed number of nodes (around 
1/100 of the original network size). This test was hence intended  to stress the ranking methods performing 
over a pool of subnetworks having   different characteristics from each other, and from the  whole original  network as well; for instance, on \data{Twitter-Kwak}, the subnetworks had 
average path length mean of 2.52  (0.86 stdev), and in/out-degree ratio mean  of 0.07  (0.13 stdev) --- this might be explained because of the adopted approach of  breadth-first traversal of the network, which led to 
connect the majority of nodes with a few source nodes having very high out-degree. 
  On \data{Twitter-Kwak}, we observed  a close behavior between the LurkerRank methods (except \algo{LR}\-\algo{out} and \algo{ac-LRout}) and \algo{AC} (around 0.19 $F$ on average), and between \algo{PR} and \algo{FB}, which however achieved a lower average $F$ (0.029) --- note that   $k$ was still set to $10^4$, hence very high for such network sizes (i.e., around 200,000 nodes). In any case, i.e., for each pair of LurkerRank method vs. competing method, the null hypothesis of equal means was rejected even at 1\% significance level, since the p-values were ranging from 1.4E-3 to 2.8E-19.  Analogous final remarks were  drawn  for \data{FriendFeed}. 

\begin{table}[t!]
\caption{Comparative performances on \data{FriendFeed} damping factor depending on the average path length.}    
\centering 
\scalebox{0.7}{
\begin{tabular}{|l||c||c|c|c|c||c||c|c|c|c|}
\hline
  & \multicolumn{5}{|c||}{$F$}      
  & \multicolumn{5}{|c|}{$Bpref$}  \\ 
    & \multicolumn{5}{|c||}{$k=10^2~//~10^3~//~10^4$}  & \multicolumn{5}{|c|}{$l=10~//~25~//~50$}  \\
  \cline{2-11}
  & \algo{DD} & \algo{IO} &  \algo{PR} & \algo{AC}   & \algo{FB}  
  & \algo{DD} & \algo{IO} &  \algo{PR} & \algo{AC}   & \algo{FB}
  \\ \hline \hline 
\!\!\algo{LRin}   & .450 & \HL{.686} & .012 & .003 & .445 & .955 & .978 & .318 & .601 &  \bfseries{\underline{.985}} \\
                                 & .422 & .582 & .079 & .078 & .341 & .914 & .975 & .567 & .800 & .974 \\
                                 & .529 & .627 & .111 & .134 & .472 & .678 & .952 & .673 & .839 & .897 \\ \hline
\!\!\algo{LRout}  & .015 & .072 & .510 & .620 & .015 & .011 & 0.0 & .689 & .672 & .027 \\
                                  & .138 & .070 & .571 & \HL{.725} & .184 & .033 & .041 & \HL{.762} & .747 & .059 \\
                                  & .154 & .208 & .508 & .704 & .189 & .155 & .121 & .738 & .737 & .199  \\ \hline
\!\!\algo{LRin-out}\!\!\  & .205 & .294 & .020 & .031 & .191 & .759 & .909 & .250 & .604 & .871 \\
                                        & .274 & .317 & .053 & .055 & .182 & .744 & \HL{.910} & .553 & .792 & .860 \\
                                        & .421 & \HL{.448} & .074 & .096 & .352 & .602 & .872 & .642 & .813 & .804 \\ \hline
\!\!\algo{ac-LRin}   & .485 &  \bfseries{\underline{.727}} & .016 & .004 & .479 & .961 & .978 & .291 & .600 & \HL{.981} \\
                                       & .450 & .623 & .088 & .090 & .367 & .916 & .975 & .545 & .800 & .967 \\
                                       & .553 & .656 & .115 & .141 & .488 & .679 & .951 & .656 & .841 & .883 \\ \hline
\!\!\algo{ac-LRout}\!\!\  & .015 & .072 & .510 & .620 & .015 & .011 & 0.0 & .689 & .672 & .027 \\
                                         & .138 & .070 & .571 & \HL{.725} & .184 & .033 & .008 & \HL{.747} & .726 & .059 \\ 
                                         & .154 & .208 & .508 & .704 & .189 & .142 & .102 & .721 & .718 & .199 \\ \hline
\!\!\algo{ac-LRin-}\!\!    & .169 & .239 & 0.0 & 0.0 & .140 & .745 & .889 & .237 & .594 & .850 \\
\!\!\algo{out}\!\!                                                        & .240 & .271 & .001 & .001 & .136 & .722 & \HL{.891} & .547 & .785 & .833 \\
                                                       & .400 & \HL{.421} & .042 & .064 & .325 & .592 & .854 & .636 & .803 & .780 \\ \hline
\end{tabular}
}

\vspace{-.5em}
\scriptsize
\begin{flushleft}
Bold values refer to the highest scores per LurkerRank method and assessment criterion. 
Underlined bold values refer to the highest scores per assessment criterion.
\end{flushleft}
\label{tab:results-friendfeed_alfa074}
\end{table}

 \begin{table}[t!]
\caption{Comparative performances on \data{GooglePlus} with damping factor depending on the average path length.}    
\centering 
\scalebox{0.7}{
\begin{tabular}{|l||c|c|c|c||c|c|c|c|}
\hline
  & \multicolumn{4}{|c||}{$F$}     
  & \multicolumn{4}{|c|}{$Bpref$}  \\ 
    & \multicolumn{4}{|c||}{$k=10^2~//~10^3~//~10^4$}  & \multicolumn{4}{|c|}{$l=10~//~25~//~50$}  \\
  \cline{2-9}
   & \algo{IO} &  \algo{PR} & \algo{AC}   & \algo{FB}  
   & \algo{IO} &  \algo{PR} & \algo{AC}   & \algo{FB}
  \\ \hline \hline
\!\!\algo{LRin}   & .729 & 0.0 & 0.0 & .551   & \bfseries{\underline{1.0}} & .438 & .584 & .985 \\
                                 & .829 & .001 & 0.0 & .631   & .989 & .585 & .700 & .963  \\
                                  & \bfseries{\underline{.864}} & .061 & .140 & .690  & .983 & .689 & .725 & .927  \\ \hline
\!\!\algo{LRout}   & .011 & .085 & 0.0 & .022   & .972 & .994 & \HL{.996} & .671 \\
                                   & .015 & .148 & .012 & .018   & .971 & .993 & .990 & .795 \\
                                   & .223 & \HL{.356} & .144 & .232  & .964 & .981 & .974 & .783 \\ \hline
\!\!\algo{LRin-out}   & .629 & 0.0 & 0.0 & .474  & \HL{.997} & .467 & .590 & .940  \\
                                         & .720 & 0.0 & 0.0 & .546  & .989 & .576 & .689 & .915 \\
                                         & \HL{.798} & .047 & .129 & .642   & .980 & .679 & .734 & .876  \\ \hline
\!\!\algo{ac-LRin}    & .732 & 0.0 & 0.0 & .551   & \bfseries{\underline{1.0}} & .459 & .579 & .986  \\
                                        & .830 & .001 & 0.0 & .629   & .990 & .550 & .702 & .963 \\
                                        & \bfseries{\underline{.864}} & .061 & .139 & .689    & .986 & .711 & .726 & .927  \\ \hline
\!\!\algo{ac-LRout}   & .011 & .083 & 0.0 & .022   & .972 & .994 & \HL{.996} & .671  \\
                                          & .015 & .148 & .012 & .018   & .971 & .993 & .990 & .796 \\
                                          & .223 & \HL{.356} & .145 & .232   & .965 & .981 & .974 & .783  \\ \hline
\!\!\algo{ac-LRin-out}\!\!     & .647 & 0.0 & 0.0 & .488   & \HL{.998} & .492 & .590 & .935  \\
                                                        & .729 & 0.0 & 0.0 & .550  & .991 & .623 & .678 & .907  \\
                                                        & \HL{.795} & .044 & .125 & .638   & .984 & .709 & .728 & .866  \\ 
 \hline 
\end{tabular}
}

\vspace{-.5em}
\scriptsize
\begin{flushleft}
Bold values refer to the highest scores per LurkerRank method and assessment criterion. 
Underlined bold values refer to the highest scores per assessment criterion.
\end{flushleft}
\label{tab:results-gplus_alfa07}
\end{table}

\paragraph{Relation between damping factor and average path length.\ }
In our proposed methods, the damping factor $\d$ is chosen to be 0.85, in analogy with the default setting of the parameter in the original PageRank algorithm. Recall this  finds an explanation based on the empirical observation that a web surfer is likely to navigate following 6 hyperlinks (before discontinuing this navigation chain and randomly jumping on another page), which corresponds to a  probability $\d= 1- (1/6) \approx 0.85$. 
On the other hand,  research on degrees-of-separation in directed network graphs has shown that for many OSNs the average path length is typically below 6 (e.g., \cite{BakhshandehSAS11,MisloveMGDB07}). Here we leverage on this result, confirmed in our network datasets as well, to understand how the ranking performance may change as the damping factor is varied in function of a network-specific structural characteristic like the average path length.      
Precisely, we set $\d$   as $\d = 1-(1/apl)$, being $apl$ the average path length of the particular network. 
For this evaluation stage,  we focused on   \data{FriendFeed} and \data{GooglePlus}, which exhibit   the lowest  average path lengths, i.e.,   3.82 and 3.32, respectively (cf. Table~\ref{tab:data}).

Comparing the results in Table~\ref{tab:results-gplus_alfa07} that correspond to $\alpha=0.7$ with the results obtained with default $\alpha$ (Table~\ref{tab:results-gplus}) 
on \data{GooglePlus}, $F$ values were slightly lower (resp. unvaried) for the \algo{in-} and \algo{in-out-}based algorithms, (resp. for the \algo{out-}based algorithms)   with respect to \algo{IO}, generally higher with respect to \algo{FB}, and equal or higher with respect to \algo{PR} and \algo{AC}. 
Again comparing with the results in Table~\ref{tab:results-gplus}, $Bpref$ slightly increased   with respect to \algo{PR} and \algo{AC} and decreased with respect to \algo{IO}.  
As for \data{FriendFeed},   comparing  Table~\ref{tab:results-friendfeed} with Table~\ref{tab:results-friendfeed_alfa074},  we found that $F$ values were generally lower when using $\alpha=0.74$ for \algo{in-} and \algo{in-out}-based algorithms, and higher for \algo{out}-based ones. A decrease in the performance of \algo{in-} and \algo{in-out-}based algorithms was observed for $Bpref$ as well, especially with respect to  \algo{DD}.  

Overall, it appears that the average path length cannot be regarded as  a good estimator of damping factor in our methods, in the sense of a necessarily better  alternative to the default 0.85. However, we would tend to take this sort of conclusion   with a grain of salt, due to the heterogeneity of such networks and the lack of more example networks with average path length significantly below 6.

\begin{figure}[t!]
\centering
\includegraphics[width=0.45\textwidth]{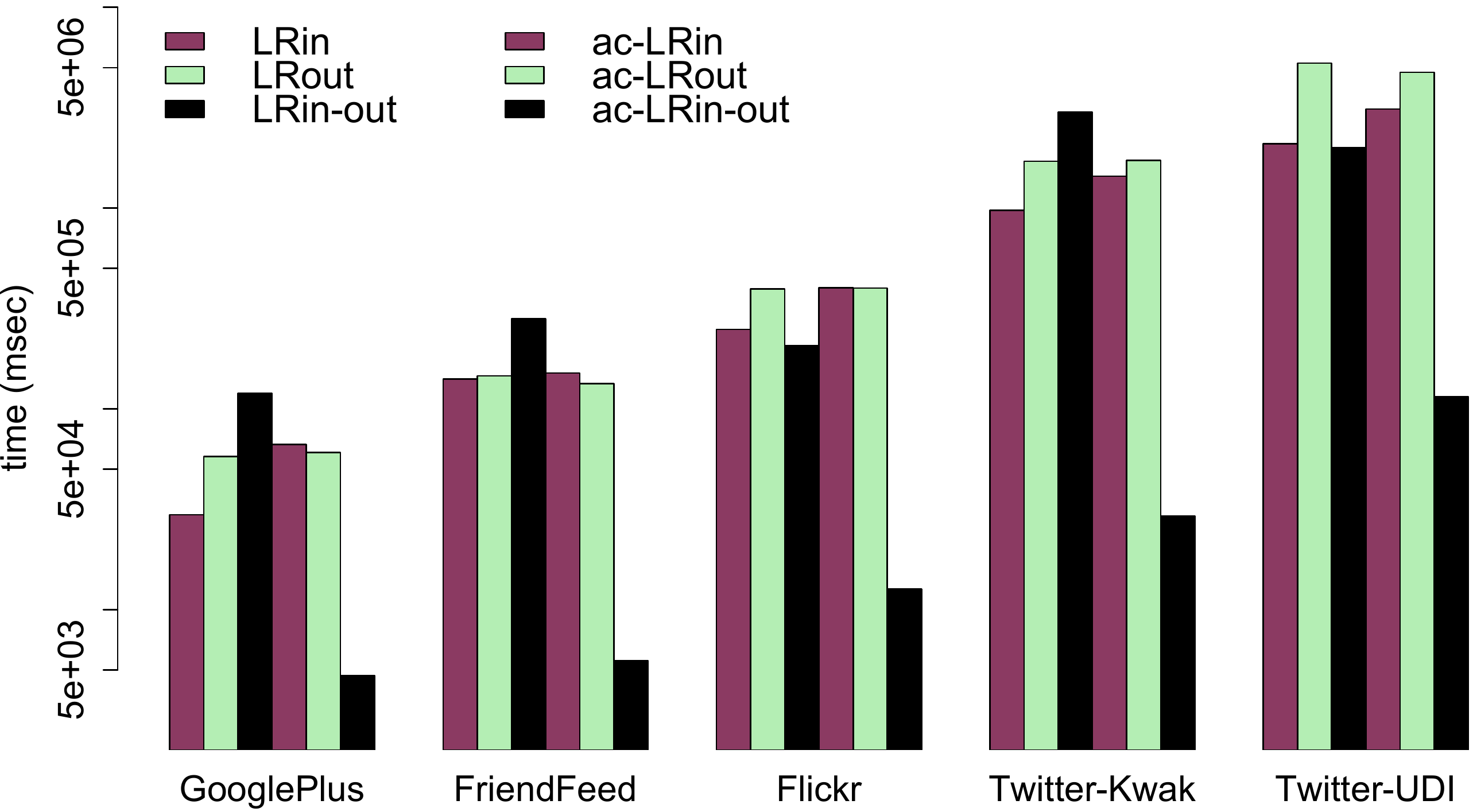}  
\caption{Runtime performance of LurkerRank methods.}
\label{fig:times}
\end{figure}

\paragraph{Efficiency results.\ }
Figure~\ref{fig:times} shows the runtime performance of LurkerRank algorithms. The times do not include the graph building step.\footnote{Experiments were carried out on an Intel Core i7-3960X CPU @ 3.30GHz, 64GB RAM machine.}     
Firstly, it was interesting to observe on all datasets that   
the  LurkerRank methods consistently reached a ranking stability very quickly, in the range 35\textdiv 75 iterations, with the exception of  \algo{ac-LRin-out} which always reached convergence with fewer iterations. The latter fact is however explained by a generally poor  diversification of the ranking scores achieved by \algo{ac-LRin-out}, which particularly affects the top of the ranking results: in fact, in most datasets, the  scores at the maximum as well as the  third quartile are of the same order of magnitude as the mean  or even as the first quartile scores.     
\algo{LRin} and \algo{LRout} mostly required pretty similar running times, while \algo{LRin-out} was slower  than the other algorithms on 3 out of 5 networks --- about twice the running time of  \algo{LRin} and \algo{LRout}, which is clearly   explained since \algo{LRin-out} needs to iterate both on the in- and   out-neighborhood of each node. 
As concerns the alpha-centrality based formulations, \algo{ac-LRin} always required a higher number of iterations to reach ranking stability than \algo{LRin},  
while \algo{ac-LRout} performed similarly and sometimes faster than \algo{LRout}, considering that in most cases both algorithms needed the same number of iterations until ranking stability.  
As a side remark, it should be noted that our power-iteration-method implementation of the \algo{LR} algorithms caused quite different performance for networks with a number of edges of the same order of magnitude, but a greater difference in the number of nodes (e.g., \data{FriendFeed} and \data{Flickr}).

\subsection{Delurking-oriented randomization}
\label{sec:delurkification}

As we discussed in the Introduction, the ultimate objective of lurker analysis is in principle to attract the lurkers to the community life, that is, to change their status to that of active players in the network.  Although  devising real delurking plans (which might rely on marketing aspects) goes beyond our study, we are still interested in conceiving a general topology-based model that can  support ``self-delurking'' of a network.

For this purpose, we introduce a novel randomization-like model, named \textit{delurking-oriented randomization}.   
Randomized models are  commonly used to monitor how varying a certain topological feature   may impact on the dynamics of the network. The most widely applied randomized model uses the concept of rewiring, so that the edges of the original (undirected) network are randomly rewired pairwise. 
The key idea behind our delurking-oriented randomization model is to simulate a mechanism of disclosure of the presence of lurkers, by letting more-likely-active users virtually hear from  less-likely-active users. 

\renewcommand{\algorithmicrequire}{\textbf{Input:}}
\renewcommand{\algorithmicensure}{\textbf{Output:}}
\begin{algorithm}[t!]
\caption{Delurking-oriented randomization\label{algo:delurkification}}
\begin{algorithmic}[1]
\REQUIRE The topology graph $\G = \langle \V, \E \rangle$ of an OSN. 
                      The ranking $L$ corresponding to a \algo{LR} solution for $\G$.  
                      Cut-off percentage thresholds $t_1, t_2$  of ranking order in $L$. 
                      Probability $p$.
                      Maximum  fraction $d$ of new edges to add to $\G$. 
\ENSURE A randomized graph $\G'$.  
\STATE $\E' \leftarrow \emptyset$
\STATE Sort $L$ by decreasing lurking score
\STATE Let $L_{\mathrm{top}}$ (resp. $L_{\mathrm{bottom}}$) be the top-$t_1$ (resp. bottom-$t_2$) of the sorted $L$
\STATE $E_{al} \leftarrow \{e=(a,l) \in \E \ | \ a \in L_{\mathrm{bottom}}, l \in L_{\mathrm{top}} \}$
\REPEAT
\STATE Pick randomly with probability $p$ an edge $(a_1,l_1) \in E_{al} \setminus \E'$
\STATE Pick randomly with probability $p$ an edge $(a_2,l_2) \in E_{al} \setminus \E'$, with $a_2 \neq a_1, l_2 \neq l_1$
\STATE $\E' \leftarrow \E' \cup \{ (l_1,a_2), (l_2,a_1)\}$     $\hfill{\mbox{\textit{/* add the new edges */}}}$
\UNTIL ($|\E'| \geq  d|E_{al}|$)
\STATE $\G' \leftarrow \langle \V, \E' \cup \E \rangle$
\end{algorithmic}
\end{algorithm}

Algorithm~\ref{algo:delurkification} shows our delurking-oriented randomization method, which substantially works by 
inserting new connections into the network each of which randomly links a vertex selected from the top of a predetermined \algo{LR} ranking solution to a vertex selected from the bottom of that ranking. The algorithm hence requires cut-off thresholds to control the selection of the head and tail of the \algo{LR} distribution, and a percentage threshold to control the degree of delurking-oriented randomization (i.e., the fraction of potentially new edges to add to the graph). 
At each step of insertion of a new pair of edges, it is to be ensured that both the new  formed edges    do not already exist in the graph --- this restriction prevents the appearance of multiple edges connecting the same pair of vertices. 
It should be noted that Algorithm~\ref{algo:delurkification} does not provide a proper randomization model in its usual definition, since both the size of the network and the degree of vertices will change.    

We applied Algorithm~\ref{algo:delurkification} to our  networks, with the following setting:  $p=0.5$,   $t_1=t_2=25\%$, and  $d$ ranging from 0.2 to 1.0 (with increment by 0.2). 
Note that this setup of the algorithm was chosen to allow us to focus mainly on the degree of delurking-oriented randomization ($d$); as for the partition of the lurker ranking list, we decided to leave the middle 50\% out and hence select one quartile both for the top ($t_1$) and the bottom ($t_2$) of the ranking list.

For this stage of evaluation, we mainly focused on two features of the network: the \algo{LR} distribution and the in/out-degree distribution (either with and without the inclusion of sink and source vertices), and analyzed   the pairwise correlations  between a \algo{LR}    (resp. in/out-degree) ranking  on a particular   network and the \algo{LR} (resp. in/out-degree) rankings obtained on the corresponding delurking-randomized networks.
 
Considering the case where all vertices were  included in the evaluation, 
we observed no clear trend both in the pairwise correlations between the \algo{LR} ranking solutions at the different degrees of delurking-oriented randomization, which were either moderate   (\data{Flickr}) or high,  and in the correlations  between an original \algo{LR} and each of the \algo{LR} solutions in the randomized networks, which were either absent  (\data{Flickr} and \data{FriendFeed}) or moderate/high. 
However, when sink and source vertices were discarded from the analysis,     trends become more  evident: 
in one case  (corresponding to the \data{Twitter} networks), the pairwise correlations between the \algo{LR} ranking solutions at the different degrees of delurking-oriented randomization were moderate, while absent or moderate with respect to  the original \algo{LR} ranking; however, in the other case (corresponding to \data{GooglePlus}, \data{Flickr}, and \data{FriendFeed}),  the \algo{LR} ranking solutions at the different degrees of delurking-oriented randomization turned out to be not or scarcely correlated to each other as well as  totally uncorrelated to the original \algo{LR} ranking. 

 \begin{figure}[t!]
\centering
\includegraphics[width=0.4\textwidth]{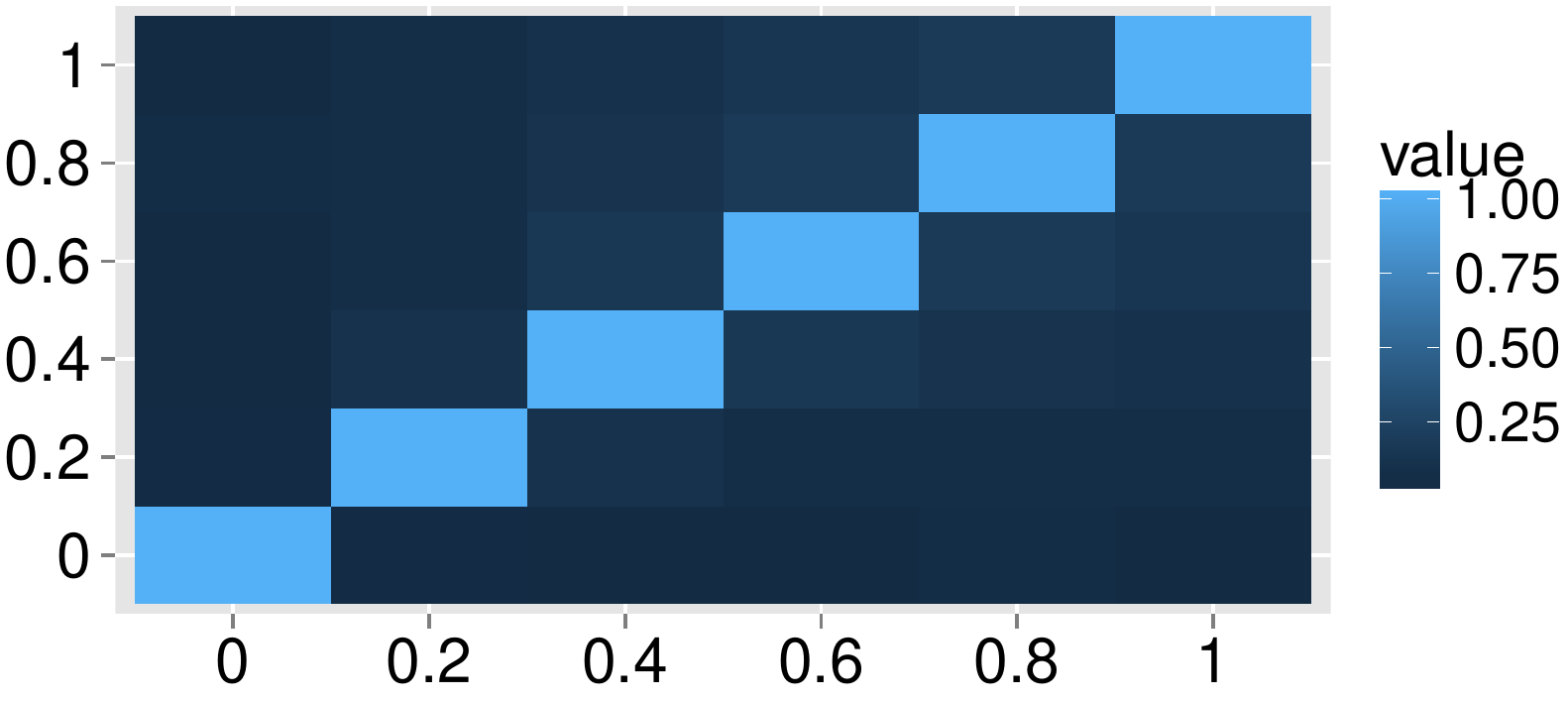} \\
\includegraphics[width=0.4\textwidth]{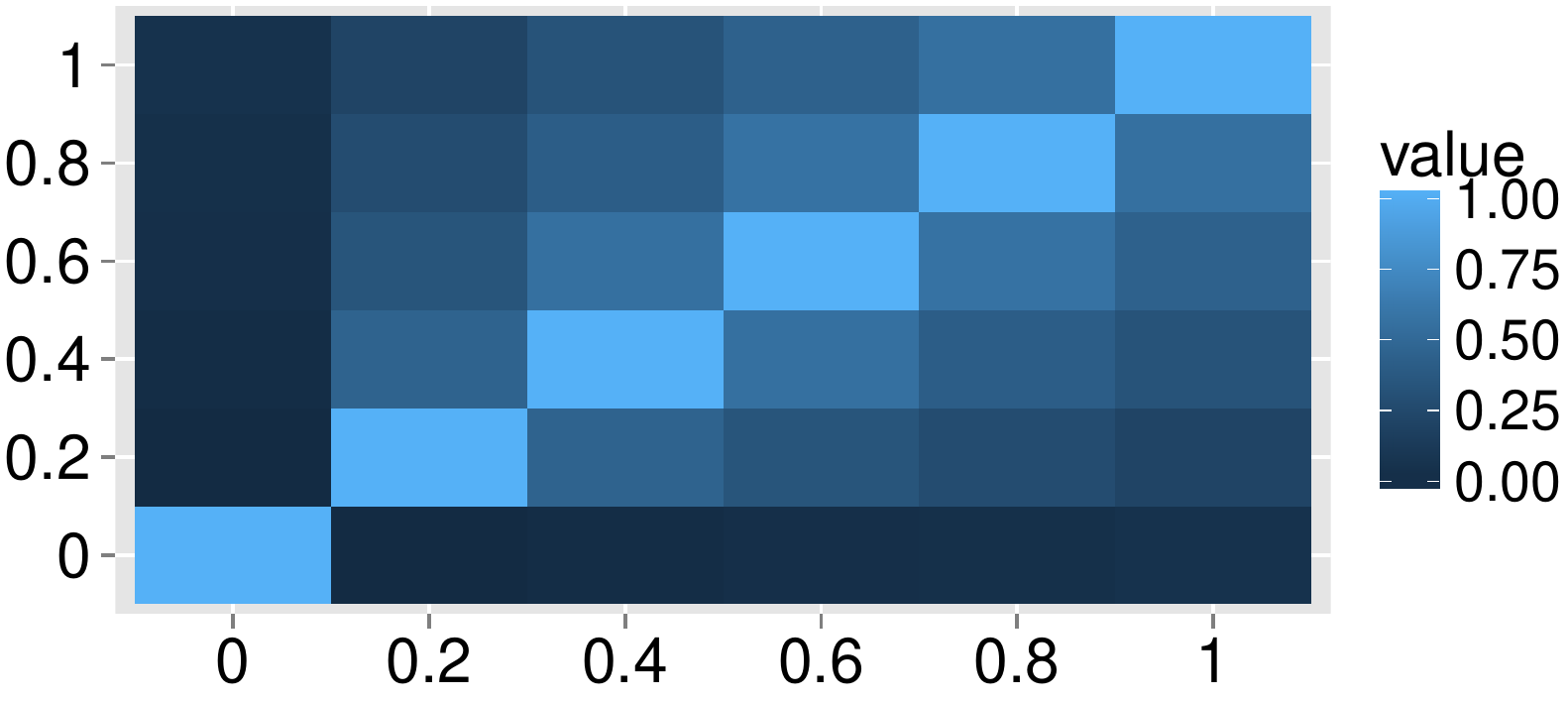} 
\caption{Delurking-oriented randomization analysis: pairwise correlation between \algo{LRin} solutions, as obtained on original network  and randomized networks, for increasing  degree of delurking-oriented randomization, on \data{Flickr} (top)  and  \data{FriendFeed} (bottom).}
\label{fig:delurkification-LRcor}
\end{figure}

Interestingly, the above  remarks indicate that  upon a delurking-oriented randomization process, the top-ranked lurkers can significantly change,  not only with respect to the original configuration of the network but also with respect to a  configuration corresponding to a different degree of delurking-oriented randomization (shown in Fig.~\ref{fig:delurkification-LRcor} for the \algo{LRin} evaluation). 
Clearly, as expected, when considered as a global feature of the network, the delurking-oriented randomization impact can be lower for larger networks  (e.g., \data{Twitter}), which have much lower (resp. higher) clustering coefficient (resp. average path length) than the other network datasets.    

\begin{figure*}[t!]
\centering
\begin{tabular}{ccc}
\includegraphics[width=0.25\textwidth]{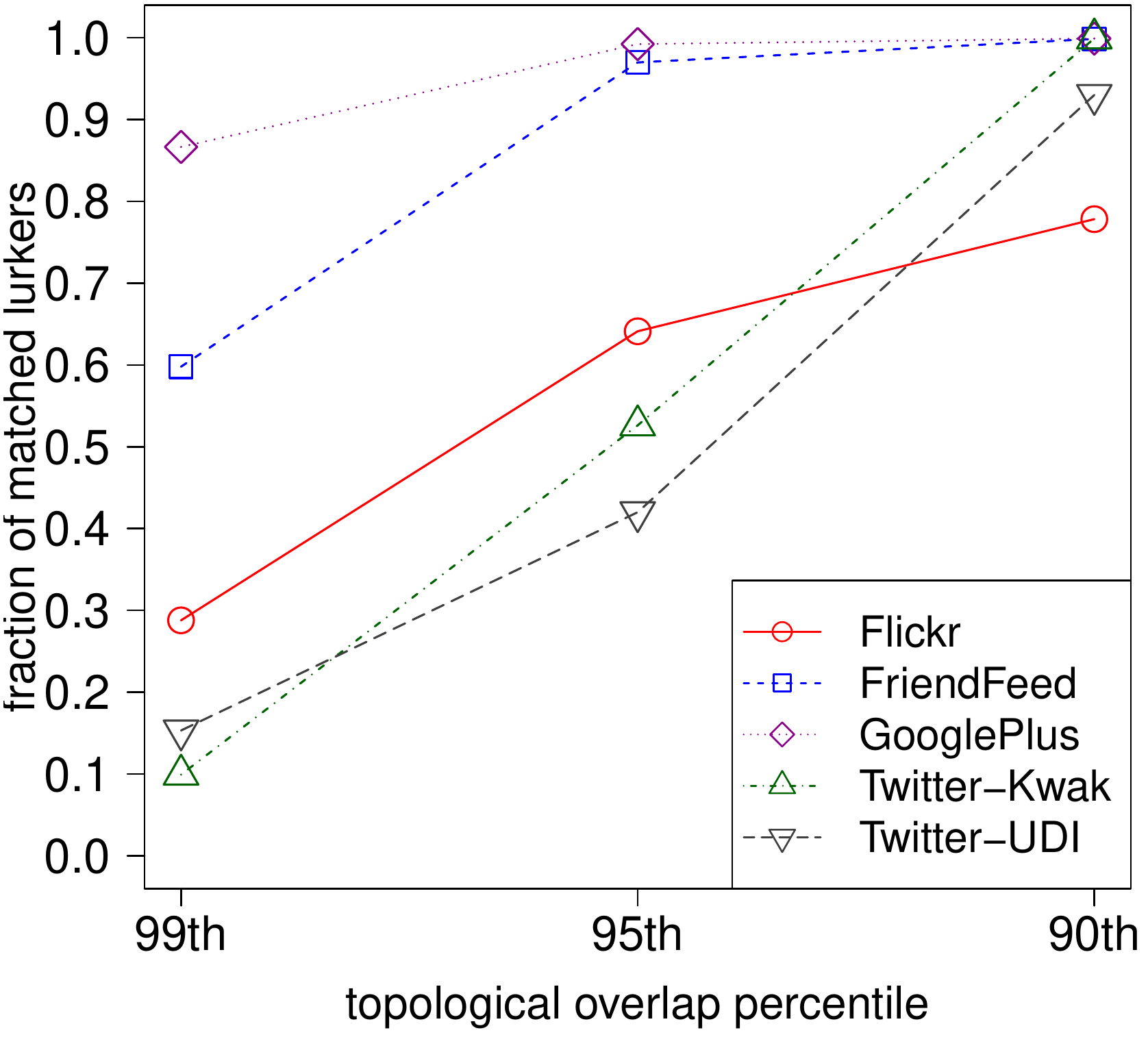} & \quad
\includegraphics[width=0.25\textwidth]{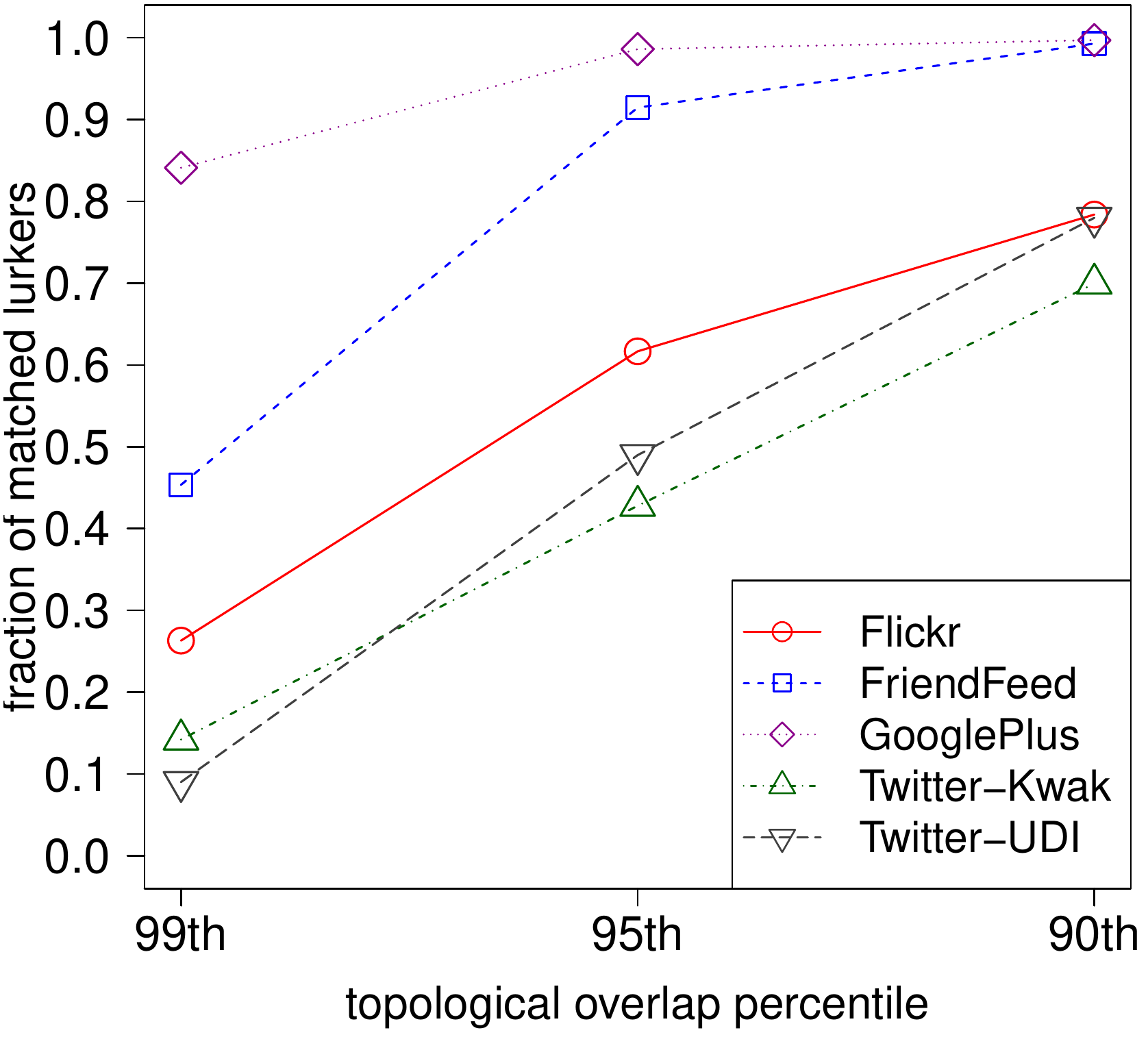}  & \qquad
\includegraphics[width=0.25\textwidth]{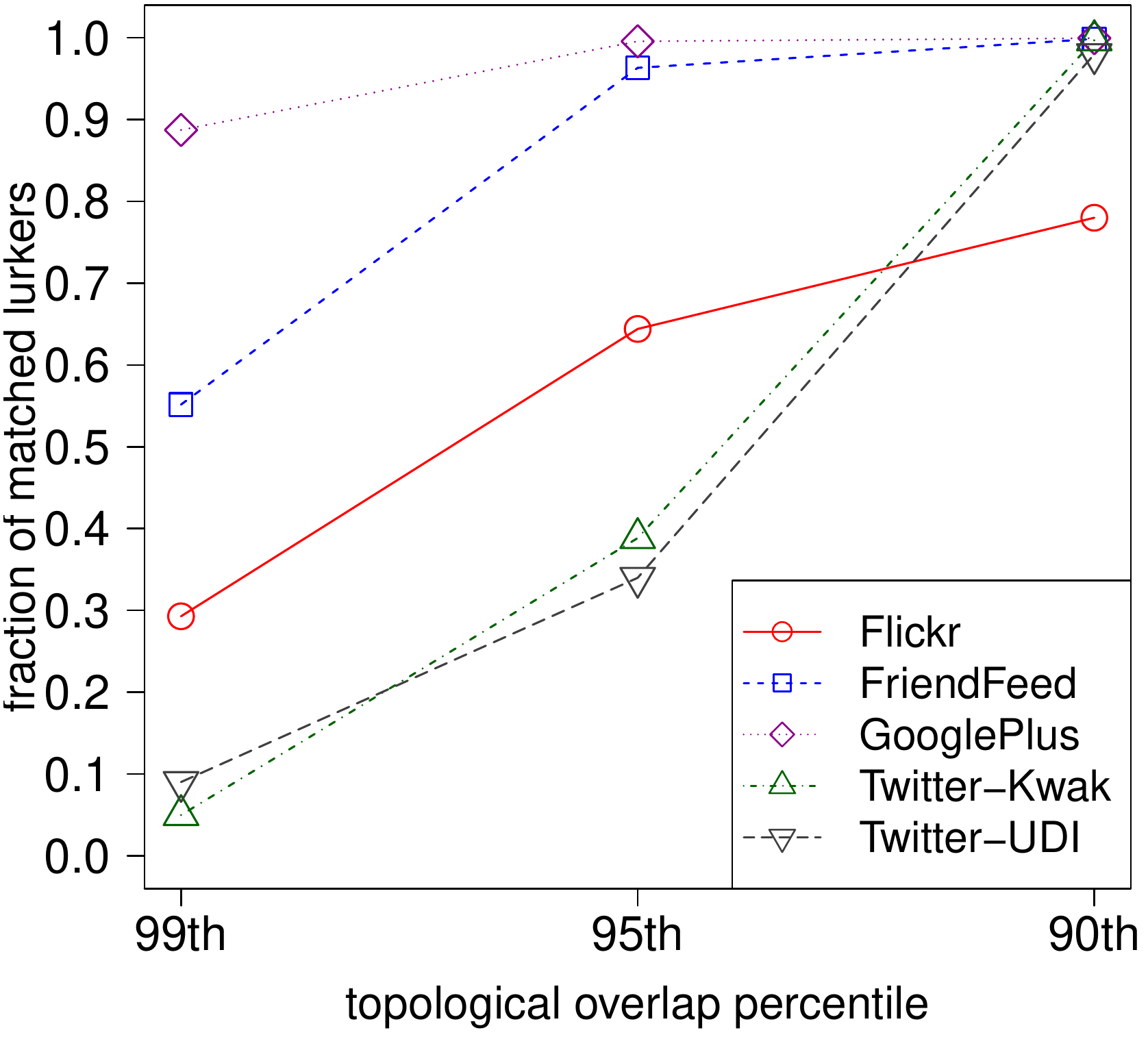} \\
(a) \algo{LRin}  & \quad  (b) \algo{LRout} & \quad (c) \algo{LRin-out} 
\end{tabular}
\caption{Percolation analysis: fraction of lurkers matched as function of the vertices removed based on directed topological overlap.}
\label{fig:percolation-match}
\end{figure*}

By contrast, the delurking-oriented randomization   seems   to negligibly  affect the in/out-degree distribution:  correlations turned out to be   moderate to high (when sinks and sources were considered) both between the in/out ranking in the original network  and each of the  in/out rankings of the randomized networks, and between the randomized in/out rankings pairwise. 
This result  would indicate that an apparently ``invasive'' alteration of the topology (through the insertion of new links) actually will not significantly change the topological features based on in- and out-degree distributions.

\subsection{Percolation analysis}
\label{sec:percolation}

Percolation analysis corresponds to studying the effect of network disruption via edge removal strategies, generally with the purpose of assessing topological integrity  properties of the network  or its vulnerability to (random) failures/attacks. An edge removal strategy is typically based on local structural properties of edges, such as topological overlap. 
Topological overlap is a measure originally introduced in~\cite{Onnela+07}  for undirected networks, which evaluates the number of neighbors shared by two given vertices $i$ and $j$. Edges between connected components are expected to have a low number of common neighbors, and hence low topological overlap. 
 
Removing edges by increasing order of topological overlap has shown to  effectively  detect the edges that act as \textit{bridges} between different  communities~\cite{GirvanNewman02,Radicchi+04}. 
Upon this we build our intuition that if we would discover a certain correlation between the result of our  lurker detection  and the result of percolation based on topological overlap, then we could claim that \textit{lurkers are likely to behave as bridges between communities}.  

Our network model however implies that edges are directed from information-producer to information-con\-sumer, therefore the   notion of bridge as highly active user must  be revised as less active user.  
Therefore,   we needed first to adapt the basic topological overlap to our setting of directed networks, whereby the neighbor sets  of any two selected vertices are partly considered according to the orientation of the edge drawn between the two vertices. 
Given edge $(i,j)$, we define the \textit{directed topological overlap} as:
\begin{multline}
O(i,j) = \frac{|R_i \cap B_j|}{(|R_i|-1) + (|B_j|-1) - |R_i \cap B_j|} 
\end{multline}

We developed a stage of evaluation in which two sets of   vertices are compared with each other: the one resulting from an edge removal strategy based on increasing order of  our directed variant of topological overlap, and the other one corresponding to the highest-ranked lurkers detected by one of our \algo{LR} algorithms. 

Figure~\ref{fig:percolation-match} plots the fraction of top-25\% of lurkers that matched the sets of vertices respectively included in the 99th, 95th and 90th percentile of the edges with lowest directed topological overlap. \algo{LRin}, \algo{LRout}, and \algo{LRin-out} were used to rank lurkers.
The  methods appear to behave very closely to each other for all data, with some 
relative differences on  the two \data{Twitter} networks. At 90th percentile of the edges with lowest directed topological overlap, almost all top-lurkers were matched on \data{FriendFeed}, \data{GooglePlus}, and only by  \algo{LRin} and \algo{LRinout}, on the two \data{Twitter} networks as well. 
Moreover, on \data{FriendFeed} and \data{GooglePlus}, most top-lurkers were matched already at 95th   percentile. 

Clearly, this relatively easy tendency of covering the set of top-lurkers  needs to be interpreted in relation to the ratio of the number of vertices removed (by increasing directed topological overlap) with respect to  the total number of vertices in the network. While on \data{FriendFeed} and \data{GooglePlus} the number of vertices removed corresponded to more than 90\% of the total vertex set (which hence explains the high rate of coverage over the top-lurkers),   on both the two \data{Twitter} networks, the above percentage was instead less than 27\%. The latter, being observed on the two largest evaluation networks, should be taken as an important finding, which would confirm the relationship between the lurkers and the bridges between communities.

\begin{figure*}[t!]
\centering
\begin{tabular}{cccc}
\includegraphics[width=0.23\textwidth]{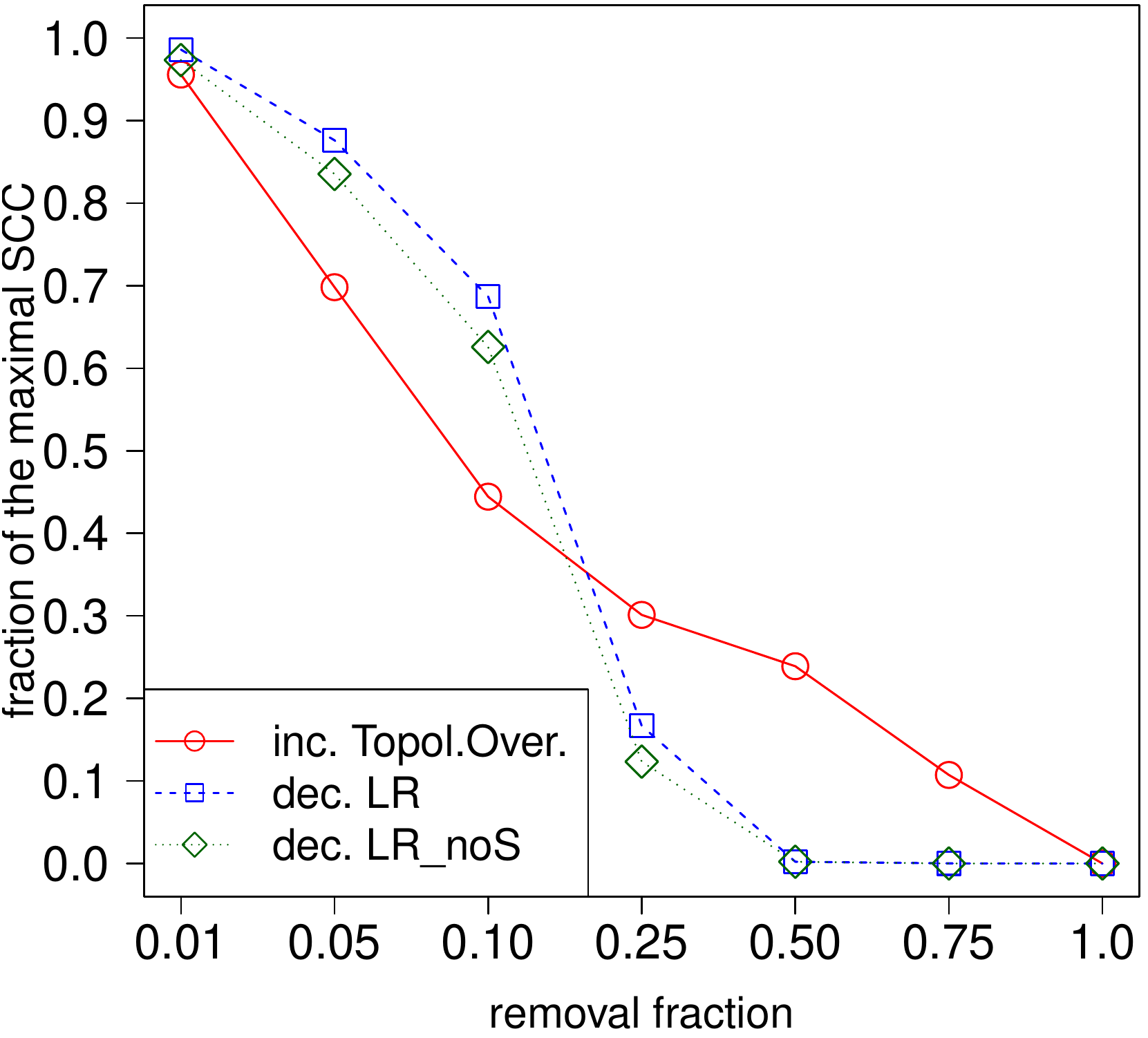} &  
\includegraphics[width=0.23\textwidth]{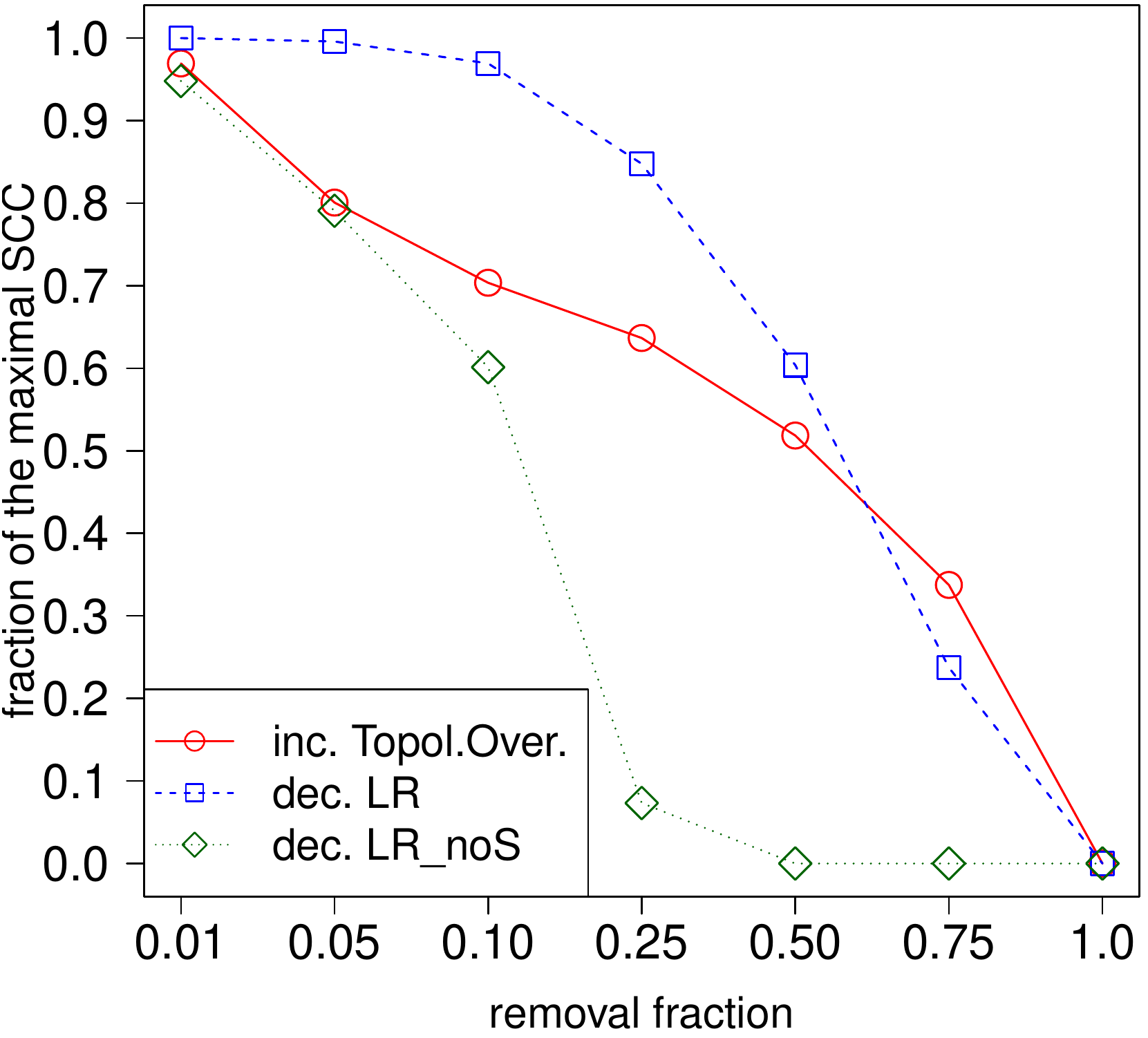}  &  
\includegraphics[width=0.23\textwidth]{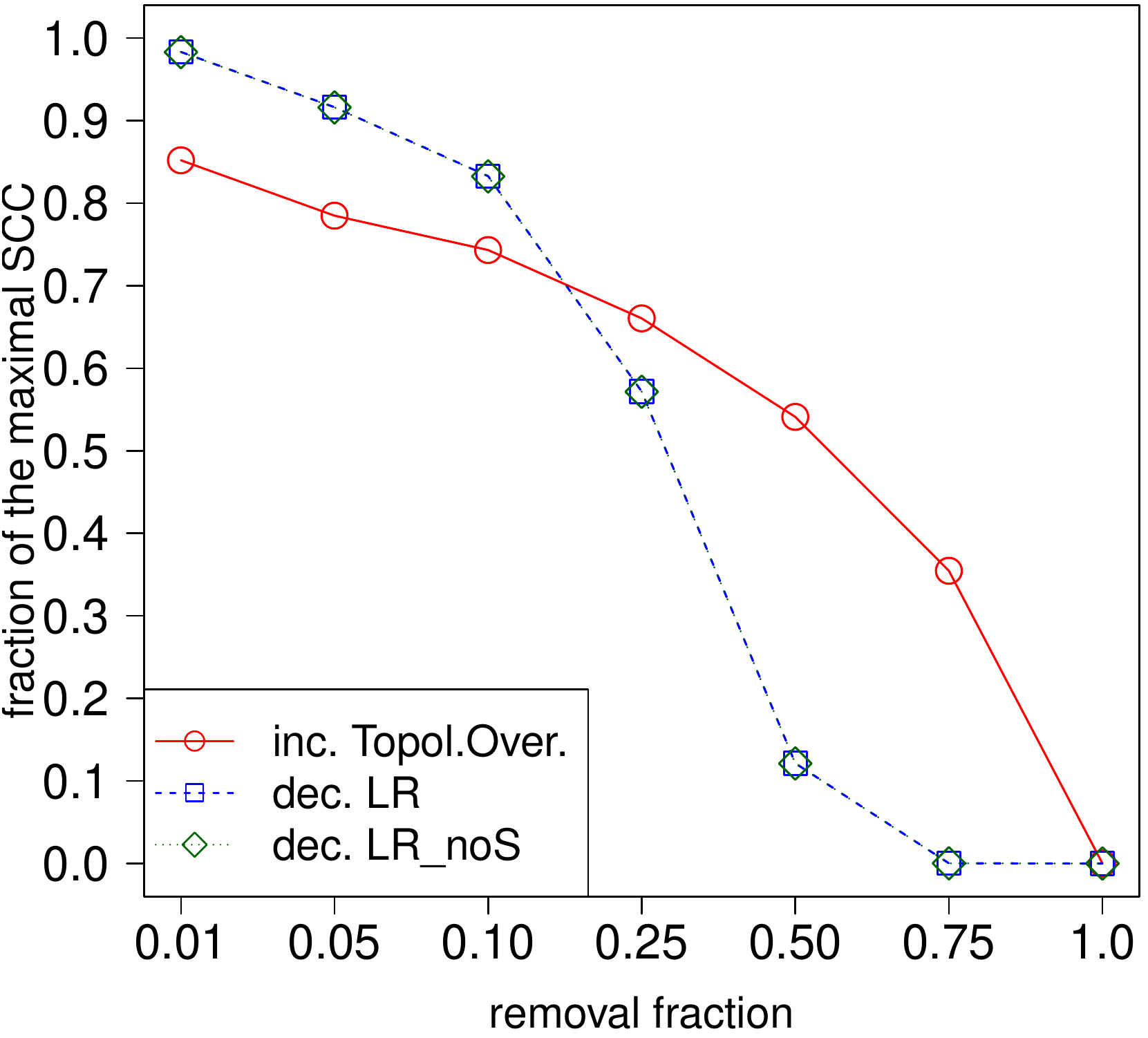}  &  
\includegraphics[width=0.23\textwidth]{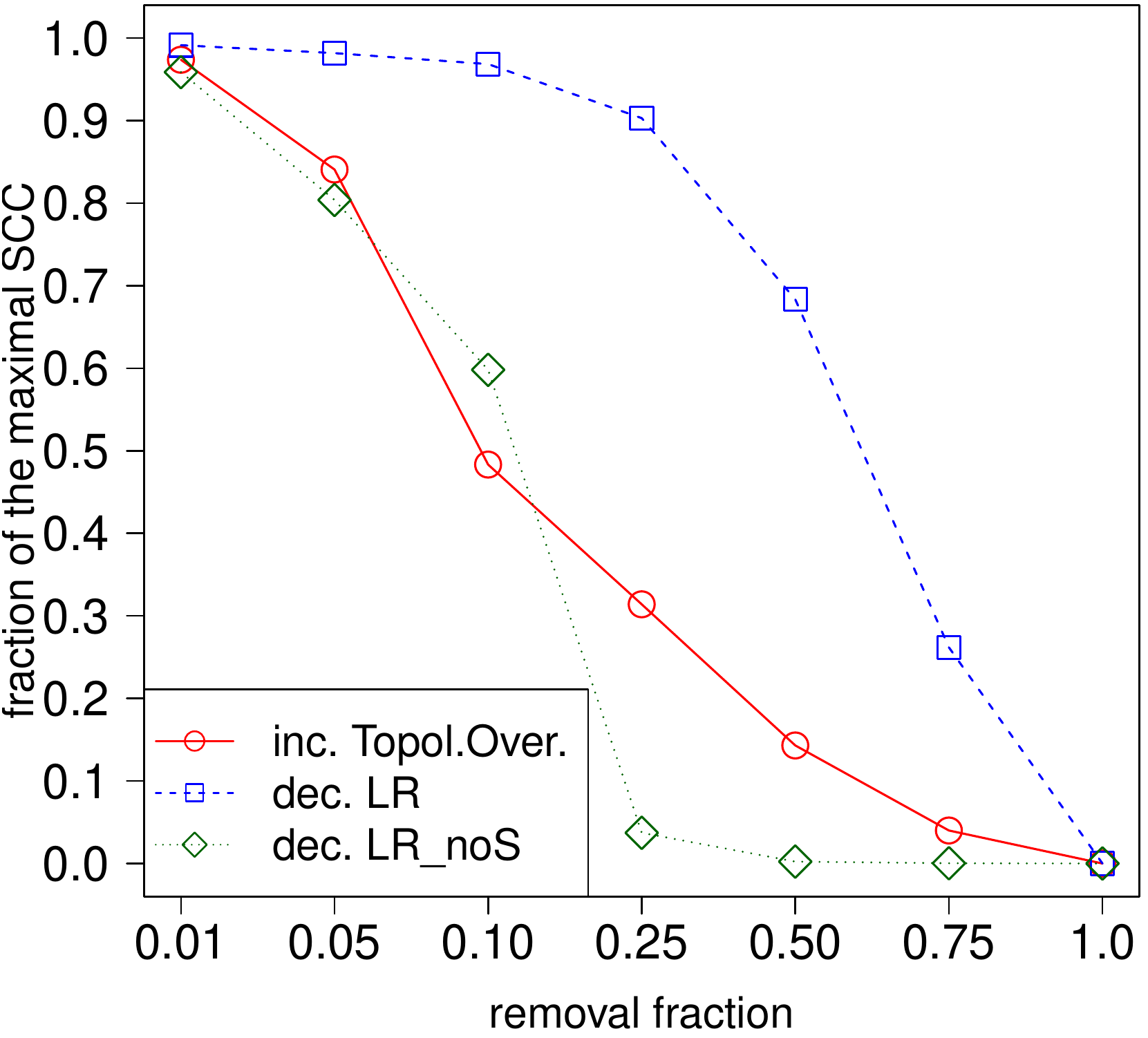} \\
(a) \data{Flickr}  &   (b)\data{FriendFeed}  & (c) \data{GooglePlus}   &  (d) \data{Twitter-Kwak}  
\end{tabular}
\caption{Percolation analysis: fraction of the maximal strongly CC as function of removed vertices.}
\label{fig:percolation-resilience}
\end{figure*}

We also analyzed the resilience of the various networks when vertices are removed by decreasing lurking order. To better evaluate the impact of sinks on the network disruption, we distinguished two cases: either sinks were preliminarily filtered out  or they were included when  selecting the fraction of lurkers to remove from the network. 
As shown in Figure~\ref{fig:percolation-resilience} for \algo{LRin},   
the removal strategy with the most disruptive effects was that based on decreasing \algo{LR} rank (with pre-filtering of sink vertices, denoted as $LR\_noS$ in the figure), 
which led to mostly dismantle the maximal strongly connected component (i.e., 80 to 90\% of its size) already  for 25\% of vertex removal   in all networks except   \data{GooglePlus} (for which 50\% of vertex removal was needed). 
By contrast, the removal strategy based on increasing topological overlap  produced   disruptive effects smoother with respect to the two \algo{LR}-based strategies, on all networks. 
Interestingly, by including sink vertices in the selection of lurkers to remove, the network resilience  was the same as in the case of sink-pre-filtering on \data{GooglePlus} and \data{Flickr}, whereas on the two \data{Twitter} and \data{FriendFeed} the resilience was higher than for  the other strategies, since a level  of dismantling below 70-60\% was reached only by  a removal fraction of 50\% or higher. 
Note that \data{Twitter} and \data{FriendFeed} are the networks with a strong presence of sink vertices, and with a sink/source ratio greater than 10.

\subsection{Qualitative evaluation} 
\label{sec:qualitative}
We   investigated the meaningfulness of the rankings produced by 
LurkerRank methods  as well as  produced by  the competing methods. 
For this analysis,   we retrieved the OSN pages of top-ranked users 
and examined the available information about their profile and neighborhoods. Our goal was to  understand whether a user actually looks like a lurker, or conversely s/he takes another role in the network.

Tables~\ref{tab:twitterranks}--\ref{tab:ffranks}  show the top-20 ranked users  obtained  on \data{Twitter-Kwak} and \data{FriendFeed} by PageRank, alpha-centrality,  Fair-Bets, and \algo{LRin}.   
  Table~\ref{tab:twitterranks} also reports the number of times a user was retweeted (\emph{\#rt}), whereas  Table~\ref{tab:ffranks} reports the total number of posts by a user (\emph{\#posts}). 
  Moreover,    we left sink nodes out of consideration  in order to avoid biasing our evaluation with trivial lurkers.  

By comparing the top-ranked lists, 
it is     evident that   \algo{LRin}  behaved differently from   the other   algorithms, 
 since  it  shared just two users  with \algo{FB} (dark-grey shaded) and no users at all with \algo{PR} and \algo{AC}. 
 Interestingly,   the \algo{LRin} top-ranked list  contains only users who have never been retweeted; 
 by retrieving the tweet post dates from Twitter, those users were all found as quite longer-time users, as in fact they joined Twitter much earlier  (e.g., \#8, \#10 and \#12 joined in 2007) than most users in the \algo{AC} and \algo{PR} top-ranked lists. Conversely, in the latter two lists   
 most users have been significantly retweeted although they joined   later (e.g., 2009).  
 
\algo{PR} and \algo{AC} showed a certain association, with ten users in common (light-grey shaded).  
Most users in both  \algo{AC}   and  \algo{PR} lists however were retweeted hundreds times, 
and hence  they should not be considered as lurkers. Our hypothesis of non-lurking for those users was fully confirmed as we   observed that those users' retweets were actually spread over a relatively short period of time (e.g., second half of 2009).   
Moreover,   \algo{AC} and \algo{PR}   ranked   the same user on top, who is also the one having the highest number of retweets in the lists; indeed,  that user is a very influential person, and in fact  s/he   has a 
followee/follower ratio much below 1: this would indicate that  both \algo{AC} and \algo{PR}  were not able to correctly handle this case (i.e., scoring it low enough), because their performance would be more affected by highly influential incoming links (i.e., followees) --- which is a clear indication of tendency to absorb valuable knowledge --- rather than by the number and type of followers. 
We also found other cases with characteristics similar to \#1, e.g.,   \#12 in the \algo{PR} list,   \#10 and \#14 in the \algo{AC} list, and the common users ``ZAP.'' (\#3 in both lists) and ``SCO.'' 
(\#17 in  \algo{PR}, \#12 in \algo{AC}).

As concerns \algo{FB}, it was surprising to find that 15 out of 20 top-ranked users  refer to 
spammers (\#4, a fashion/cosmetic marketing spammer, \#9, in advertising, and \#15, a porn spammer), 
or in general to suspended accounts  (\#2-3, \#5, \#8, \#10-11, \#13-14, \#17-20). 
Only \#6, \#12 and   \#16   appear to be  lurkers, which might   be confirmed by their   high  in/out-degree ratio coupled with a zero retweet-count. By contrast, \#1 is an art director and designer, and  \#7  refers to an account   actively used for  academic advising purposes; 
probably,   the high number of followees (e.g., about 1800 for \#7)  has misled  the method. 
Therefore, like \algo{PR} and \algo{AC}, \algo{FB} might also fail  to correctly recognize real  lurkers.

In \data{FriendFeed} (Table~\ref{tab:ffranks}), a large intersection was found among the  top-20 users not only between \algo{PR} and \algo{AC} (like in \data{Twitter-Kwak}) but also between \algo{FB} and \algo{LRin}. 
Looking at the users' profiles and at the contents of their posts, we can state that most of the users shared by the top20 lists of \algo{FB} and \algo{LRin}  are recognized   either as content spammers (i.e.,   users that have produced spamming contents, regardless of the popularity and number of their posts), or  as professionals who aim to improve their visibility  while staying as observers in the community (e.g., \#9 in \algo{LRin}/\#7 in \algo{FB} is a marketing expert, \#7 in \algo{LRin}/\#5 in \algo{FB} is a graphic designer). 
Some distinct profiles are also found to be clones, as they are associated to   the same spamming contents (e.g., \#1 and \#4 in \algo{LRin}, which correspond to  \#1 and \#17 in \algo{FB}, both probably  related to a Russian commercial site).  
A reason for this massive presence of spammers probably can be found in the nature of the \data{FriendFeed} social network: being   a real-time cross-network feed aggregator makes it a desirable and user-friendly means for spammers to reach high visibility, producing a number of user profiles for spamming attempts having very similar characteristics to lurker ones (e.g., high in/out degree ratio, low interaction with other members).
Looking at \algo{PR} and \algo{AC} top-20 users we found that, as in the \data{Twitter-Kwak} case, most of them are not recognizable as lurkers, but rather as active and authoritative users (e.g.,\#7 in \algo{PR}/\#3 in \algo{AC} is a finance blogger, \#1 in both \algo{PR} and \algo{AC} is an industrial designer, \#5 in \algo{PR}/\#2  in \algo{AC} represents a philanthropic foundation). 
We also found a user shared by \algo{PR} and \algo{FB} top-$20$s: although the account does not exist  anymore, its name would hint  that the user was probably a spammer for a hosting solutions company.

Concerning \data{GooglePlus} (results not shown), the top-ranked list by \algo{LRin} is mainly comprised of  users that show  poor public activity,  and  that    added a lot of people to their circles although scarcely reciprocated.   
\algo{FB}   showed a behavior nearly similar to \algo{LRin}, however its top-20 list contains less real lurkers than those detected by \algo{LRin}. \algo{PR} and \algo{AC} ranked high   users that are  likely to be pretty influential,  such as a classical guitarist with more than 60 thousand followers (ranked \#1 by \algo{PR}), a landscape photographer with more than 42 thousand followers (ranked \#1 by \algo{AC}), and even a social media director with nearly 400 thousand followers (ranked \#6 by \algo{PR}). In contrast to the other network datasets, there were no shared users among \algo{PR} and \algo{AC} top-20s, while \algo{FB} shared 2 users with \algo{PR} and 7 with \algo{LRin}.

\begin{table}[t!]
\caption{Top-20 \data{Twitter-Kwak} users   by   lurking score.}
{\centering 
\scalebox{0.6}{
\begin{tabular}{|c||c|c|c|c|c|c|c|c|}
\hline
\!\!\!\emph{rank}\!\!\! & \multicolumn{2}{|c|}{\algo{PR}} & \multicolumn{2}{|c|}{\algo{AC}} & 
\multicolumn{2}{|c|}{\algo{FB}} & \multicolumn{2}{|c|}{\algo{LRin}} \\ 
\cline{2-9}
 & \emph{user}  & \emph{\#rt} & \emph{user} & \emph{\#rt} 
  & \emph{user} & \emph{\#rt} & \emph{user} & \emph{\#rt}
\\\hline \hline
1 &   \colorbox{gray!30}{B.O.}   &  17811  &   \colorbox{gray!30}{B.O.}   &  17811   &  \colorbox{gray!70}{D.W.S.}   & 0 &    R.F.   &   0 \\
2 &   \colorbox{gray!30}{W.F.}   &    1676   &   \colorbox{gray!30}{ZAI.}   &    10902    &  n.a.  & 0 &      R.J.   &   0 \\
3 &   \colorbox{gray!30}{ZAP.}   &    8707    &   \colorbox{gray!30}{ZAP.}   &    8707    &  APA.  & 0 &       R.M.K.   &   0 \\
4 &   TH.   &    7169    &   \colorbox{gray!30}{AS.}   &    1172    &  T.S.C.  & 1 &      \colorbox{gray!70}{B.B.P.}    &   0 \\
5 &   L.E.   & 683 &   \colorbox{gray!30}{M.M.}   & 7 &   n.a.  & 0 &      TR.   &   0 \\
6 &   J.B.   &    1248   &   \colorbox{gray!30}{W.F.}   &    1676    &  CON.  & 0 &      MU.   &   0 \\
7 &   M.S.   & 476 &    \colorbox{gray!30}{M.K.}   & 48 &   K.T.  & 0 &      B.R.   &   0 \\
8 &   \colorbox{gray!30}{AS.}   &    1172    &   P.B.   & 328 &  n.a.  & 0 &      AZ.   &   0 \\
9 &   OH.   & 1009 &   \colorbox{gray!30}{W.A.}   & 2814 &   S.M.  & 0 &      O.L.   &   0 \\
10 &   H.T.   & 43 &   C.B.   &    11943    &  n.a.  & 0 &      N.T.   &   0 \\
11 &   E.T.   &   2435   &   EL.   & 902 &  n.a.  & 0 &      FR.   &   0 \\
12 &    SCH.   &    3277    &   \colorbox{gray!30}{SCO.}   &   6970   &   M.P.  & 0 &       \colorbox{gray!70}{D.W.S.}   &   0 \\
13 &   RE.   &   1467    &   \colorbox{gray!30}{WI.}   & 811 &  n.a.  & 0 &      AW.   &   0 \\
14 &   H.S.   &   1346   &   O.W.   &    1803   &  n.a.  & 0 &      O.B.   &   0 \\
15 &   \colorbox{gray!30}{M.M.}   & 7 &    T.B.B.   & 102 &  M.E.   & 0 &      N.C.   &   0 \\
16 &   \colorbox{gray!30}{ZAI.}   &   10902    &   T.S.   & 74 &  \colorbox{gray!70}{B.B.P.}  & 0 &      D.P.   &   0 \\
17 &    \colorbox{gray!30}{SCO.}   &    6970   &   S.S.   & 789 &   n.a.  & 0 &      AU.   &   0 \\
18 &   \colorbox{gray!30}{M.K.}   & 48 &   M.W.   & 363 &  n.a.  & 0 &      EM.   &   0 \\
19 &   \colorbox{gray!30}{WI.}   & 811 &   H.R.   & 750 &  n.a.  & 0 &      DI.   &   0 \\
 20   &   \colorbox{gray!30}{W.A.}   &    2814   &   A.K.   & 1572 &   n.a.  & 0 &      M.A.   &   0 \\
\hline  
\end{tabular}
}
}

\vspace{-.5em}
\scriptsize
\begin{flushleft}
For privacy reasons, users' names were replaced with their initials or abbreviations.
\end{flushleft}
\label{tab:twitterranks}
\end{table}

\begin{table}[t!]
\caption{Top-20 \data{FriendFeed} users   by   lurking score.}
{\centering 
\scalebox{0.6}{
\begin{tabular}{|c||c|c|c|c|c|c|c|c|}
\hline
\!\!\!\emph{rank}\!\!\! & \multicolumn{2}{|c|}{\algo{PR}} & \multicolumn{2}{|c|}{\algo{AC}} & 
\multicolumn{2}{|c|}{\algo{FB}} & \multicolumn{2}{|c|}{\algo{LRin}} \\ 
\cline{2-9}
 & \emph{user} & \emph{\#posts} & \emph{user}  & \emph{\#posts} 
  & \emph{user} & \emph{\#posts} & \emph{user} & \emph{\#posts}
\\\hline \hline
1 &  \colorbox{gray!30}{N.D.P.}  & 350 &  \colorbox{gray!30}{N.D.P.}  & 350 &  \colorbox{gray!70}{M.C.D.}  & 11 &  \colorbox{gray!70}{M.C.D.}  &  11 \\
2 &  FRE.  & 3 &  \colorbox{gray!30}{C.T.}  & 5 &  \colorbox{gray!70}{BOG.}  & 367 &  \colorbox{gray!70}{BOG.}  &  367 \\
3 &  \colorbox{gray!30}{BR.}  & 71 &  \colorbox{gray!30}{J.D.A.}  & 282 &  \colorbox{gray!70}{L.H.}  & 1 &  \colorbox{gray!70}{B.I.}  &  61 \\
4 &  A.C.  & 142 &  \colorbox{gray!30}{MBL.}  & 37 &  DIM.  & 1 &  \colorbox{gray!70}{N.D.}  &  13 \\
5 &  \colorbox{gray!30}{C.T.}  & 5 &  \colorbox{gray!30}{BR.}  & 71 &  \colorbox{gray!70}{B.I.}  & 61 &  \colorbox{gray!70}{G.A.}  &  11 \\
6 &  \colorbox{gray!30}{MBL.}  & 37 &  \colorbox{gray!30}{U.R.}  & 52 &  \colorbox{gray!70}{G.A.}  & 11 &  \colorbox{gray!70}{L.H.}  &  1 \\
7 &  \colorbox{gray!30}{J.D.A.}  & 282 &  TAV.  & 65 &  \colorbox{gray!70}{A.C.}  & 2 &  \colorbox{gray!70}{R.W.}  &  7 \\
8 &  \colorbox{gray!30}{U.R.}  & 52 &  D.H.  & 89 &  \colorbox{gray!50}{W.H.O.}  & 10 &  ZAH.  &  3 \\
9 &  S.M.  & 106 &  P.B.  & 13 &  ASR.  & 0 &  \colorbox{gray!70}{A.C.}  &  2 \\
10 &  \colorbox{gray!50}{W.H.O.}  & 10 &  C.E.  & 447 &  \colorbox{gray!70}{H.P.B.}  & 3 &  \colorbox{gray!70}{E.J.S.}  &  24 \\
11 &  \colorbox{gray!30}{RID.}  & 886 &  \colorbox{gray!30}{RID.}  & 886 &  MUA.  & 5 &  M.P.  &  2 \\
12 &  D.G.  & 35 &  W.B.  & 5 &  \colorbox{gray!70}{E.J.S.}  & 24 &  Y.P.  &  1 \\
13 &  \colorbox{gray!30}{L.A.C.}  & 4 &  R.T.  & 68 &  SVL.  & 1 &  S.E.  &  72 \\
14 &  \colorbox{gray!30}{JSI.}  & 49 &  \colorbox{gray!30}{K.K.}  & 134 &  \colorbox{gray!70}{R.W.}  & 7 &  J.N.  &  110 \\
15 &  \colorbox{gray!30}{K.K.}  & 134 &  D.S.  & 105 &  S.F.T.  & 4 &  \colorbox{gray!70}{H.P.B.}  &  3 \\
16 &  S.O.  & 12 &  \colorbox{gray!30}{L.A.C.}  & 4 &  D.G.  & 5 &  P.C.  &  3 \\
17 &  W.M.  & 108 &  \colorbox{gray!30}{JSI.}  & 49 &  \colorbox{gray!70}{N.D.}  & 13 &  MRT.  &  3 \\
18 &  STR.  & 2 &  B.C.  & 14 &  I.P.G.  & 10 &  I.K.G.  &  2 \\
19 &  C.F.  & 3 &  D.V.  & 85 &  ARG.  & 2 &  N.L.  &  1 \\
20 &  R.T.  & 68 &  M.M.H.  & 34 &  E.E.M.  & 5 &  F.F.  &  764 \\
\hline  
\end{tabular}
}
}

\vspace{-.5em}
\scriptsize
\begin{flushleft}
For privacy reasons, users' names were replaced with their initials or abbreviations.
\end{flushleft}
\label{tab:ffranks}
\end{table}

\subsection{Some lessons learned} 
Our study so far allows us to draw some interesting conclusions, which are briefly summarized as follows.  

Quantitative and qualitative results have demonstrated the ability of our  approach in unveiling lurking cases that are intuitive yet non-trivial.  The best-performing ranking methods  are those based on  in-neigh\-bors-driven and in-out-neighbors-driven lurking, i.e., the models emphasizing  the first two principles underlying our lurking definition. These methods have shown high correlation with the data-driven ranking, and outperform   competing methods, i.e.,  PageRank, alpha-central\-ity, Fair-Bets model, and the baseline in/out-degree ranking.  
Moreover, results tend to be relatively consistent over   the PageRank-based and the alpha-centrality-based formulations of the lurker ranking methods. (We expect however that  a different setting in the damping factor along with the introduction of  a term modeling personalization or exogenous information in the respective formulas would bring to a more evident differentiation of the two ranking approaches.) 
From a runtime efficiency viewpoint, \algo{LRin} tends to perform faster than \algo{ac-LRin}, while \algo{ac-LRin-out} achieves the highest rate of convergence although at the cost of much less   diversified   ranking  scores.    
Furthermore, our qualitative analysis of the OSN pages of the top-ranked users has  provided clear evidence that: (i) our approach successfully detects  lurkers  in an OSN, and conversely (ii) the competing methods fail in doing this --- PageRank and alpha-centrality still detect influential users, whereas Fair-Bets tends rather to identify spammers. 

From a pure network-analysis perspective, lurkers are not very prone to reciprocate  each other, whereas preferential attachment is likely to occur between lurkers  and the active users they are linked to. 
 Under a percolation analysis framework,  lurkers tend to be  matched by users that are involved in links with low (directed) topological overlap: this would hint at a relation existing between lurkers and users playing the role of bridges between communities, under the assumption of lurking-oriented topological graph of an OSN. 
Finally,   our proposed  delurking-oriented randomization strategy reveals that self-delurking can be useful to change the top-ranked lurkers in the network, while scarcely affecting  the in/out degree distribution.

 \section{Related Work}
 \label{sec:related}

The topic of lurking has been long studied in social science and  recently has gained renewed interest in the computer-human interaction community.   
 \cite{SorokaR06}  investigates relations between lurking and cultural capital, i.e., a member's level of community-oriented knowledge. Cultural capital is found positively correlated with both the degree of active participation  and, except for longer-time lurkers, with de-lurking. 
\cite{CranefieldYH11} leverages  the significance of conceptualizing the lurking roles in relation to their boundary spanning and knowledge brokering activities across multiple community engagement spaces.    
The study proposed in~\cite{ChenC11} raises the opportunity of rethinking of the nature of lurking from a group learning perspective, whereby the engagement  of intentional lurkers is considered within  the collective knowledge construction activity.  
The interactive/interpassive connotation  of social media users' behavior is studied in~\cite{KapplerQ11}, under a qualitative and grounded-theory-based approach.  
In the context of multiple  online communities in an   enterprise community service,   
  lurking is found as only partially driven by the member's engagement but significantly affected by the member's disposition  toward a topic, work task or social group~\cite{Muller12}.   
Exploring epistemological motivations behind lurking dynamics is the main focus of the study in~\cite{SchneiderKJ13},   
which indeed reviews major relevant literature on epistemic curiosity in the context of  online communities 
 and provides a set of propositions on the propensity to lurk and de-lurk. However, as with~\cite{CranefieldYH11}, the paper only offers  insights  that    might be useful to guide an empirical evaluation of lurkers' emotional traits. 
The study in~\cite{HalfakerKT13} examines   peripheral participation in  Wikipedia, and designs a system to elicit lightweight editing contributions from Wikipedia readers. 

To the best of our knowledge, there has been no study other than ours that provides a formal computational  methodology for  lurker ranking. 
The study in~\cite{FazeenDG11}, which aims to develop classification methods for the various OSN actors,  
actually treats the lurking problem margin\-ally, and in fact lurking cases are left out of experimental evaluation. Similarly, \cite{LangW12} analyzes various factors  that influence lifetime of OSN users, 
also distinguishing between active and passive lifetime; however, analyzing passive lifetime is made possible only when the user's last login date is known,  which is a rarely available information.

We finally mention some research studies that have focused on latent relationships or side-effect benefits in an OSN. 
For instance,~\cite{AnandCSV13} defines a Stackelberg game  to maximize the benefit each user gains extending help to other users, hence to determine the advantages of being  altruistic. Some interesting remarks relate the altruism of users to their level of capabilities, and indicate that  the benefit derived from  being altruistic  is larger than that reaped by selfish users or free riders.  
\cite{MalliarosV13} also builds upon game theory to study the property of users' departure dynamics, i.e.,  the tendency of individuals to leave the community. 
\cite{XieLZLG12} studies the problem of identifying the off-line real-life social community of a given user, by analyzing the topological structure in an on-line social network like Twitter. To the purpose, user interactions are modeled in the form followee-to-follower (like in our setting), and a  PageRank-like algorithm is applied over a probability transition matrix that embeds three key principles underlying the notion of off-line community, namely mutual reachability, friendship retainability, and community affinity. It should be noted that mutual reachability  is not a peculiar characteristic of lurkers, i.e., it can hold for active users as well. Moreover, as for the community affinity principle, lurkers are usually not grouped into communities such that each community members are (indirectly) connected to each other;  rather, as we have discussed  in this paper, 
lurkers may lay on the boundary of a component and bridge over other components.

\paragraph{Relations with existing definitions of  lurking.\ }
 Our definition of lurking is substantially consistent with the various existing  perspectives on lurking, previously mentioned in the Introduction.    
 It can in general recognize and measure behaviors that rely on 
 phenomena of lack of information production (i.e., inactivity or occasional  activity) as well as on  phenomena of   information hoarding or overconsumption, like free-riding and leeching. 
 
 It is worth emphasizing that taking into account the authoritativeness of the information received as well as the non-authoritativeness of the information produced by lurkers is essential to the correct scoring of lurkers.  Therefore, our definition of lurking can also explain more complex perspectives, such as legitimate peripheral participation. In this case,  a lurker is regarded as a novice, for which it's   legitimate to learn from experts as a form of cognitive apprenticeship. Indeed, by applying our LurkerRank methods, in~\cite{ICADL}  we have  addressed an exemplary  form of legitimate peripheral participation, known as vicariously learning, in the context of research collaboration networks.    

Finally, note  that other interpretations  of lurking, such as  microlearning  and knowledge sharing barriers, actually  aim to understand the various reasons for lurking, and to what extent they might be perceived as   fruitful, rather than neutral or harmful,  for the knowledge sharing in the online community. Therefore, they mostly involve sociological and psychological aspects whose study is  beyond the objective of our work.

 \section{Challenges and future directions}
\label{sec:challenges}

The inherent complexity of lurking would advise that more information 
besides the network topology needs to be considered for an enhanced  detection and 
 ranking  of lurkers.  Some of the most challenging issues for research in this context are discussed next.  
 
\paragraph{Temporal, context-biased lurking.\ }
Starting as visitors and newcomers, members of a community naturally evolve over time playing different roles, thus showing a  stronger or weaker tendency toward lurking on different times. Lurkers have unusual frequency of online presence, and hence any 
knowledge on the online participation frequency of the users  could guide the identification of critical time intervals to reveal lurking behaviors.   
Moreover, the user's engagement level in the community clearly depends also on the number and type of contexts in which the user is   involved. 

\paragraph{Boundary-spanning and cross-network lurking.\ }
Some of the members that lay on the boundary of a component may bridge over other  components. In Section~\ref{sec:percolation}, we have found out that indeed relations may exist between lurkers and users that act as bridges over different components of an OSN graph.  
To a larger extent, and given the increased interest towards cross-network services (see the latest  examples of YouTube and GooglePlus),   members who lurk inside an OSN may not lurk, or even take on the role of experts, in other OSNs.  An analysis of the lurker ranking problem across different OSNs would  represent a great potential  to get a more complete picture of their users.  

\paragraph{Lurking and trust contexts.\ }
Active users tend to avoid wasting their time with people who are very likely to not reply or show slow responsiveness, or who have few/bad feedbacks; as a consequence, lurkers could in principle be perceived as \textit{untrustworthy} users. 
Another challenge would hence be modeling the dynamics of lurking behaviors in  trust contexts~\cite{Adali13},   and ultimately   understanding   relations between lurkers and trustworthy/untrustworthy users in ranking problems.

While we believe this represents an important issue that deserves much attention in future  studies, we nevertheless provide here a preliminary insight into a comparison of our LurkerRank methods with a classic method for ranking pages/users according to their trustworthiness, namely \textit{TrustRank}~\cite{GyongyiGP04}. Moreover, we further propose to integrate the ability of detecting trustworthy users (featured by TrustRank) into our LurkerRank in order to improve the \textit{trustworthiness} of the lurkers to be detected. 
The result is a new set of methods, we call \textit{TrustRank-biased LurkerRank} methods, in which the uniform personalization vector of a LurkerRank  method is replaced by the ranking vector produced by  TrustRank over the same network.

We recall that TrustRank is substantially a biased PageRank in which  the teleportation set corresponds to the ``good part'' of an a priori selected seed set. The seed set is comprised of a relatively small subset of nodes in the graph, each of which is labeled as either trustworthy or untrustworthy by some \textit{oracle} function. 
Note that unlike   trust network data, OSNs do not contain explicit trust assessments  among users. 
However, behavioral trust information in social media networks can be inferred from some forms of user interaction that would provide an intuitive way of indicating trust in another user~\cite{AdaliEGHMSWW10}. 
Here we leverage information  on the number of  favorite markings  received by a user's  photographs in \data{Flickr} as implicit trust statements. (We will refer to \data{Flickr} as case in point for this evaluation, although the   approach we shall present can straightforwardly be  generalized to  any social media network).  
In order to define the oracle function based on the above indicators of trust, 
we simply postulate that the higher the number of users that indicate trust in a user $i$, the more likely is the trustworthiness  of $i$. 
We formalize this intuition as an entropy-based oracle function $H$, in such a way that for any user $i$: 
$$
H(i) = - \frac{1}{\log |\V_i|}  \sum_{j \in \V_i}  p_j \log p_j 
$$ 
with $p_j = ET(j,i)/ (\sum_{k \in \V_i} ET(k,i))$, where $\V_i$ is the set of neighbors of node $i$, and $ET(j,i)$ is the  empirical trust function measuring the number of implicit trust statements (i.e., favorites) assigned by node $j$ to node $i$.   
A user $i$ will be regarded as ``good'' if  the corresponding $H(i)$ belongs to the  third quartile  of the distribution of $H$ values over all users.\footnote{Inferring and modeling trust in OSNs is a  challenging topic per se: more refined alternatives  to our entropy-based inference of trust  can certainly be found.}

It is  important to point out that TrustRank requires a graph model with edge orientation that is inverse with respect to LurkerRank. That is, if  $i$ likes a post by $j$, an edge from $j$ to $i$ ($j \rightarrow i$)  is created in the LurkerRank graph, whereas the opposite ($i \rightarrow j$)  is created in the TrustRank graph, as $i$ indicates trust in $j$.

\begin{table}[t!]
\caption{Comparative performance (Kendall tau rank correlation) of TrustRank-biased LurkerRank methods against original TrustRank and LurkerRank methods, on \data{Flickr}.}
\centering 
\scalebox{0.7}{
\begin{tabular}{|l||c||c|c|}
\hline
  &  \algo{LR}  &  \algo{trust-LR}   &  \algo{trust-LR}         \\
    &    vs. \algo{TrustRank}    &   vs. \algo{TrustRank}  &     vs. \algo{LR}
\\ \hline \hline
\algo{LRin}         & .393 & .436 & .639 \\
\algo{LRout}      & \HL{.562}  & .556 & \HL{.980} \\
\algo{LRin-out} & .441 & .640 & .688 \\
\algo{ac-LRin}   & .445 & .434 & .728 \\
\algo{ac-LRout} & .561 & .559 & .945 \\
\algo{ac-LRin-out} & .402   & \HL{.724} & .498 \\ 
\hline
\end{tabular}
}

\vspace{-.5em}
\scriptsize
\begin{flushleft}
Bold values refer to the highest scores per method. 
\end{flushleft}
\label{tab:kendallTrustRank}
\end{table}

Table~\ref{tab:kendallTrustRank} summarizes  Kendall correlation values obtained on \data{Flickr} by a pairwise comparison between our LurkerRank methods, their TrustRank-biased versions (denoted as \algo{trust-LR}), and the original Trust\-Rank. 
Several observations stand out. 
First, looking at the first-column group of results, all LurkerRank methods showed positive correlation with TrustRank. This is   interesting   as it would indicate that the trustworthiness of users is likely to be considered when ranking  lurkers;  
note that the LurkerRank behavior against untrustworthy users or spammers was already observed in our qualitative evaluation (cf. Section~\ref{sec:qualitative}).  
By personalizing a LurkerRank method with TrustRank, the   correlation with TrustRank itself generally increased (up to 0.72), as we expected. More interestingly, \algo{trust-LR} methods still showed a strong correlation with their respective original LurkerRank methods. This   suggests   that introducing a trust-oriented bias in LurkerRank methods    would not significantly decrease their  lurker ranking effectiveness while also accounting for the user trustworthiness.

\section{Conclusion}
\label{sec:conclusion}
We addressed the previously unexplored   problem of     ranking lurkers in an OSN. 
We introduced a topology-driven lurking definition that rely on three basic principles to model lurking in a network, namely overconsumption, authoritativeness of the information received, and non-authoritativeness of the information   produced.  
We proposed various lurker ranking models,   for which we provided a complete specification in terms of the well-known PageRank and alpha-centrality.   
 We have been positively impressed by  results achieved on a number of real-world networks      
 by some of our lurker ranking methods, especially in terms of significance  and higher meaningfulness with respect to other competing methods.  
 Future directions of research have also been issued.  
 
\section*{Acknowledgements}
The final publication is available at Springer via \\ http://dx.doi.org/10.1007/s13278-014-0230-4.

\bibliographystyle{spmpsci}
\bibliography{LRrefs}

\end{document}